\documentclass[aps,prl,twocolumn,showpacs,superscriptaddress,groupedaddress]{revtex4-2}  
\usepackage[T1]{fontenc}
\usepackage{graphicx}  
\usepackage{dcolumn}   
\usepackage{bm}        
\usepackage{amssymb}   
\usepackage{amsmath}
\usepackage{bbold}
\usepackage{array}
\usepackage{makecell}
\usepackage{braket}
\usepackage{titlesec} 
\usepackage[colorlinks,citecolor=red,urlcolor=blue,hypertexnames=true]{hyperref}
\usepackage{subfigure}

\usepackage[usenames,dvipsnames,svgnames,table]{xcolor} 

\begin{document}

\widetext


\title{\textcolor{Sepia}{\textbf{\Large Circuit Complexity as a novel probe of Quantum Entanglement:  \\A study with Black Hole Gas in arbitrary dimensions}}}
\author{Kiran Adhikari${}^{1}$,~Sayantan Choudhury${}^{2,3*}$, \\~Satyaki Chowdhury${}^{2,3}$, ~ K. Shirish ${}^{4}$, ~Abinash Swain ${}^{5}$}
\thanks{{\it Corresponding author,}\\
	{{ E-mail:sayantan.choudhury@niser.ac.in,  sayanphysicsisi@gmail.com}}}
~~~~~~~~	\\


\affiliation{${}^{1}$ 
Institute for Theoretical Particle Physics and Cosmology(TTK), RWTH Aachen University, D-52056, Aachen, Germany}
\affiliation{${}^{2}$National Institute of Science Education and Research, Jatni, Bhubaneswar, Odisha - 752050, India.}
\affiliation{${}^{3}$Homi Bhabha National Institute, Training School Complex, Anushakti Nagar, Mumbai - 400085,
	India.}
\affiliation{${}^{4}$Visvesvaraya National Institute of Technology, Nagpur, Maharashtra, 440010, India}
\affiliation{${}^{5}$Department of Physics, Indian Institute of Technology Gandhinagar, Palaj, Gandhinagar- 382355, India
}

\begin{abstract}
In this article,  we investigate the quantum circuit complexity and entanglement entropy in the recently studied black hole gas framework using the two-mode squeezed states formalism written in arbitrary dimensional spatially flat cosmological Friedmann-Lema$\hat{i}$tre-Robertson-Walker (FLRW) background space-time.  We compute the various complexity measures and study the evolution of these complexities by following two different prescriptions viz.  Covariant matrix method and Nielsen’s method.  Independently,  using the two-mode squeezed states formalism we also compute the Rényi and Von-Neumann entanglement entropy,  which show an inherent connection between the entanglement entropy and quantum circuit complexity.  We study the behaviour of the complexity measures and entanglement entropy separately for three different spatial dimensions and observe various significant different features in three spatial dimensions on the evolution of these quantities with respect to the scale factor.  Furthermore,  we also study the underlying behaviour of the equilibrium temperature with two of the most essential quantities i.e.  rate of change of complexity with scale factor and the entanglement entropy. We observe that irrespective of the spatial dimension, the equilibrium temperature depends quartically on entanglement entropy.

\end{abstract}

\pacs{}
\maketitle
\section*{\textcolor{Sepia}{\textbf{ \Large Introduction}}}
\label{sec:introduction}
Circuit Complexity has become a helping hand to not only the high energy physics community but also to the people from other branches as well \cite{Chapman:2018dem,Chapman:2018lsv,Cano:2018aqi,Barbon:2018mxk,Flory:2018akz,Chapman:2018bqj,Agon:2018zso,Goto:2018iay,Bernamonti_2020,Caceres_2020,Bernamonti_2019,Goto_2019,Guo_2018,Bhattacharyya:2018bbv,Khan:2018rzm,Hackl:2018ptj,Alves:2018qfv,Camargo:2018eof,Camargo:2019isp,Chapman:2018hou,Chapman:2017rqy,Doroudiani:2019llj,Hashemi:2019aop,Choudhury:2020lja}. This quantum information theory technique has been significantly used recently to probe many features which were previously difficult. Though this concept is a computation tool, its contribution in the field of physics of late is massive. It provides a way to probe physics behind the horizon of black holes through the use of the "Complexity=Volume" and "Complexity=Action" conjectures  \cite{Susskind:2018pmk,Stanford:2014jda,Susskind:2014rva,Roberts:2014isa,Susskind:2014jwa}. Since then it has been extensively used in quantum field theory  and in studies involving AdS/CFT correspondence \cite{Jefferson:2017sdb}. These holographic approaches connect a probe on the gravity side with a concept of quantum information theory. 

In the recent past, along with the out-of-time ordered correlation functions \cite{Choudhury:2020yaa,Bhagat:2020pcd, Choudhury:2021tuu,Hashimoto:2017oit,Hashimoto:2020xfr}, it has formed the web of quantum chaos. It has been found to reveal essential information like Lyapunov exponent, scrambling time etc. required to diagnose chaos in a system. Many interesting works have been done using this tool in wide areas of physics. It was studied for cosmological islands in \cite{Choudhury:2020hil}, where the authors showed that entanglement entropy from circuit complexity via a famous relation proposed in \cite{Susskind:2018pmk} resembles the page curve in some particular regime where a universal relation between circuit complexity, OTOC and entanglement entropy can be written.  It was used to study early universe chaos within the framework of bouncing cosmology \cite{Bhargava:2020fhl}. People have also computed circuit complexity in the context of supersymmetric quantum field theory \cite{Choudhury:2021qod}. The connection between entanglement and emergence of space-time has been an active area of study, where the entanglement entropy is the minimum cross-sectional area of an Einstein-Rosen Bridge(ERB). However, classically the ERB continues to grow for a very long time, whereas the dual thermodynamic system comes to a thermal equilibrium quickly. This led Susskind to introduce a new variable namely Complexity which could be responsible for the ERB growth \cite{Susskind:2018pmk}. Complexity of a quantum system is a real quantity and its growth rate has been conjectured to be proportional to the entropy of the black hole based on these observations. It was very recently shown in \cite{Eisert:2021mjg} that there exists some relationship between entangling power and circuit complexity.  Most importantly, if the entanglement entropy grows linearly with time, the geometric circuit complexity also grows linearly.

In this paper, we will study the evolution of complexity w.r.t. the entanglement entropy of the blackhole gas model \cite{Mathur:2020ivc}. 
An important feature of this model is that the total entropy of the blackhole gas is directly proportional to the volume of the system instead of the area and the system behaves like a thermodynamic gas. We will compute the most common measures for determining the entanglement between the squeezed states of blackhole gas, namely the von Neumann and Rényi entropy, which quantifies the amount of uncertainty linked to the density matrix. We will measure the entanglement entropy by constructing an effective thermal representation for the reduced single-mode state from the two-mode squeezed state which varies linearly w.r.t. to squeezing parameter $r$. Then by using the scale factor predicted by the blackhole gas model in the flat space-time metric as a dynamical variable we will study the evolution of complexity in three spatial dimensions and compare it with the entanglement entropy in terms of squeezed state parameters. One of the reasons we are interested in the blackhole gas is that its equation of state describes a universe right after the big bang and before the start of inflation, if one wants to avoid such an equation of state governing radiations of very high densities then one needs to have inflation right from the Planck scale.
\\
\underline{\textbf{Key Highlights}}

\begin{itemize}
	\item The behaviour of circuit complexity calculated from two different approaches viz. Covariance matrix method and Nielsen's wave function method has been studied w.r.t. scale factor for the blackhole gas model. We observe interesting behaviours for different spatial dimensions.
	\item The behaviour of the Von Neumann entropy and the Rényi entropy has been studied w.r.t. the scale factor for different spatial dimensions. We observe similar features for $d = 1,2$ whereas for $d=3$ we observe slightly different behaviour.
	\item The behaviour of $dC/da$ w.r.t. von Neumann and Rényi entropy has been studied for different spatial dimensions. It is shown that by no means is it a linear function.
	\item The behaviour of the equilibrium temperature for the blackhole gas model is identical compared to entropy for different spatial dimensions whereas it seems dependent on spatial dimension when compared to $dC/da$.
	
\end{itemize}
\underline{\textbf{Organization of the paper}}
The organization of the paper is as follows:
\begin{itemize}
	\item We begin by providing a review of the Black Hole gas given by Samir Mathur in \cite{Mathur:2020ivc} in the section  \hyperref[sec:blackholegas]{A Short Note on Black Hole Gas}. Solving the Friedmann equation for blackhole gas, we relate the scale factor $a(t)$ with spatial dimension. 
	
	\item In sec:2 \hyperref[sec:blackholeper]{Black hole gas perturbation theory in d+1 dimensions}, we investigate the squeezed state formalism by perturbing the black hole gas geometry in FLRW spatially flat spacetime..
	
	\item In sec:3 \hyperref[sec:CircuitComplexity]{A Short Note on Circuit Complexity}, We review of the Circuit Complexity in the section. We discuss the geometric framework of circuit complexity developed by Nielsen and collaborators. 
	
	\item In sec:4 \hyperref[sec:twomode]{Circuit Complexity of two-mode
		squeezed states}, after introducing the notion of squeezed states, we calculate it's circuit complexity using two approaches: Complexity using the covariance matrix and Nielsen's method of wave functions.
	
	\item In sec:5 \hyperref[sec:ee]{Entanglement entropy of two-mode
		squeezed states}, we compute the entanglement entropy of two-mode squeezed states. We compute Rényi-entropy, Von-Neumann entropy and Rényi-2 entropy. We find entanglement entropy grows linearly with increasing squeezing parameters for the short time period. Then, we compare entanglement entropy with circuit complexity obtained in sec:1.
	
	\item In sec:6 \hyperref[sec:numerical]{Numerical Results}, we numerically study the behaviour of the complexity measures and entanglement entropy separately for three spatial dimensions (d= 1,2 and 3).
	
	\item In sec:7 \hyperref[sec:Conclusions]{Conclusions}, we conclude with some discussions. 
\end{itemize}
\section*{\textcolor{Sepia}{\textbf{ \Large A Short Note on Black Hole Gas}}}
\label{sec:blackholegas}
In this section, we review a model proposed in ref.~\cite{Mathur:2020ivc}, where the author have studied the state of a system moving towards maximal entropy $S$.  Note that this system is unlike the inflationary model where we have a low entropy state after the inflation because the positive energy of the matter content is compensated by the gravitational potential. Here, we provide a quick derivation of the equation of state that describes the pre-inflationary universe. We will consider a configuration where we find entropy $S(E, V)$ of a system in a toridal box of volume $V$ in the limit $E\rightarrow \infty$. At low enough energies the matter phase corresponds to radiation whose entropy as a function of $E$ and $V$ in dimension $d$ can be given by $S \approx V\rho^{\frac{d}{d + 1}}$. By putting more energy into the box one can look for a configuration where the system turns into a blackhole of radius $R$ whose entropy is given by
\begin{equation}
S_{hole} = \frac{A}{G}\label{1}
\end{equation}
One might suspect that for a given box of radius $R$, eqn. \eqref{1} describes the state with maximum entropy for energy $E = E_{bh}$, since throwing more energy into the black hole will only result in increasing the hubble expansion. However if we let go the constraint that the energy inside the box shouldn't be greater than the mass of the blackhole inside the box then one could arrive at a configuration where the entropy of the system is greater than \eqref{1}.  This could be achieved by putting $N$ number of black hole each of radius $R$ in a lattice instead of a single box of volume $V$. The number of black holes in such a configuration is given as
\begin{equation}
N_{hole} = \biggl(\frac{V}{R^{d}}\biggr)
\end{equation}
Therefore the total entropy of the system is
\begin{equation}
S = N_{hole}~S_{hole} = \biggl(\frac{V}{R^{d}}\biggr)\biggl(\frac{R^{d-1}}{G}\biggr) = \frac{V}{RG}\label{2}
\end{equation}
We notice that \eqref{2} is in contrast with \eqref{1} where the entropy is directly proportional to the volume rather than the area of the horizon. The energy leading to such a state is undoubtedly greater than $E_{bh}$ and could be expressed as follows
\begin{equation}
E = N_{hole}~E_{hole} = \biggl(\frac{V}{R^{d}}\biggr)\biggl(\frac{R^{d-2}}{G}\biggr) = \frac{V}{R^{2}G}\label{3}
\end{equation}
substituting the value of $R$ from \eqref{3} to \eqref{2} we get
\begin{equation}
S = \biggl(\frac{V}{EG}\biggr)^{-\frac{1}{2}} 
\end{equation}
for $\rho =E/V$ we get, 
\begin{equation}
S = K\sqrt{\frac{\rho}{G}}V \label{6}
\end{equation}
This contrast in the definition of entropy is subjected to the constraint that microstates cannot expand freely to larger size unlike in asymptotically flat space where the entropy is given by the area law. Also the resulting lattice configuration with $E > E_{bh}$, having $N$ number of black holes wouldn't collapse to form one large blackhole as the entropy corresponding to the lattice configuration is larger than a single black hole state in a box of volume $V$. 

Now from the first law of thermodynamics we can show
\begin{equation}
T = \biggl(\frac{\partial S}{\partial E}\biggr)^{-1} = \frac{2}{K}\sqrt{\frac{EG}{V}}
\end{equation}
\begin{equation}
p = T\biggl(\frac{\partial S}{\partial V}\biggr) = \frac{E}{V} = \rho \label{4}
\end{equation}
Now from \eqref{4} we see that equation of state takes the form $\rho = wp$ with $w = 1$. 
The solution for the black hole gas model can be obtained by starting with the FLRW flat metric in 1+d dimensions which is given by the following line element
\begin{equation}
ds^2 = - dt^2 + a^2(t) d\vec{x}^2
\end{equation}
Solving the Friedmann equation for black hole gas in the $d + 1$ dimensional flat metric  with scale factor $a(t)$ we get
\begin{eqnarray}
a(t) = a_{0}t^{\displaystyle 1/d}
\end{eqnarray}
The above FLRW flat metric can be written in terms of the conformal time scale by using the following conversion relation
\begin{equation}
d\tau = \frac{dt}{a(t)}
\end{equation}
Further integrating the both sides of the above equation we get the following relationship between the physical time $t$ and the conformal time $\tau$ in the arbitrary $d+1$ dimension black hole gas:

\begin{equation}
t =\left\{
\begin{array}{ll}
\displaystyle \exp(a_0 \tau)  & \text{$d$ = 1}\\
\displaystyle \left(\frac{a_0 (1-d) }{d}\right)^{\displaystyle \frac{d}{d-1}}~\tau^{\displaystyle \frac{d}{d-1}}  & \text{$d$ > 1}	 
\end{array}
\right. 
\end{equation}


This relationship is extremely useful for the present computation which helps us the directly translate the information in terms of desired conformal time from physical time.

In this conformal time coordinates, the flat FLRW line element gets transformed as 
\begin{equation}
ds^2 = a^2(\tau) (-d\tau^2 + d\vec{x}^2 )
\end{equation}

Hence,  the above solution of the black hole gas scale factor can be written in terms of conformal time as follows 

\begin{equation}
a(\tau) =\left\{
\begin{array}{ll}
\displaystyle a_0 \exp(a_0 \tau)  & \text{$d$ = 1}\\
\displaystyle	a_0 \left(\frac{a_0 (1-d) }{d}\right)^{\displaystyle \frac{1}{d-1}}~\tau^{\displaystyle \frac{1}{d-1}}  & \text{$d$ > 1}	 
\end{array}
\right. 
\end{equation}
The corresponding Hubble parameter with respect to the conformal time scale can also be written as:
\begin{equation}
{\cal H}(\tau) =\left\{
\begin{array}{ll}
\displaystyle a_0  & \text{$d$ = 1}\\
\displaystyle	\frac{1}{d-1}~\frac{1}{\tau}~~~~~~~~~ & \text{$d$ > 1}	 
\end{array}
\right.
\end{equation}
Now we will describe an equivalent scenario inside a black hole gas where one can obtain the above mentioned scale factor and Hubble parameter in any arbitrary $d+1$ dimensions.  In this scenario we embed a scalar field within the framework of Einstein gravity having the previously mentioned spatially flat FLRW space-time,  where the solution of the scale factor is exactly same as mentioned earlier. In this framework the representative action of the scenario is described as:
\begin{eqnarray}
S=\frac{1}{2}\int d^{d+1}x~\sqrt{-g}~\bigg[R-(\partial \phi)^2-V(\phi)\bigg]
\end{eqnarray}
where we have fixed the Planck mass $M_p=1$ for the present computation.  First term in the action represent the usual Einstein Hilbert term,  second term represents the kinetic term of the embedded scalar filed and the last term represents the potential function in a black hole gas $V(\phi)$ in any arbitrary $d+1$ dimensions.  We have found the the following two possibilities of the potential functions are allowed in the present context which can finally give rise to same same scale factor which we have mentioned earlier:
\begin{equation}
\label{Vphi}
V(\phi) =\left\{
\begin{array}{ll}
\displaystyle \frac{3}{d^2}~\exp(-\sqrt{2d}~\phi)  & \text{Choice~I} \\
\displaystyle	\frac{3}{d^2}~\exp(\sqrt{2d}~\phi) & \text{Choice~II}
\end{array}
\right.
\end{equation}
Here it is important to note that,  since we have embedded the scalar field in the homogeneous and isotropic spatially flat FLRW background it turns out to be the field is only function of the time coordinate.

Further solving the Klein Gordon field equation in $d+1$ dimension spatially flat FLRW background the dynamical solution of the homogeneous and isotropic background scalar field $\phi$ in terms of the conformal time coordinate can be expressed as: 
\begin{eqnarray}
\phi(\tau)=\mp \frac{\sqrt{2}}{d-1}~\ln\bigg(\frac{a_0 (d-1)}{d}~\tau\bigg). 
\end{eqnarray}
This solution actually representing the dynamics of the field inside the black hole gas in $d+1$ dimension spatially flat FLRW background.
\begin{figure}[h!]
	\centering
	\includegraphics[width=9cm,height=8.5cm]{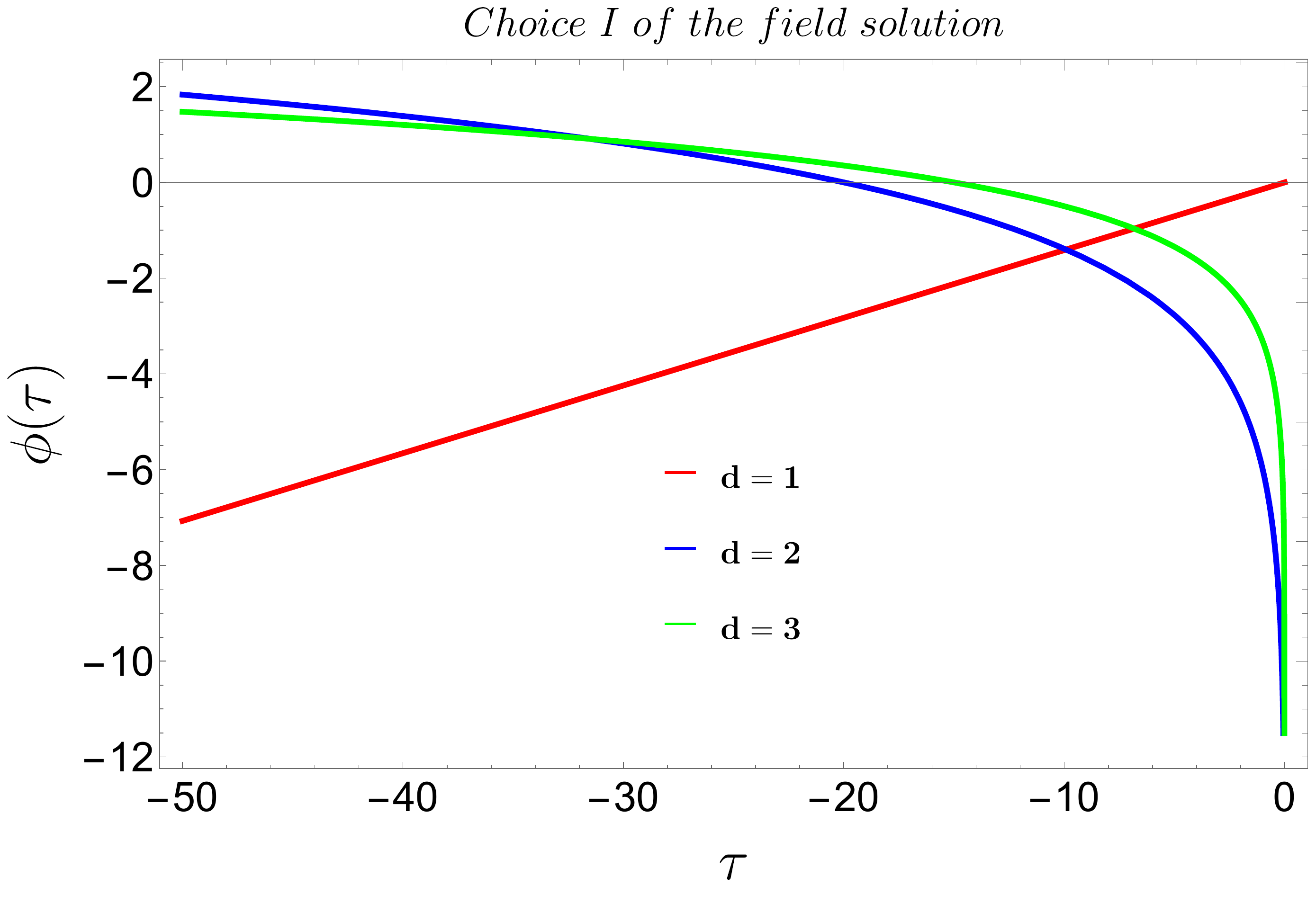}
	\caption{Behaviour of the first choice of the field solution w.r.t conformal time for different spatial dimensions.}
	\label{fig_field1}
\end{figure}

\begin{figure}[h!]
	\centering
	\includegraphics[width=9cm,height=8.5cm]{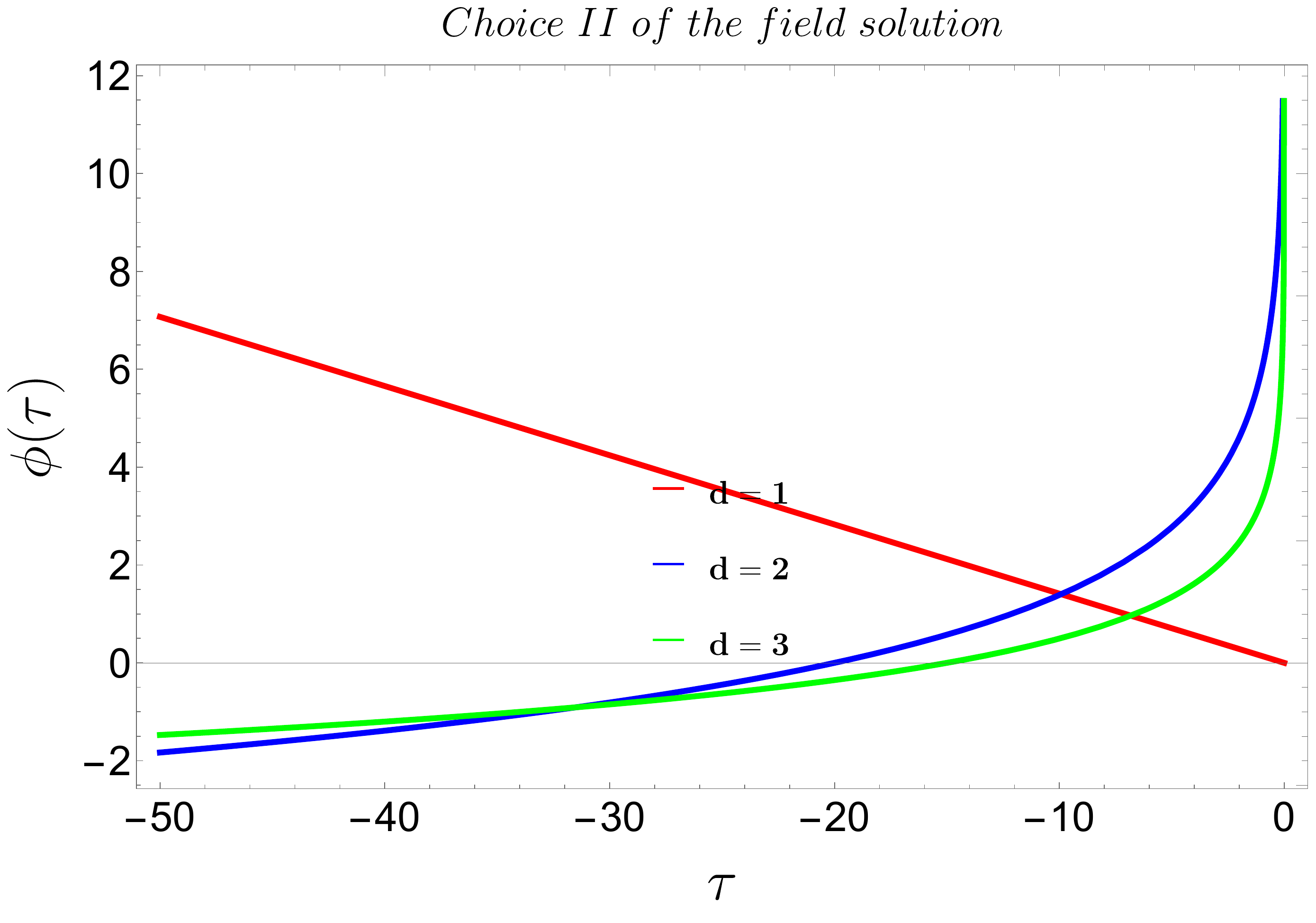}
	\caption{Behaviour of the second choice of the field solution w.r.t conformal time for different spatial dimensions.}
	\label{fig_field2}
\end{figure}
Now using this solution one can find out the following conformal time dependence of the potential function in the context of black hole gas:
\begin{equation}
V(\tau) =\left\{
\begin{array}{ll}
\displaystyle 3 \exp(-2 a_0 \tau) & \text{d=1} \\ \\
\displaystyle \frac{3}{d^2}~\bigg(\frac{a_0 (1-d)}{d}~\tau\bigg)^{\displaystyle-~\frac{2d}{d-1}}  & \text{d>1}
\end{array}
\right.
\end{equation}
In Fig. \ref{fig_field1} and \ref{fig_field2}, we have plotted the two choices of fields with respect to the conformal time. In choice I, we see that for $d=1$ the field keeps on increasing monotonously and for higher dimensions $i.e.$ $d=2,3$ the field value decreases slowly initially w.r.t. conformal time and then rapidly as we increase the value of conformal time. The exact opposite behaviour can be seen for choice II. For $d=1$ the field value keeps on decreasing monotonously, where as for $d=2,3$ field value grows steadily and then at a faster rate with conformal time.

\begin{figure}[h!]
	\centering
	\includegraphics[width=9cm,height=8.5cm]{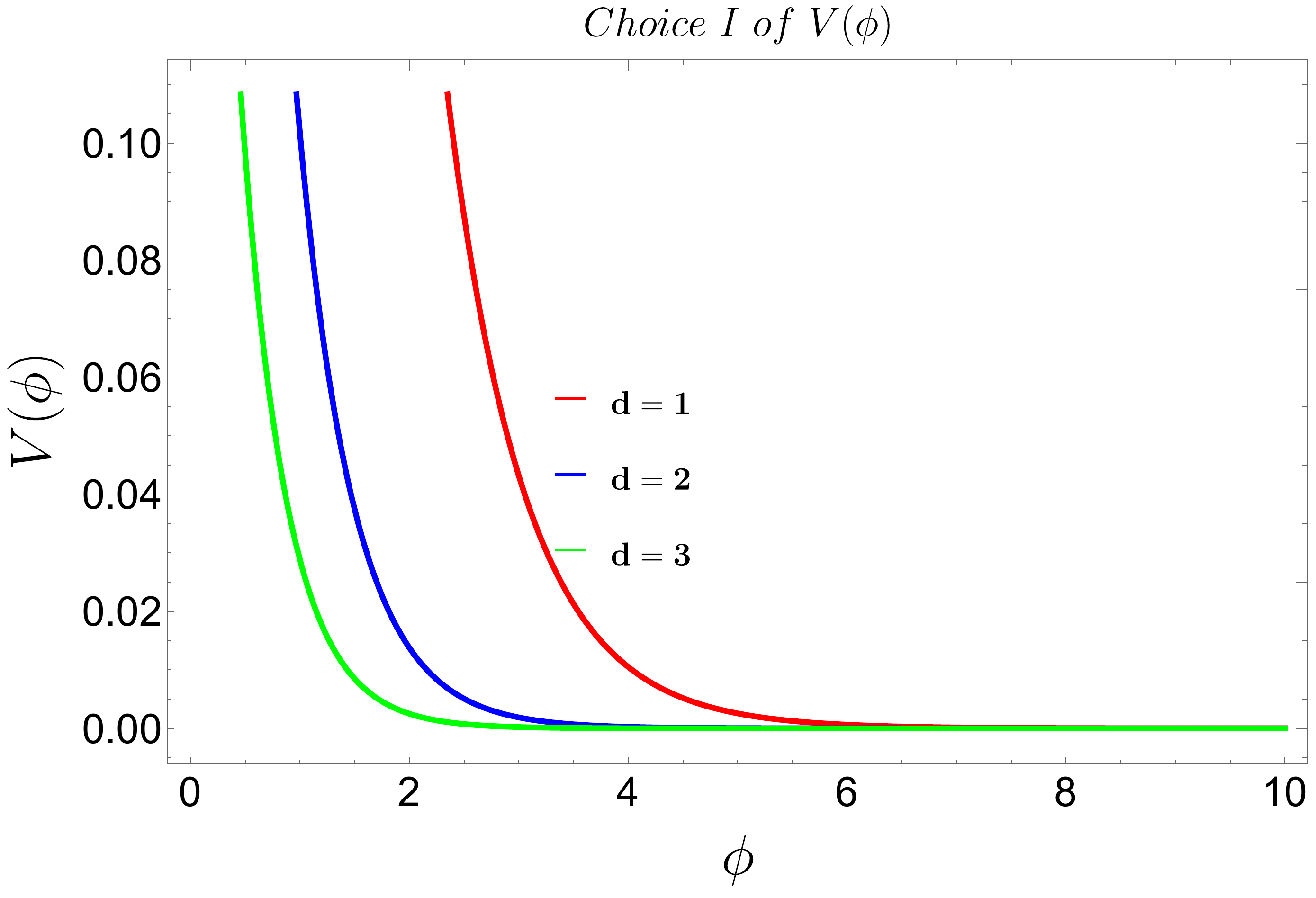}
	\caption{Behaviour of the first choice of the potential w.r.t. the field for different spatial dimensions.}
	\label{fig_pot1}
\end{figure}

\begin{figure}[h!]
	\centering
	\includegraphics[width=9cm,height=8.5cm]{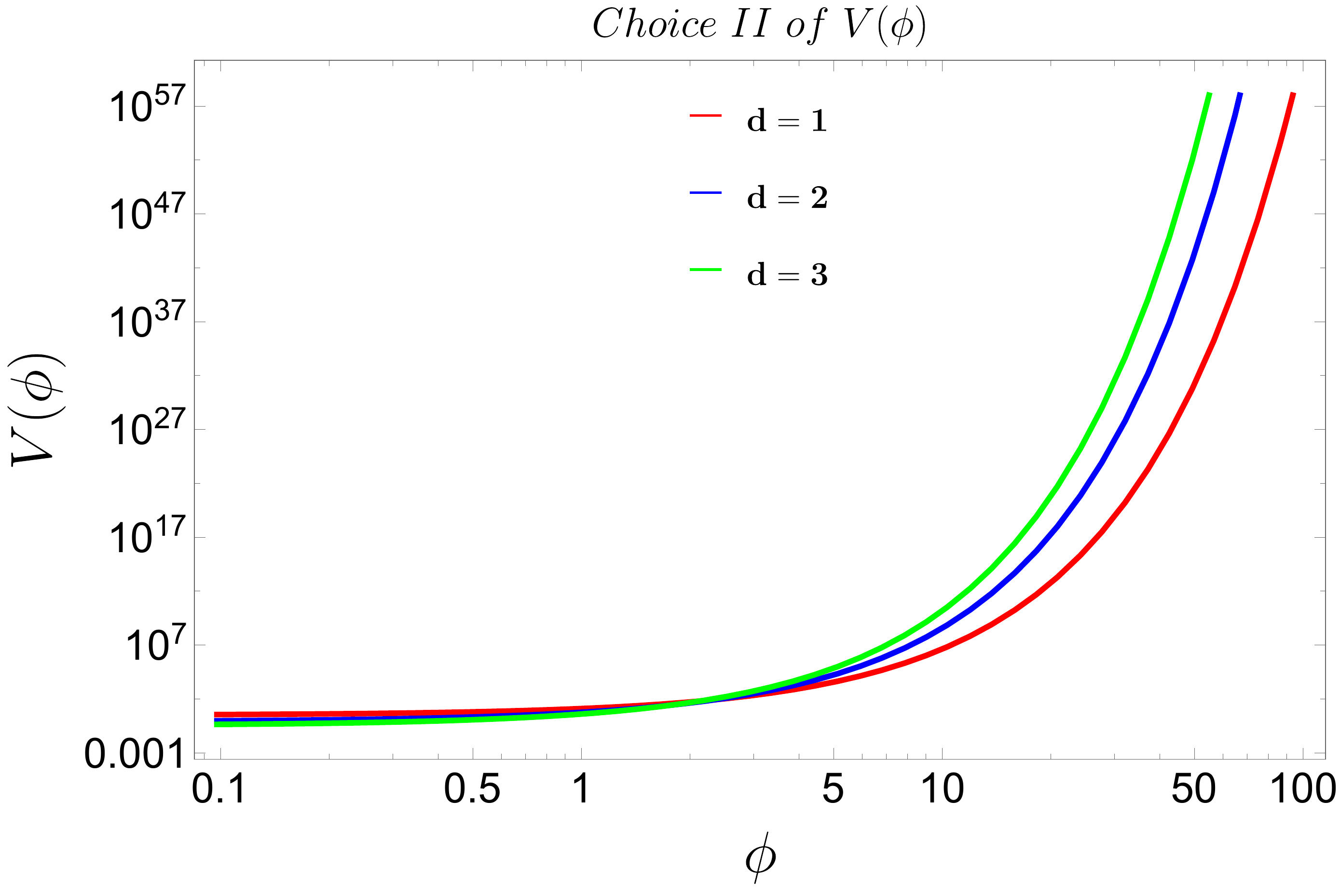}
	\caption{Behaviour of the second choice of the potential w.r.t. the field for different spatial dimensions.}
	\label{fig_pot2}
\end{figure}

In Fig. \ref{fig_pot1} and \ref{fig_pot2}, we have plotted potential for the two choices against the field. As can be seen from the Eq. \eqref{Vphi}, the potential function takes the form of exponentially decreasing and increasing function for first and second choices respectively.

\begin{figure}[h!]
	\centering
	\includegraphics[width=9cm,height=8.5cm]{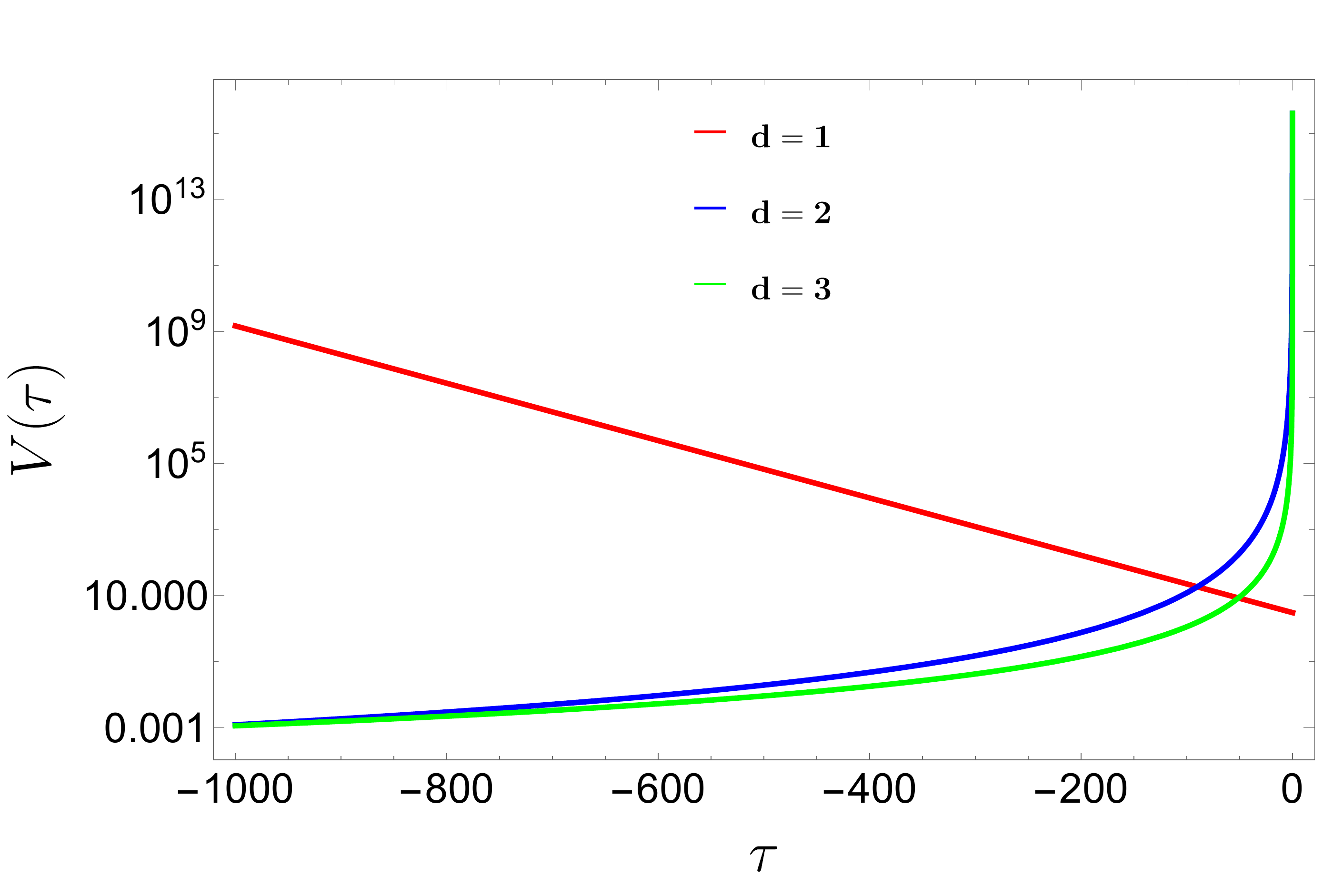} 
	\caption{Behaviour of the potential w.r.t. the conformal time. Here we have taken logarithm along the vertical axis}
	\label{fig_pottau}
\end{figure}

In Fig. \ref{fig_pottau}, we have drawn potential function with respect to conformal time. We have taken the logarithm for the potential. It can be seen that for $d=1$, the potential goes on decreasing continuously in a straight line where as for $d=2,3$ the potential rises quickly after going through a slow rise period in the beginning.

\begin{figure}[h!]
	\centering
	\includegraphics[width=9cm,height=8.5cm]{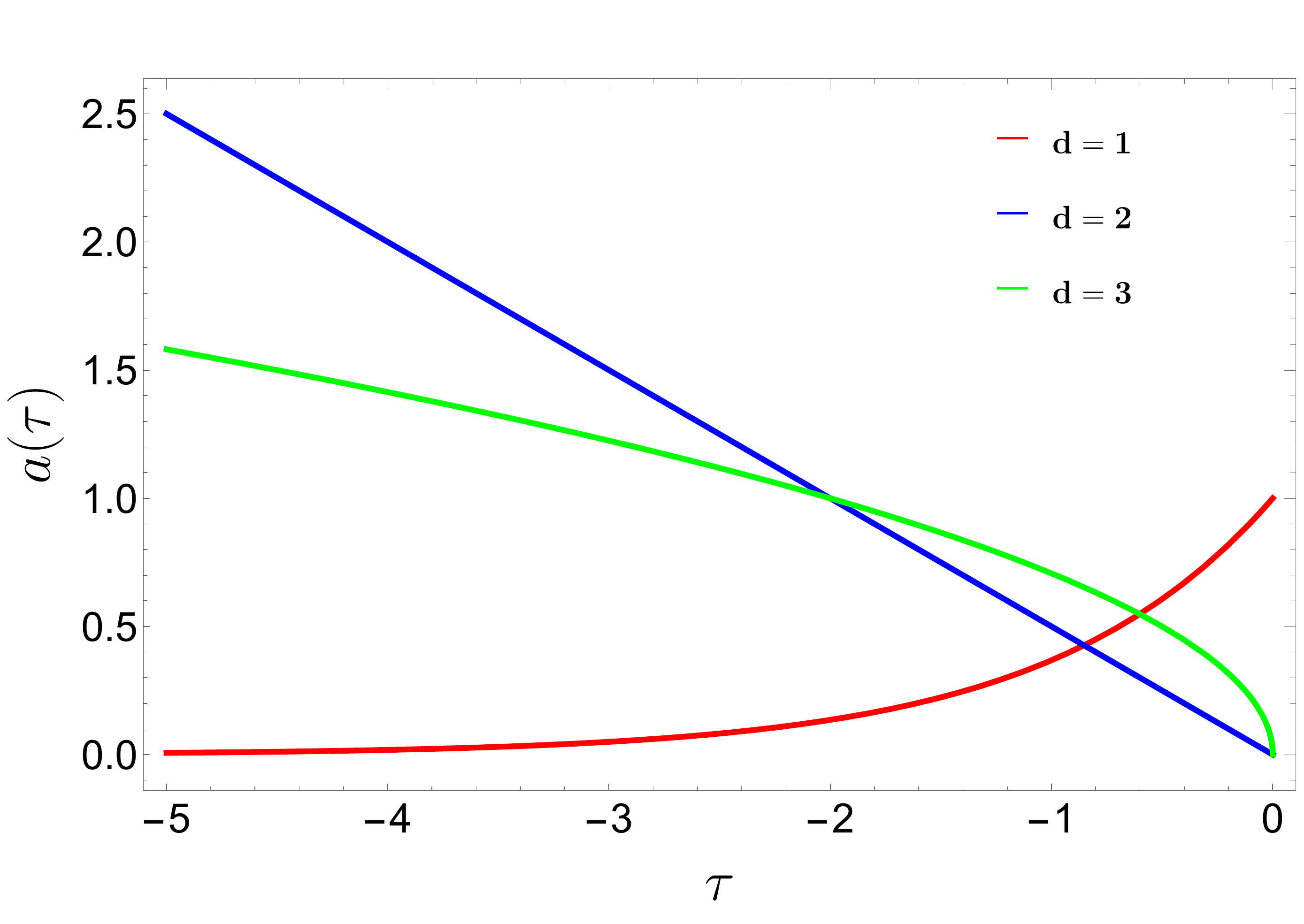}
	\caption{Behaviour of the scale factor $a(\tau)$ w.r.t. the conformal time for different spatial dimensions. }
	\label{fig_scale}
\end{figure}
\begin{figure}[h!]
	\centering
	\includegraphics[width=9cm,height=8.5cm]{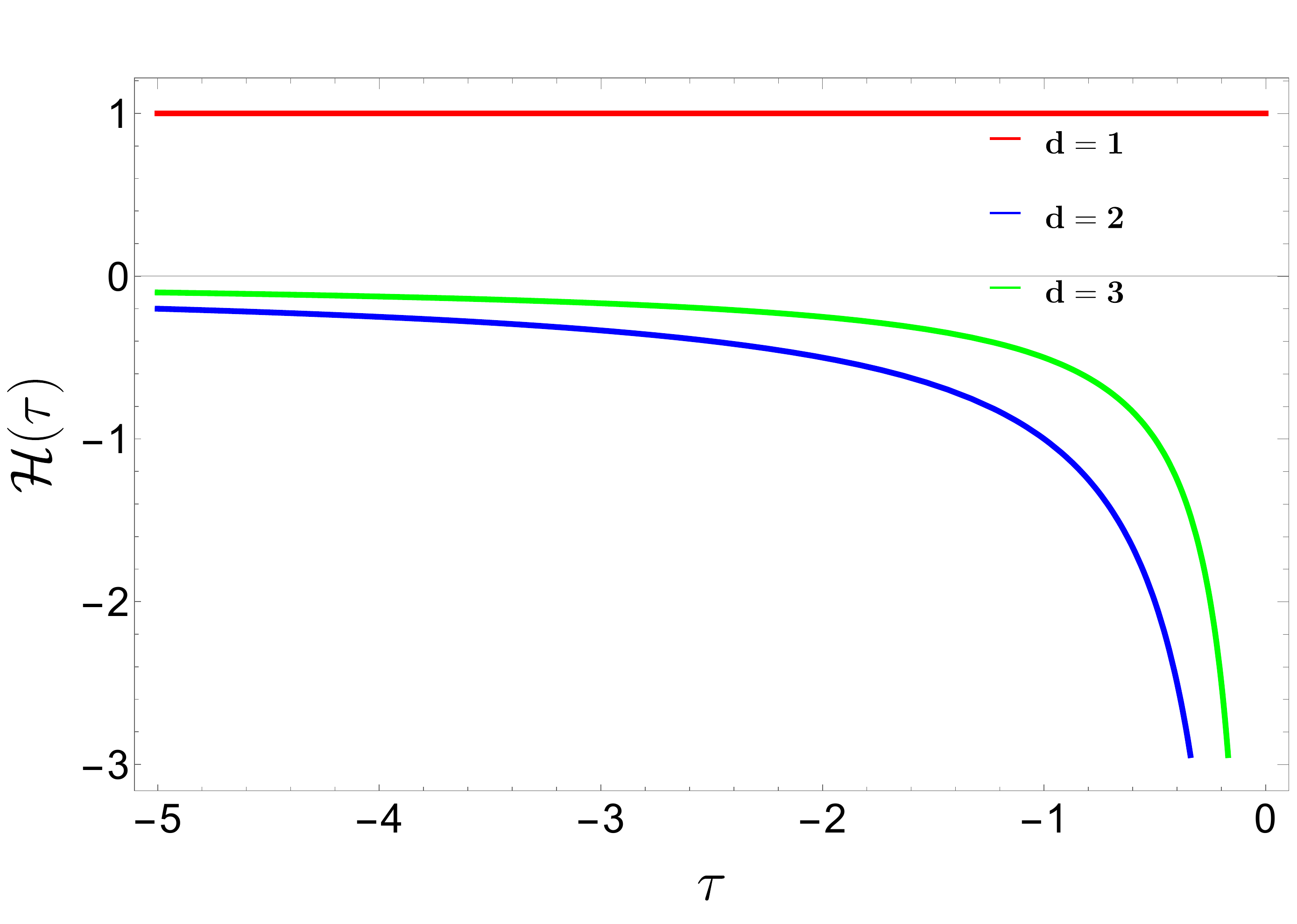}
	\caption{Behaviour of the Hubble parameter $\mathcal{H}(\tau)$ w.r.t. the conformal time for different spatial dimensions.}
	\label{fig_script}
\end{figure}
In fig: \ref{fig_scale}, we have plotted the scale factor of the black hole gas model w.r.t the conformal time scale. The plots have been done by fixing the value of the constant $a_0$ to 1. We observe a significant difference in the behavior of the scale factor for the spatial dimension d=1 and the higher spatial dimensions. The scale factor corresponding to the spatial dimension d=1 shows an increasing behavior in the late time scales. The scale factor for higher spatial dimension shows a decreasing behavior. The decrease is linear for the spatial dimension d=2 whereas it is non-linear for spatial dimension d=3.

In fig: \ref{fig_script}, we have plotted the behavior of the Hubble parameter \bigg($\mathcal{H}(\tau)=\frac{a'(\tau)}{a(\tau)}$\bigg) w.r.t the conformal time scale. It can be seen that the Hubble parameter is just a positive constant for the spatial dimension d=1, whereas it exhibits a decreasing behavior approaching negative infinity at late conformal time scales (near $\tau$=0) for the other spatial dimensions. Moreover, the value taken by the Hubble parameter for the higher spatial dimensions is always negative.

One could also arrive to \eqref{6} by using the T and S duality symmetries of the string theory, this seems interesting because it doesn't require the state of blackhole gas to be state near the big bang since microstates of stringy theory could  also describe the microstates of blackhole.
\section*{\textcolor{Sepia}{\textbf{ \Large Black hole gas perturbation theory in d+1 dimensions}}}
\label{sec:blackholeper}
In this section we will study squeezed state formalism within the framework of black hole gas theory for FLRW spatially flat background. 
In this context one needs to consider the following perturbation in the scalar field:
\begin{equation} 
\phi({{\vec{x}}},\tau)=\phi(\tau)+\delta\phi({{\vec{x}}},\tau)~~~~~~~~~
\end{equation}
and to express the whole dynamics in terms of a gauge invariant description through a variable:
\begin{equation}
\zeta({{\vec{x}}},\tau)=-\frac{{\cal H}(\tau)}{\displaystyle\left(\frac{d\phi(\tau)}{d\tau}\right)}\delta\phi({{\vec{x}}},t).~~~~~~~~\end{equation}
At the level of first order perturbation theory in a spatially flat FLRW background metric, we fix the following gauge constraints:
\begin{eqnarray}
\nonumber
&&\delta\phi({{\vec{x}}},\tau)=0,\\ \nonumber
&&g_{ij}({{\vec{x}}},\tau)= a^2(\tau)\left[\left(1+2\zeta({{\vec{x}}},\tau)\right)\delta_{ij}+h_{ij}({{\vec{x}}},\tau)\right],\\
&&\partial_{i}h_{ij}({{\vec{x}}},\tau)=0=h^{i}_{i}({{\vec{x}}},\tau),~~~~~~~~
\end{eqnarray}
which fix the space-time re-parametrization. In this gauge, the spatial curvature of constant hyper-surface vanishes, which implies curvature perturbation variable is conserved outside the horizon. 
Applying the ADM formalism one can further compute the second-order perturbed action for scalar modes. The action, after gauge fixing, can then be expressed by the following:
	\begin{eqnarray}
	\nonumber 
	\delta^{(2)}S&=&\frac{1}{2}\int d\tau~d^d{{\vec{x}}}~\frac{a^{d-1}(\tau)}{{\cal H}^2}\left(\frac{d{\phi}(\tau)}{d\tau}\right)^2 \\ && ~~~\left[\left(\partial_{\tau}\zeta({{\vec{x}}},\tau)\right)^2-\left(\partial_{i}\zeta({{\vec{x}}},\tau)\right)^2\right].
	\end{eqnarray}
Now introducing the \textit{Mukhanov variable}, defined as $ v({{\vec{x}}},\tau)=z(\tau)~\zeta({{\vec{x}}},\tau)$ where, $~~z(\tau)=a^{\frac{d-1}{2}}(\tau)\sqrt{2\epsilon(\tau)}$, the second order perturbed action can be rewritten as:
\begin{widetext}
	\begin{equation} 
	\delta^{(2)}S = \int d\tau~d^d{{\vec{x}}} \biggl[ v'^{2}({{\vec{x}}},\tau)-(\partial_{i}v({{\vec{x}}},\tau))^{2} +\biggl(\frac{z'(\tau)}{z(\tau)}\biggr)^{2}v^{2}({{\vec{x}}},\tau) - 2\biggl(\frac{z'(\tau)}{z(\tau)}\biggr)v'({{\vec{x}}},\tau)v({{\vec{x}}},\tau) \biggr],
	\end{equation}
\end{widetext}
where the quantity $\epsilon(\tau)$ is known as the conformal time dependent slow roll parameter and is defined as
\begin{equation}
\epsilon(\tau) = 1- \frac{\mathcal{H}'}{\mathcal{H}^2}=\frac{1}{{\cal H}^2}\left(\frac{d{\phi}(\tau)}{d\tau}\right)^2
\end{equation}
For the black hole gas model, it is very easy to verify that the slow roll parameter is equal to the dimension in which the black hole model is being considered i.e.
\begin{equation}
\epsilon(\tau)=d.
\end{equation}
which is finally independent of conformal time coordinate $\tau$. 

By implementing the following ansatz for the Fourier transformation:
\begin{equation}
\label{eq:fouriermodes}
v({{\vec{x}}},\tau):=\int \frac{d^{d}{{\vec{k}}}}{(2\pi)^{d}}v_{{{\vec{k}}}}(\tau)~\exp(-i{{\vec{k}}}.{{\vec{x}}}),
\end{equation}
the second-order perturbation for the scalar modes in Fourier space can be further recast as:
\begin{eqnarray}
\nonumber 
\delta^{(2)}S &=& \int d\tau~d^d{{\vec{k}}}~ \biggl[ |v'_{{\vec{k}}}(\tau)|^2+\Biggl(k^2+\biggl(\frac{z'(\tau)}{z(\tau)}\biggr)^{2}\Biggr)\\ &&|v_{{\vec{k}}}(\tau)|^{2} - 2\biggl(\frac{z'(\tau)}{z(\tau)}\biggr)v'_{{\vec{k}}}(\tau)v_{-{{\vec{k}}}}(\tau) \biggr],~~~
\end{eqnarray}
Now, varying the above second order action, we get the following equation of motion 
\begin{equation} 
v''_{{\vec{k}}}(\tau)+\omega^2(k,\tau)v_{{\vec{k}}}(\tau)=0.
\end{equation}
The above equation is known as the \textit{Mukhanov-Sassaki} equation with the frequency of the oscillator given by
\begin{equation}
\omega^2(k,\tau):=k^2+m^2_{\rm eff}(\tau)
\end{equation}
The conformal time dependent effective mass in the present computation is given by
\begin{equation}
m^2_{\rm eff}(\tau)=-\frac{z''(\tau)}{z(\tau)}=\frac{1}{\tau^2}\left(\nu^2_{\rm BHG}(\tau)-\frac{1}{4}\right),
\end{equation}
with the conformal time dependent mass parameter for the black hole gas in arbitrary dimension upto leading order terms is given by
\begin{eqnarray}
\nu_{\rm BHG}(\tau)& \approx & \frac{d}{2}-1+\bigg(1-\frac{1}{\epsilon(\tau)}\bigg)\frac{\mathcal{H}''}{\mathcal{H}^2}\nonumber\\
&=& \frac{d}{2}-1-\frac{2}{d}(1-d)^3~\mathcal{H},
\end{eqnarray}
where we can clearly observe that the slowly varying conformal time dependence is appearing from the third term where we have truncated the expansion.  During computing this mass parameter have used the following facts:
\begin{eqnarray}
&&\frac{{\cal H}^{\prime}}{{\cal H}^2}=(1-d),\\
&&\frac{{\cal H}^{\prime\prime}}{{\cal H}^2}=2(1-d)^2~{\cal H}
\end{eqnarray}
where the explicit expression for the Hubble parameter in conformal time coordinate is quoted in the previous section.  Finally using this result we get the following simplified expression for the mass parameter for black hole gas:
\begin{equation}
\label{nubhg}
\nu_{\rm BHG}(\tau)  =\left\{
\begin{array}{ll}
\displaystyle \frac{d}{2}-1-\frac{2}{d}(1-d)^3~a_0 & \text{$d$ = 1} \\ \\
\displaystyle \frac{d}{2}-1+\frac{2}{d}(1-d)^2~\frac{1}{\tau}~~~~~~~~~ & \text{$d$ > 1}
\end{array}
\right.
\end{equation}
A general solution to the equation of motion is written as
\begin{equation}
v_{{\vec{k}}}(\tau):=\sqrt{-\tau}\left[{\cal C}_1~{\cal H}^{(1)}_{\nu_{\rm BHG}}(-k\tau)+{\cal C}_2~{\cal H}^{(2)}_{\nu_{\rm BHG}}(-k\tau)\right]
\end{equation}
where ${\cal H}^{(1)}_{\nu_{\rm BHG}}(-k\tau)$ and ${\cal H}^{(2)}_{\nu_{\rm BHG}}(-k\tau)$ are Hankel functions of the first and second kind,respectively, with argument $-k\tau$ and order $\nu_{\rm BHG}$. The two integration constants can be fixed by the choice of various initial conditions. In this paper, we restrict our choice to only the \textit{Bunch-Davies} vacuum case in which one fixes the initial conditions, ${\cal C}_1$ as 1 and  ${\cal C}_2$ as 0. However, it is generally difficult to work with these full solutions and one takes the asymptotic limits of the solutions which is given by
\begin{eqnarray}
\nonumber
v_{{{\vec{k}}}}(\tau)&=&\frac{2^{\nu_{\rm BHG}-\frac{d}{2}}(-k\tau)^{\frac{d}{2}-\nu_{\rm BHG}}}{\sqrt{2k}}\left|\frac{\Gamma(\nu_{\rm BHG})}{\Gamma\left(\frac{d}{2}\right)}\right|~\left(1-\frac{i}{k\tau}\right)\\ &&\exp\left(-i\left\{k\tau+\frac{\pi}{2}\left(\nu_{\rm BHG}-\frac{d}{2}\right)\right\}\right)f_d(k,\tau).~~~~~~~
\end{eqnarray}
where the function $f_d(k,\tau)$ takes the following form
\begin{eqnarray}
\nonumber
f_d(k,\tau)&=& 2^{\frac{d-3}{2}} (-k \tau)^{\frac{3-d}{2}}\bigg|\frac{\Gamma(d/2)}{\Gamma(3/2)}\bigg|\\ &&~~~~~~~~~~~~~\exp\bigg\{-i\bigg({\frac{\pi}{2}\bigg(\frac{d-3}{2}\bigg)}\bigg)\bigg\}.
\end{eqnarray}
The normalization with the factor $f_d(k,\tau)$ has been done in such a way that for spatial dimension $d=3$,  it becomes $f_3(k,\tau)=1$.
\begin{figure}[h!]
	\centering
	\includegraphics[width=9cm,height=8.5cm]{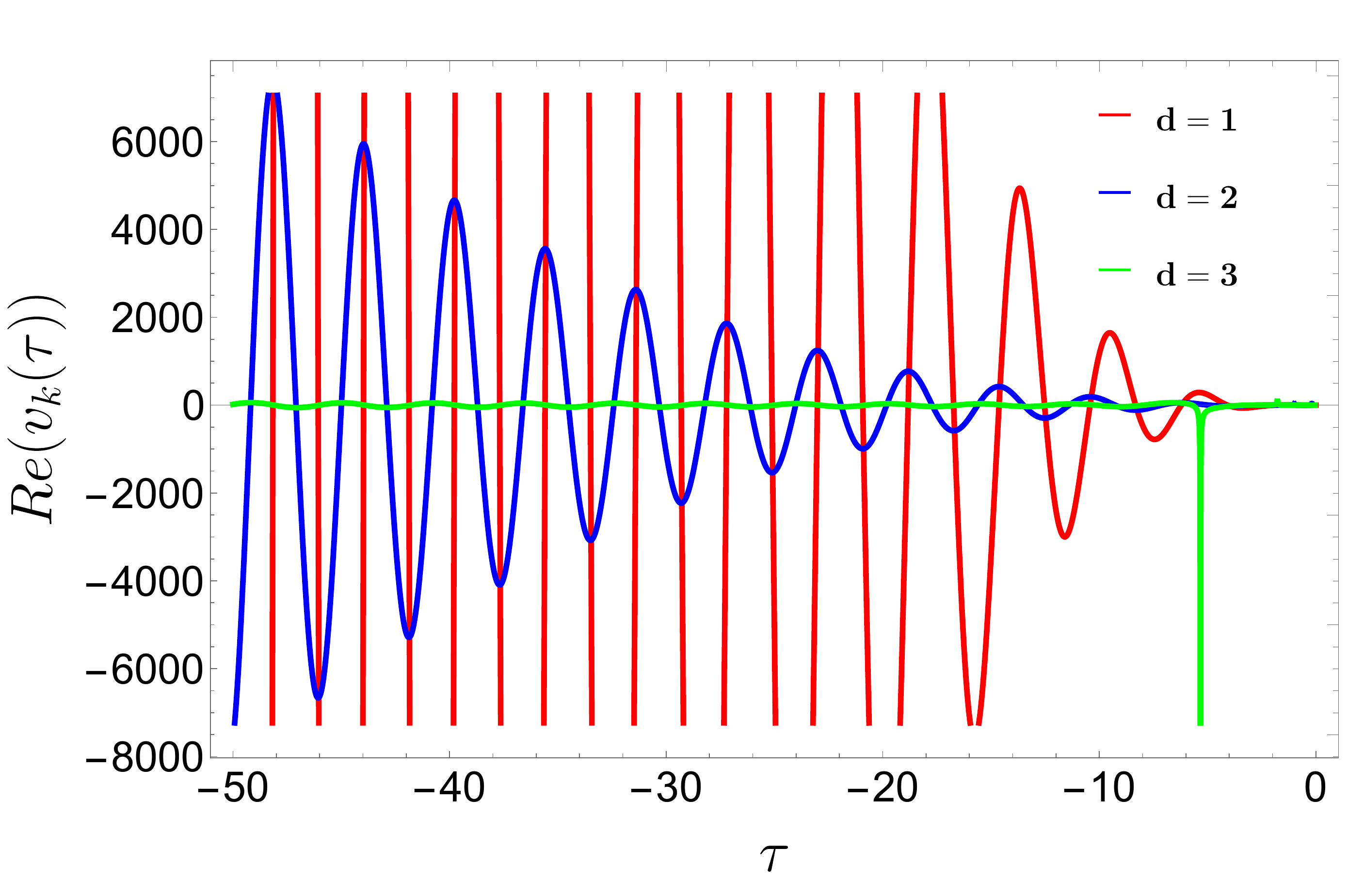}
	\caption{Behaviour of the real part of the mode solution $v_k$ with the scale factor. }
	\label{fig_realsol}
\end{figure}
\begin{figure}[h!]
	\centering
	\includegraphics[width=9cm,height=8.5cm]{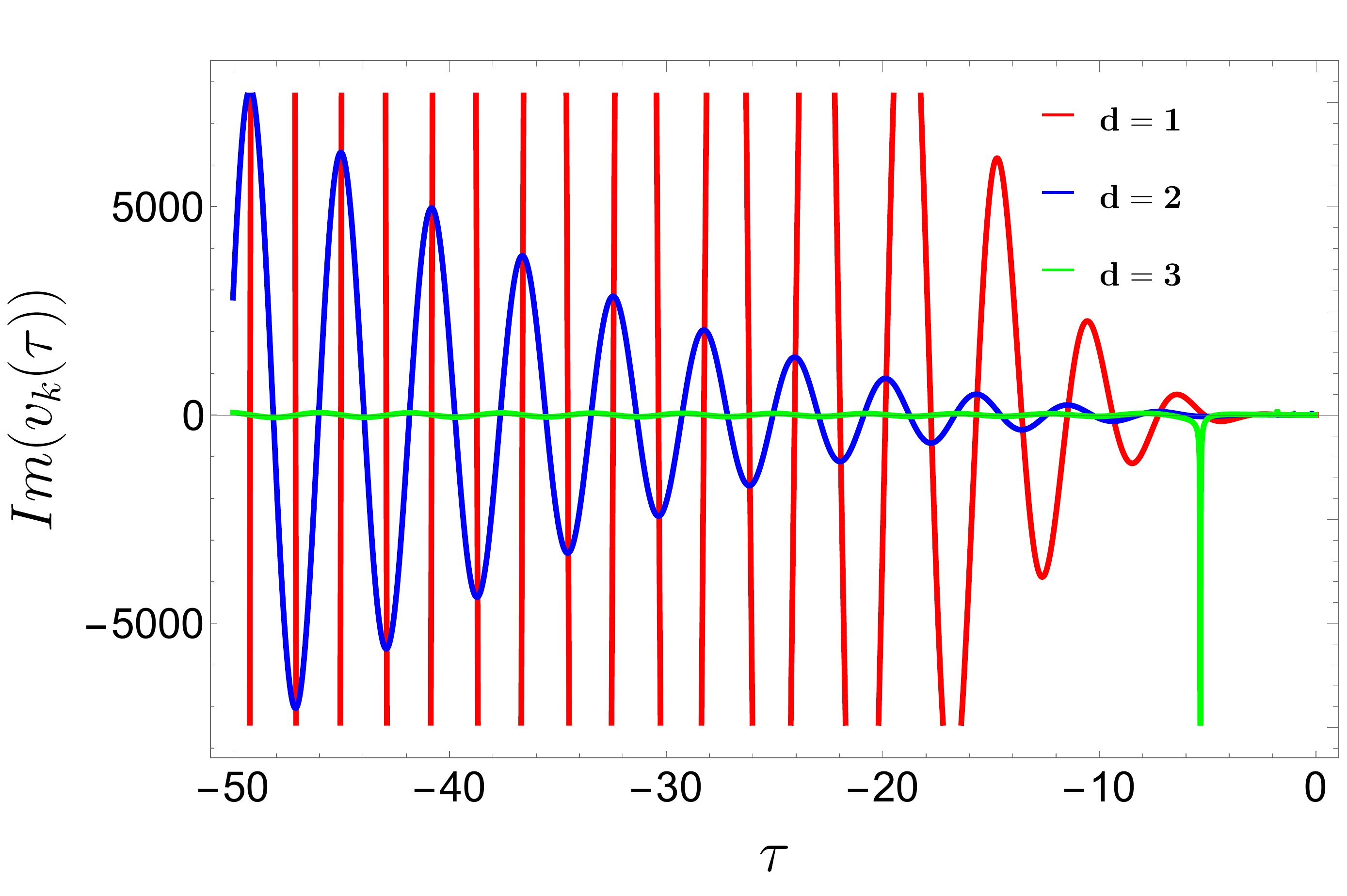}
	\caption{Behaviour of the imaginary part of the mode solution $v_k$ with the scale factor.}
	\label{fig_imsol}
\end{figure}
In fig:\ref{fig_realsol} and fig: \ref{fig_imsol}, we have plotted the behaviour of the real and the imaginary part of the mode functions with respect to the conformal time. We observe an identical behavior for both the real and imaginary part of the mode functions. It shows an oscillatory behavior with respect to the conformal time and the amplitude of oscillation decreases with the conformal time evolution. Also, it can be observed that the amplitude of oscillation decreases with the increase in the number of spatial dimension.
From the mode functions, one calculates the conjugate momentum to the mode functions and thus constructs the classical Hamiltonian function. By promotion the mode function and the conjugate momentum to quantum mechanical operators in the Heisenberg picture, one quantizes the Hamiltonian which is written as
\begin{eqnarray}
&&\widehat{H}(\tau)=\frac{1}{2}\int d^3{{\vec{k}}}\Biggl[\Omega_{{\vec{k}}}(\tau)\left(\hat{c}^{\dagger}_{{{\vec{k}}}}\hat{c}_{{\vec{k}}}+\hat{c}^{\dagger}_{-{{\vec{k}}}}\hat{c}_{-{{\vec{k}}}}+1\right)\nonumber\\
&&~~~+i\lambda_{{\vec{k}}}(\tau)\Biggl(e^{-2i\phi_{{\vec{k}}}(\tau)}\hat{c}_{{\vec{k}}}\hat{c}_{-{{\vec{k}}}}-e^{2i\phi_{{\vec{k}}}(\tau)}\hat{c}^{\dagger}_{{\vec{k}}}\hat{c}^{\dagger}_{-{{\vec{k}}}}\Biggr)~\Biggr],
\end{eqnarray}
where, the origin of the creation and the annihilation operators can be understood, when one promotes the mode functions and its conjugate momentum to quantum mechanical operators.
where the symbols $\Omega_{\bf k}(\tau)$ and $\lambda_{\bf k}(\tau)$ are defined by the following expressions:
\begin{eqnarray} 
\Omega_{{\vec{k}}}(\tau):&=&\Biggl\{\left|v^{'}_{{{\vec{k}}}}(\tau)\right|^2+\mu^2(k,\tau)\left|v_{{{\vec{k}}}}(\tau)\right|^2\Biggr\}, \\ 
\lambda_{{\vec{k}}}(\tau):&=&\Biggl(\frac{z'(\tau)}{z(\tau)}\Biggr).
\end{eqnarray}
where the quantity $\mu^2(k,\tau)$ is given by
\begin{equation}
\mu^2(k,\tau) = k^2-\lambda_{k}^2(\tau)
\end{equation}
Imposing the initial condition at the horizon crossing ($k/\mathcal{H}=1$), given by time scale ($\tau=\tau_0$) which is given by the following conditions
\begin{equation}
\label{horizon}
\text{Horizon~crossing}  =\left\{
\begin{array}{ll}
\displaystyle \frac{k}{a_0}=1 & \text{$d$ = 1} \\ \\
\displaystyle k(d-1)\tau_0=1~~~~~~~~~ & \text{$d$ > 1}
\end{array}
\right.
\end{equation}
one can calculate the quantum operators at any arbitrary scale in the Heisenberg picture. Our next job is to determine the expression of the unitary operator in the context of cosmological primordial perturbations of the scalar modes where the concept of squeezed state formalism plays a significant role. In this approach, the unitary operator is factorized as follows :
\begin{equation}
\mathcal{U}(\tau,\tau_0) = \hat{\mathcal{S}}(r_{\vec{k}}(\tau,\tau_0) ,\phi_{\vec{k}}(\tau) )\hat{\mathcal{R}}(\theta_{\vec{k}}(\tau) ),
\end{equation} 
where $\mathcal{R}$ is the two mode rotation operator which is defined as:
\begin{equation}
\label{eq:rotationoperator}
\hat{\mathcal{R}}(\theta_{\vec{k}}(\tau) ) = \exp\left( -i\theta_{k}(\tau)\big( \hat{c}_{\vec{k}}\hat{c}_{\vec{k}}^{\dagger} + \hat{c}_{-{\vec{k}}}^{\dagger}\hat{c}_{-{\vec{k}}} \big) \right),
\end{equation}
and $\hat{\mathcal{S}}$ is the two-mode squeezing operator, defined as:
\begin{eqnarray}
\label{eq:Squeezedoperator}
\nonumber
\hat{\mathcal{S}}(r_{\vec{ k}}(\tau) ,\phi_{\vec{k}}(\tau) )&=& \exp\bigg(r_{\vec{k}}(\tau) \big[ e^{-i \phi_{\vec{k}}(\tau)}\hat{c}_{\vec{k}}\hat{c}_{-{\vec{ k}}}  \\ &&~~~~~~~~~~~~~~~~~~- e^{i \phi_{\vec{k}}(\tau)}\hat{c}_{-{\vec{ k}}}^{\dagger}\hat{c}_{\vec{ k}}^{\dagger} \big]\bigg).~~~~~
\end{eqnarray}
Here the squeezing amplitude is represented by the time-dependent parameter, $r_{\vec{ k}}(\tau)$ ,and the squeezing angle or the phase is represented by the time-dependent parameter $\phi_{\vec{ k}}(\tau)$. The two-mode rotation operator, $\hat{\mathcal{R}}$ produces an irrelevant phase contribution and is ignored. 
The ground state of the free part of the above Hamiltonian is taken as the initial quantum state, whereas the two mode squeezed quantum vacuum state obtained by acting the squeezed operator on the initial vacuum is taken as the final target state.  
The time evolution of the conformal time dependent quantum operators $\hat{\mathcal{R}}$ and $\hat{\mathcal{S}}$, described by the Schr\"{o}dinger equation, gives the following set of differential equations for the squeezing parameters:
\begin{align}
\label{eq:evolution}
\frac{dr_{\vec{k}}(\tau)}{d\tau} &=-\lambda_{\vec{k}}(\tau)\cos(2\phi_{\vec{k}}(\tau)) \\
\frac{d\phi_{\vec{k}}(\tau)}{d\tau} &=\Omega_{{\vec{k}}}(\tau)+\lambda_{{\vec{k}}}(\tau)\coth(2r_{{\vec{k}}}(\tau))\sin(2\phi_{{\vec{k}}}(\tau))
\end{align}
\section*{\textcolor{Sepia}{\textbf{ \Large A Short Note on Circuit Complexity}}}
\label{sec:CircuitComplexity}
One of the challenges in Quantum Information processing is it to find out the efficient circuit for implementing a unitary operation $U$ which can be used to solve a computational problem like Search Algorithm or Shor's factoring \cite{shor1, Shor:1994jg, nielsen_chuang_2010}. In computer science, similar term called complexity \cite{cscomplexity1, cscomplexity2} exists, which can be defined as the minimum number of computational gates required to implement a certain algorithm. If we extend this definition to quantum version as minimum number of quantum gates out of basic unitary gates \cite{Barenco:1995na} in order to implement a unitary operation $U$, we get quantum complexity \cite{Aaronson:2016vto, DiVincenzo_2000}. Therefore, understanding the difficulty of implementing such unitary operation $U$, as a sequence of logical gates, is very helpful and at the same time challenging.

In refs \cite{nielsen2005geometric,Nielsen2,nielsen2006geometric}, the authors introduced a geometric approach to compute quantum circuit complexity based on the idea that finding optimal quantum circuit is equivalent to problems  of computing geodesics in Riemannian geometry. Here, we define a Riemannian metric on the space of $n$-qubit operations, and distance $d(I,U)$ between identity and target unitary operation $U$ is equivalent to the number of quantum gates which is then identified as the circuit complexity. It was shown that minimizing this distance $d(I,U)$ i.e. finding geodesic length, gives a good measure of complexity. Thus, one can employ well developed tools of Riemannian geometry such as Levi-Civita connection, geodesic, curvature, etc. to analyze the quantum circuit complexity. It is important to note that, even though in refs.\cite{nielsen2005geometric,Nielsen2,nielsen2006geometric} geometric techniques were introduced to study complexity, in ref. \cite{khaneja2}, the authors have previously used techniques from theory of symmetric spaces to study time-optimal control of quantum evolution. 

Let $U$ be a transformation which transforms reference state $\ket{\psi_{R}}$ to the target state $\ket{\psi_{T}}$ via:
\begin{equation}
\label{eq:unitary}
\ket{\psi_{T}} = U \ket{\psi_{R}}
\end{equation}
The unitary transform $U$, in the language of quantum computation has a order of unitary gates $Q_i$ such that $U = Q_1Q_2...Q_d$ where $d$ is the depth of the circuit. We can also introduce the tolerance $\epsilon$ which tells us whether the transformation is successful:
\begin{equation}
\label{}
||\ket{\psi_{T}} -  U \ket{\psi_{R}}||^2 \leq \epsilon
\end{equation} 
This makes sense because in any practical implications, it is difficult to represent the unitary transformation $U$ exactly as a combination of discrete unitary $Q_i$ operations.

Obviously, there exists infinite number of ways to achieve this target state $\ket{\psi_{T}}$ from the reference state. The circuit complexity is then the depth of the optimal circuit out of this infinite possibilities.

Motivated from the theory of Hamiltonian control problem, authors in refs. \cite{nielsen2005geometric,Nielsen2,nielsen2006geometric} introduced a geometric approach to compute this circuit complexity which was later used to compute complexity in various quantum mechanical and quantum field theoretic models. Instead of directly counting discrete set of gates required for constructing $U$, Nielsen's approach is geometric. In this method, with a time-dependent Hamiltonian $H(t)$ one constructs unitary $U$ as:
\begin{equation}
\label{eq:controlHamiltonian}
U = \overleftarrow{\mathcal{P}} \text{ exp}\left[ -i\int_{0}^{1}d\tau H(\tau) \right] \text{ where } H(\tau) = \sum_{I} Y^I(\tau)\mathcal{O}_I
\end{equation}
Here, the hermitian operators $\mathcal{O}_I$ forms the basis for time dependent Hamiltonian $H(\tau)$. Path ordering operator $\mathcal{P}$ is another version of time ordering operator which indicates that the circuit, made out of non-commuting operators, is built from right to left. The right to left application of operators is a choice of convention. 
The control functions $Y^I(\tau)$ can be thought as particular gates added at a time $s$  represented by Eq. \eqref{eq:controlHamiltonian}.

One can then construct paths in the space of unitaries as:
\begin{equation}
\label{eq:generalTrajectory}
U = \overleftarrow{\mathcal{P}} \text{ exp}\left[ -i\int_{0}^{\tau}d\tau' H(\tau') \right] 
\end{equation}
The most interesting case is when trajectory satisfy the boundary conditions $U(\tau = 0)= \mathrm{1}$ and $U(\tau = 1)= U$. These $\mathcal{O}_I$ and $Y^I(\tau)$ satisfy:
\begin{equation}
\label{eq:velocity}
Y^I(\tau)\mathcal{O}_I = \partial _{\tau} U(\tau)U^{-1}(\tau)
\end{equation}
Eq. \eqref{eq:velocity} is actually just a time-dependent Schrödinger equation in the form when one solves it via time ordered exponentials. 

As we discussed before, there are infinite number of ways of achieving unitary transformation $U$. However, not all processes are optimal. In order to find out the optimal transformation, a cost function $F(U, \vec{Y}(\tau))$ is defined. This cost function $F(U, \vec{Y}(\tau))$ is a local functional along the trajectory of the $U(\tau)$ and tangent vectors $\vec{Y}(\tau))$. Now, for each path the cost is defined as:
\begin{equation}
\label{eq:geodesiclength}
\mathcal{D}(U(t)) = \int_{0}^{1}dt F(U(t),\dot{U}(t))
\end{equation}
Nielsen showed that, the variational geometric approach of minimizing this functional is basically finding the optimized quantum circuit. 
Standing on the physical grounds, the cost function $F$ should satisfy certain properties. Those are appended below. 
\begin{itemize}
\item {\textbf{Continuity}:}
 $F \in C^0$ i.e. $F$ should be continuous. It is reasonable to assume continuity on physical grounds.
\item{\textbf{Positivity}}:
Based on the definition of cost function $F$, it is reasonable to expect 
\begin{equation}
F(U,v) \geq 0,
\end{equation}
 where equality holds if and only if $v = 0$. The equality condition also implies that the reference and target is basically same.
\item{\textbf{Positive homogeneity}}:
For any positive real number $\alpha$ and any vector $v$, we get $F(\alpha v) = \alpha F(v)$. 
\item{\textbf{Triangle Inequality}}:
$F$ satisfies triangle inequality, 
\begin{equation}
F(U,v+v') \leq F(U,v) + F(U,v'),
\end{equation}
 for all tangent vectors $v$ and $v'$. Special case $F(U,v+v') = F(U,v) + F(U,v')$ is satisfied if and only if $v$ and $v'$ are along the same ray coming out from the origin. 
\end{itemize}
\par
If one extends the continuity condition  $F \in C^0$ with $F \in C^\infty$ i.e $F$ is smooth, then the manifold is known as Finsler Manifold. Nielsen's geometric approach of determining complexity is computing the geodesic in Finsler geometry, and the length of this geodesic gives the complexity. This definition of complexity is also called the geometric circuit complexity. 
\par
In literature, there are different choices of these cost functions $F(U,v)$. These choices depends on how one defines the complexity, and the elementary gate sets for their setup. Some of the simple examples are:
\begin{align}
\begin{split}
  \label{eq:cost-functional}
F_1(U,Y) &= \sum_I |Y^I| \\ F_p(U,Y) &= \sum_I p_I|Y^I| \\
F_2(U,Y) &= \sqrt{\sum_I |Y^I|^2 }   \\ F_q(U,Y) &= \sqrt{\sum_I q_I|Y^I|^2 } 
\end{split}
\end{align}
We can now give comments on various choices of these cost functions. $F_1$, linear cost functional, measure is the nearest concept that is close to counting individual gates in the circuit. $F_2$, quadratic cost functional, can be considered as the proper distance in the manifold. $F_{1p}$ can be thought of as a cost function where penalty factors $p_I$ are used to favor certain directions over others. This becomes reasonable when one consider elementary gates coupled only to the neighboring qubits and discard those qubits which are non-local. Depending on the system, one can choose different cost functions to study the circuit complexity.

One can also introduce a general class of inhomogeneous and homogeneous family of functionals represented by:
\begin{equation}
\begin{split}
    \label{eq:k_cost_functional}
    F_k(U,Y) &= \sum_I |Y^I|^k \\
    F_{\frac{1}{k}}(U,Y) &= \sum_I |Y^I|^{\frac{1}{k}}
\end{split}
\end{equation}
where $k \geq 1$ represents the degree of homogeneity. $F_k$ was introduced in the context of hologrpahy to match the results obtained from "Complexity = Action" and "Complexity = Volume" conjectures.

\section*{\textcolor{Sepia}{\textbf{ \Large Circuit Complexity of two mode squeezed states}}} \label{sec:twomode}
A simple but a very rich example of entangled multi-mode field states is two-mode squeezed vacuum state. More details about two-mode squeezed states can be found in \cite{gerry_knight_2004}. As already defined in the previous section, the two mode squeezing operator is given by:
\begin{equation}
    \label{eq:two-mode-squeezing}
    \hat{S}_{\vec{k}}(\xi) = \exp(\xi^* \hat{c}_{\vec{k}}\hat{c}_{\vec{-k}}-\xi\hat{c}^\dagger_{\vec{-k}} \hat{c}_{\vec{k}}^\dagger)
\end{equation}
where, $\xi = r_ke^{i\phi_k}$, $r_k$ and $\phi_k$ is known as squeezing parameters and $0 \leq r_k < \infty$ and $0 \leq \phi _k \leq 2\pi$. The two mode squeezed vacuum state, which will act as our target state is given by the action of two mode squeezing operator $\hat{S}_{\vec{k}}(\xi)$ on two-mode vacuum state (initial state), $\ket{0}_{\vec{k}}\ket{0}_{\vec{-k}} = \ket{0,0}$:
\begin{align}
\nonumber
    \ket{\psi_{\text{sq}}}_{\vec{k},\vec{-k}} &= \hat{S}_{\vec{k}}(\xi) \ket{0,0} \\
    &= \exp(\xi^* \hat{c}_{\vec{k}}\hat{c}_{\vec{-k}}-\xi\hat{c}^\dagger_{\vec{-k}} \hat{c}_{\vec{k}}^\dagger)\ket{0,0}
\end{align}
In terms of number states, one can show that the state of two-mode squeezed states is given by:
\begin{equation}
\label{eq:two-mode state}
    \ket{\psi_{\text{sq}}}_{\vec{k},\vec{-k}} = \frac{1}{\text{cosh}r_k} \sum_{n= 0}^\infty (-1)^n e^{in\theta}(\text{tanh}r_k)^n \ket{n_k,n_{-k}})
\end{equation}
We are now in the position to compute circuit complexity of the reference and target state of two mode squeezed states and compare it to the entanglement entropy. 
For this, we need to write our reference and target quantum state as gaussian wave functions. The auxiliary position and momentum variables are:
\begin{align}
 \hat{q}_{\vec{k}} &= \frac{1}{\sqrt{2\Omega_k}} \big (\hat{c}^\dagger_{\vec{k}} + \hat{c}_{\vec{k}} \big)\\
 \hat{p}_{\vec{k}} &=i \sqrt{\frac{\Omega_k}{2} }\big (\hat{c}^\dagger_{\vec{k}} - \hat{c}_{\vec{k}} \big)
\end{align}
with, $[\hat{q}_{\vec{k}},\hat{p}_{\vec{k'}}] = i\delta^3(\vec{k}-\vec{k'})$. The reference state i.e the two mode vacuum state, in the position space can be expressed as a gaussian wave function as follows:
\begin{align}
\label{eq:referenceGaussianState}
\nonumber
   \psi_R(q_{\vec{k}},q_{-\vec{k}}) &= \langle q_{\vec{k}}, q_{-\vec{k}} |0\rangle_{\vec{k},-\vec{k}}\\
    &= \left( \frac{\Omega_k}{\pi} \right)^{\frac{1}{4}}\text{exp}\left( -\frac{\Omega_k}{2}\big(q_{\vec{k}}^2+ q_{-\vec{k}}^2 \big)\right) 
\end{align}
The target state, two mode squeezed state, in the position space has the wavefunction:
\begin{align}
\label{eq:targetGaussianState}
\nonumber
     \psi_{\text{sq}}(q_{\vec{k}},q_{-\vec{k}}) &= \langle q_{\vec{k}}, q_{-\vec{k}} |\psi_{\text{sq}}\rangle_{\vec{k}} \\
    &= \frac{e^{A\big(q_{\vec{k}}^2+ q_{-\vec{k}}^2 \big) - Bq_{\vec{k}}q_{-\vec{k}}}}{\text{cosh}r_k\sqrt{\pi}\sqrt{1-e^{-4i\phi_k}\text{tanh}^2r_k}}   
\end{align}
where, $A$ and $B$ are the coefficients and are functions of squeezing parameters $r_k$ and $\phi_k$ :
\begin{align}
\begin{split}
    A &= \frac{\Omega_k}{2} \frac{e^{-4i\phi_k}\text{tanh}^2r_k+1}{e^{-4i\phi_k}\text{tanh}^2r_k-1} \\
    B &= 2\Omega_k \frac{e^{-2i\phi_k} \text{tanh}r_k}{e^{-4i\phi_k}\text{tanh}^2r_k-1}
\end{split}
\end{align}
It is helpful to define three other terms for simplifying complexity calculation:
\begin{align}
\label{eq:omegas}
    \Sigma _{\vec{k}} &= -2A +B,\Sigma _{-\vec{k}} = -2A -B,\omega _{\vec{k}}= \omega_{-\vec{k}}= \frac{\Omega_k}{2}
\end{align}
Three methods of computing complexity was discussed in \cite{Ali:2018fcz}. For our case, two methods i.e. computing complexity via covariance matrix method and Nielsen's method are relevant. As we discussed before in Eq. \eqref{eq:cost-functional} and \eqref{eq:k_cost_functional}, complexity depends on the choice of cost functions. Let $C_1$ be the circuit complexity corresponding to linear cost functional $F_1$, $C_2$ to quadratic cost functional $F_2$ and $C_k$ to k family of functionals $F_k$.

\subsection*{\textcolor{Sepia}{\textbf{ Complexity via Covariance matrix method}}}
This method is interesting because complexity from the covariance matrix method is independent of squeezing angle $\phi_k$. We will see later that entanglement entropy obtained is also independent of the squeezing angle $\phi_k$, so the comparison between circuit complexity and entanglement entropy is more visible in this approach.
Since our reference and target states \eqref{eq:referenceGaussianState} and \eqref{eq:targetGaussianState} are in Gaussian form, we can express it as covariance matrix. The covariance matrix for the reference state is given as:
\begin{eqnarray}
    G_k^{s=0}=\displaystyle 
\begin{bmatrix}
\displaystyle \frac{1}{\Omega_k} & 0 & 0 & 0 \\
0 & \displaystyle\Omega_k & 0 & 0\\
0 & 0 &\displaystyle \frac{1}{\Omega_k} & 0 \\
0 & 0 & 0 & \displaystyle\Omega_k
\end{bmatrix}
\end{eqnarray}

The covariance matrix for the target state is given as:
\begin{eqnarray}
    G_k^{s=1} = 
\begin{bmatrix}
\displaystyle\frac{1}{\text{Re}(\Sigma _{\vec{k}})} & \displaystyle-\frac{\text{Im}(\Sigma _{\vec{k}})}{\text{Re}(\Sigma _{\vec{k}})} & 0 & 0 \\
\displaystyle-\frac{\text{Im}(\Sigma _{\vec{k}})}{\text{Re}(\Sigma _{\vec{k}})} & \displaystyle \frac{|\Sigma_{\vec{k}}|^2}{\text{Re}(\Sigma _{\vec{k}})} & 0 & 0 \\
0 & 0 & \displaystyle \frac{1}{\text{Re}(\Sigma _{-\vec{k}})} &\displaystyle -\frac{\text{Im}(\Sigma _{-\vec{k}})}{\text{Re}(\Sigma _{-\vec{k}})} \\
0 & 0 & \displaystyle-\frac{\text{Im}(\Sigma _{-\vec{k}})}{\text{Re}(-\Sigma _{\vec{k}})} & \displaystyle\frac{|\Sigma_{-\vec{k}}|^2}{\text{Re}(\Sigma _{-\vec{k}})}
\end{bmatrix}\nonumber\\
&&
\end{eqnarray}

where $\Sigma _{\vec{k}}$ and $\Sigma _{-\vec{k}}$ are defined in \eqref{eq:omegas}. The covariance matrix basically carries the same information as the wave function. 
In the context of covariance matrix approach, circuit complexity quantifies the number of gates to take covariance matrix of reference state to the covariance matrix of target state. We will factorize the covariance matrix $G$ into two $2 \times 2$ matrices. This gives us the benefit that we can compute complexity for each block and sum over all $\Omega _k$ to give total complexity. The two symmetric blocks for the reference states are:
\begin{eqnarray}
 G_{k=0}^{s=0} &= 
\begin{bmatrix}
\displaystyle \frac{1}{\Omega_k} & 0 \\
0 & \displaystyle\Omega_k \\
\end{bmatrix}, 
 G_{k=1}^{s=0} = 
\begin{bmatrix}
\displaystyle\frac{1}{\Omega_k} & 0 \\
0 & \displaystyle\Omega_k \\
\end{bmatrix}
\end{eqnarray}
While, the two symmetric blocks for the target states are:
\begin{eqnarray}
G_{k=0}^{s=1} &=&
\begin{bmatrix}
\displaystyle\frac{1}{\text{Re}(\Sigma _{\vec{k}})} & \displaystyle-\frac{\text{Im}(\Sigma _{\vec{k}})}{\text{Re}(\Sigma _{\vec{k}})} \\
\displaystyle-\frac{\text{Im}(\Sigma _{\vec{k}})}{\text{Re}(\Sigma _{\vec{k}})} & \displaystyle\frac{|\Sigma_{\vec{k}}|^2}{\text{Re}(\Sigma _{\vec{k}})}
\end{bmatrix}, 
\\
    G_{k=1}^{s=1} &=&
\begin{bmatrix}
\displaystyle\frac{1}{\text{Re}(\Sigma _{\vec{-k}})} &\displaystyle -\frac{\text{Im}(\Sigma _{\vec{-k}})}{\text{Re}(\Sigma _{\vec{-k}})} \\
\displaystyle-\frac{\text{Im}(\Sigma _{\vec{-k}})}{\text{Re}(\Sigma _{\vec{-k}})} & \displaystyle\frac{|\Sigma_{\vec{-k}}|^2}{\text{Re}(\Sigma _{\vec{-k}})}
\end{bmatrix}, 
\end{eqnarray}
The basis for each block is changed to make the calculation easier as follows:
\begin{equation}
    \Tilde{G}^{s=1} = SG^{s=1}S^T, \Tilde{G}^{s=0} = SG^{s=0}S^T
\end{equation}
where, $S$ is a specifically choosen matrix such that $\Tilde{G}^{s=0} = \mathbb{1}$. In our case, matrix $S$ is given by:
\begin{eqnarray}
    S =
    \begin{bmatrix}
   \displaystyle \sqrt{\Omega_k} & 0\\
    0 & \displaystyle\frac{1}{\sqrt{\Omega_k} }
    \end{bmatrix}
\end{eqnarray}
 This implies $\Tilde{G}^{s=0} = \mathbb{1}$ and 
\begin{eqnarray}
    \Tilde{G}^{s=1} = 
    \begin{bmatrix}
\displaystyle\frac{\Omega_k}{\text{Re}(\Sigma _{\vec{k}})} & \displaystyle-\frac{\text{Im}(\Sigma _{\vec{k}})}{\text{Re}(\Sigma _{\vec{k}})} \\
 & \\
\displaystyle-\frac{\text{Im}(\Sigma _{\vec{k}})}{\text{Re}(\Sigma _{\vec{k}})} & \displaystyle \frac{|\Sigma_{\vec{k}}|^2}{\Omega_k \text{Re}(\Sigma _{\vec{k}})}
\end{bmatrix}, 
\end{eqnarray}
We can assume that $k$ is real. In the language of covariance matrix, the unitary transformation of wave functions can be expressed as:
 \begin{equation}
 \Tilde{G}^s = \Tilde{U}(\tau)\Tilde{G}^{s=0}\Tilde{U}(\tau)^T.
 \end{equation}
 The unitary transformations is then parametrized with gates satisfying $SL(2,R)$ algebra as:
\begin{widetext}
\begin{eqnarray}
    \Tilde{U}(\tau) = 
    \begin{bmatrix}
   \displaystyle \text{cos}(\mu(\tau))\text{cosh}(\rho(\tau))-\text{sin}(\theta(\tau))\text{sinh}(\rho(\tau))~~~~~ &~~~~~ \displaystyle-\text{sin}(\mu(\tau))\text{cosh}(\rho(\tau))+\text{cos}(\theta(\tau))\text{sinh}(\rho(\tau)) \\
    &    \\
  \displaystyle  \text{sin}(\mu(\tau))\text{cosh}(\rho(\tau))+\text{cos}(\theta(\tau))\text{sinh}(\rho(\tau)) & 
   \displaystyle \text{cos}(\mu(\tau))\text{cosh}(\rho(\tau))+\text{sin}(\theta(\tau))\text{sinh}(\rho(\tau))
    \end{bmatrix}
\end{eqnarray}
\end{widetext}
where, $\mu, \rho, \theta$ are the coordinates on the $SL(2,R)$ group.
Now, we will set following boundary conditions:
\begin{equation}
    \begin{aligned}
        \Tilde{G}^{s=1} &= \Tilde{U}(\tau = 1)\Tilde{G}^{s=0}\Tilde{U}(\tau =1)^T \\
        \Tilde{G}^{s=0} &= \Tilde{U}(\tau = 0)\Tilde{G}^{s=0}\Tilde{U}(\tau =0)^T
    \end{aligned}
\end{equation}
This boundary conditions applied with the parametrized unitary transformations gives:
\begin{equation}
\label{eq:boundaryConditions}
    \begin{aligned}
    \left(\text{cosh}(2\rho(1)),\text{tan}(\theta(1)+\mu(1))\right) &= \left( \frac{\Omega_k^2 + |\Sigma_k|^2}{2\Omega \text{Re}(\Sigma_k)}, \frac{\Omega_k^2 - |\Sigma_k|^2}{2\Omega \text{Im}(\Sigma_k)} \right) \\
    \left( \rho(0), \theta(0) + \mu(0) \right) &= \left(0,c\right)
    \end{aligned}
\end{equation}
In order to make the calculation simpler,  we choose:
\begin{enumerate}
\item $\displaystyle \mu(\tau = 1) = \mu(\tau = 0)= 0$.
\item $\displaystyle \theta(\tau = 0) =\theta(\tau = 1) = c = \text{tan}^{-1}\left(\frac{\Omega_k^2 - |\Sigma_k|^2}{2\Omega \text{Im}(\Sigma_k)} \right)$.
\end{enumerate}

Given this conditions the metric for $\Tilde{U}$ becomes:

\begin{multline}
    ds^2 = d\rho^2 + \text{cosh}(2\rho)\text{cosh}^2\rho d\mu^2 +\text{cosh}(2\rho)\text{sinh}^2\rho d\theta^2 \\
    -\text{sinh}(2\rho)^2d\mu d\theta
\end{multline}
    
The simple geodesic is a straight line on this geometry which is given by $\rho(\tau) = \rho(1)\tau$.
From, the boundary conditions \eqref{eq:boundaryConditions}, we get:
\begin{equation}
    \rho_k(\tau = 1) = \frac{1}{2}\text{cosh}^{-1} \left[ \frac{\Omega_k^2 + |\Sigma _{\vec{k}}|^2 }{2\Omega_k\text{Re}(\Sigma _{\vec{k}})} \right]
\end{equation}
In order to get total circuit complexity, we have to sum over both values of $k$ i.e. $k$ and $-k$. Therefore, for linear and quadratic cost functions $C_1$ and $C_2$, we will get: 
\begin{widetext}
\begin{align}
        \label{eq:circuitComplexityCovariance}
         C_1(\Omega_k) &=\rho_k(\tau = 1) +\rho_{-k}(\tau = 1)= \frac{1}{2}\Bigg[\text{cosh}^{-1} \left[ \frac{\Omega_k^2 + |\Sigma _{\vec{k}}|^2 }{2\Omega_k\text{Re}(\Sigma _{\vec{k}})} \right]  +\text{cosh}^{-1} \left[ \frac{\Omega_{-k}^2 + |\Sigma _{-\vec{k}}|^2 }{2\Omega_{-k}\text{Re}(\Sigma _{-\vec{k}})} \right] \Bigg] ,\\
 C_2(\Omega_k) &= \sqrt{\rho_k(\tau = 1)^2 +\rho_{-k}(\tau = 1)^2} = \frac{1}{2} \sqrt{\left( \text{cosh}^{-1} \left[ \frac{\Omega_k^2 + |\Sigma _{\vec{k}}|^2 }{2\Omega_k\text{Re}(\Sigma _{\vec{k}})} \right]  \right)^2 +\left( \text{cosh}^{-1} \left[ \frac{\Omega_{-k}^2 + |\Sigma _{-\vec{k}}|^2 }{2\Omega_{-k}\text{Re}(\Sigma _{-\vec{k}})} \right]  \right)^2 }
\end{align}
\end{widetext}

Using the explicit form of $\Sigma _{\vec{k}}$, $\Sigma _{-\vec{k}}$, the circuit complexity reduces to simple form which is independent of squeezing angle $\phi_k$:
\begin{align}
    \label{eq:circuitComplexityCovariance}
    C_1(\Omega_k) &= 4r_k \\
    C_2(\Omega_k) &= 2\sqrt{2}r_k
\end{align}
These two cost functions are then related by: $C_1(\Omega_k) = \sqrt{2}C_2(\Omega_k)$.  Naturally, for small squeezing parameter $r_k \rightarrow 0$, $C_1 \approx 0$ and $C_2 \approx 0$. This makes sense as for small squeezing parameter $r_k$, the reference and target states are basically same. 

\subsection*{\textcolor{Sepia}{\textbf{ Complexity via Nielsen's wave-function method}}}
Unlike covariance matrix method, Nielsen's approach using wavefunctions gives circuit complexity of two mode squeezed states that is sensitive to both squeezing parameters: $r_k$ and $\phi_k$. The general philosophy of computing circuit complexity is basically same as in covariance approach. However, instead of representing the wave function as a covariance matrix, we will directly compute the complexity using reference and target two-mode squeezed states, i.e. eq: \eqref{eq:referenceGaussianState} and eq: \eqref{eq:targetGaussianState} respectively. Then, we will be able to write the circuit complexity in terms of squeezing parameters $r_k$ and $\phi_k$. 

The exponent of the target state i.e. two-mode squeezed states eq: \eqref{eq:targetGaussianState} can be diagonalized as: 
\begin{eqnarray}
\label{eq:nieslenTarget}
    \psi_{\text{sq}} = \mathcal{N}\text{exp}\left(-\frac{1}{2} \Tilde{\mathcal{M}}^{ab}q_aq_b\right)
\end{eqnarray}
where, $\mathcal{N}$ is the normalization constant i.e. denominator in \ref{eq:targetGaussianState} and, 
\begin{eqnarray}
    \Tilde{\mathcal{M}} &=& 
    \begin{bmatrix}
     \displaystyle  -2A+B & 0\\
      & \\
    0 &  \displaystyle  -2A -B
    \end{bmatrix} \nonumber\\
    &=& 
    \begin{bmatrix}
    \displaystyle  \Sigma _{\vec{k}} & 0\\
      & \\
    0 &  \displaystyle  \Sigma _{\vec{-k}} 
    \end{bmatrix}
\end{eqnarray}
The unsqueezed state, reference state is also a guassian wave function represented by:
\begin{align}
\label{eq:nielsenReference}
\begin{split}
        \psi_{\text{R}} &= \mathcal{N}\text{exp}\left( -\frac{\Omega_k}{2}\big(q_{\vec{k}}^2+ q_{-\vec{k}}^2 \big)\right) \\
        &=  \mathcal{N}\text{exp}\left( \frac{1}{2} \sum_{k,-k}\Omega_k\vec{k}^2 \right)
\end{split}
\end{align}
Our two gaussian wave functions has a form:
\begin{equation}
    \psi^\tau = \mathcal{N}\text{exp}\left( -\frac{1}{2}\left( v_a. \mathcal{A}_{ab}^\tau.v_b  \right) \right)
\end{equation}
where, $v = (q_{\vec{k}}, q_{\vec{-k}})$ and $\mathcal{A}^\tau$ is an $2 \times 2$ diagonal matrix. For the target state eq: \eqref{eq:nieslenTarget},
\begin{eqnarray}
    \mathcal{A}^{\tau=1} = \mathcal{M} = 
    \begin{bmatrix}
   \displaystyle   \Sigma _{\vec{k}} & 0\\
    &  \\
    0 &  \displaystyle  \Sigma _{\vec{-k}} 
    \end{bmatrix}
\end{eqnarray}
while for our reference state eq: \eqref{eq:nielsenReference}, matrix $\mathcal{A}$ is $\mathcal{A}^{\tau=0}$. So,
\begin{eqnarray}
    \mathcal{A}^{\tau=0} =
    \begin{bmatrix}
    \displaystyle   \Omega_k & 0\\
         &   \\
    0 &  \displaystyle  \Omega_{-k}
    \end{bmatrix}
\end{eqnarray}
The unitary transformation eq: \eqref{eq:generalTrajectory} acts like,
\begin{equation}
    \mathcal{A}^\tau = \mathcal{U}(\tau).\mathcal{A}^{\tau = 0 }.\mathcal{U}^T(\tau)
\end{equation}
The boundary conditions is given by:
\begin{align}
\begin{split}
     \mathcal{A}^{\tau = 1} &= \mathcal{U}(\tau = 1).\mathcal{A}^{\tau = 0 }.\mathcal{U}^T(\tau = 1) \\
     \mathcal{A}^{\tau = 0} &= \mathcal{U}(\tau = 0).\mathcal{A}^{\tau = 0 }.\mathcal{U}^T(\tau = 0)
\end{split}    
\end{align}
$\mathcal{U}$ can be parametrized as in eq: \eqref{eq:controlHamiltonian} such that at $\tau = 1$, the required target state is achieved. Since, $\mathcal{A}^{\tau = 1}$ and $\mathcal{A}^{\tau = 0}$ can have complex elements, elementary gates are restricted to $GL(2,C)$ unitaries. Tangent vector components $Y^I$ in eq: \eqref{eq:velocity} are complex parameters while $\mathcal{O}_I$ are the generators. Eq: \eqref{eq:velocity} can also be expressed as:
\begin{equation}
    Y^I = \text{Tr}(\partial _{\tau} U(\tau)U^{-1}(\tau)(\mathcal{O}_I)^T)
\end{equation}
where,  we have note that:
\begin{eqnarray}
\text{Tr}(\mathcal{O}_I.\mathcal{O}_J^T) = \delta^{IJ},
\end{eqnarray} 
and $I,J = 0,1,2,3$. The metric is then given by: 
\begin{eqnarray}
ds^2 = G_{IJ}dY^IdY^{*J}.
\end{eqnarray}
For simplicity, we will choose penalty factors $G_{IJ} = \delta^{IJ}$ where we fix it to unity. The off-diagonal elements in $GL(2,C)$ can be set to zero as they increase the distance between states. The $U(\tau)$ will become:
\begin{equation}
    U(\tau) = \text{exp}\left(\sum_{i \in (k,-k)}\alpha^i(\tau)\mathcal{O}_i^{diagonal}\right)
\end{equation}
where, $\alpha^i(\tau)$ are complex parameters and $\mathcal{O}_i^{diagonal}$ are generators with identity at $i$ diagonal elements. The metric takes a simple form:
\begin{equation}
    ds^2 = \sum_{i \in (k,-k)} (d\alpha^{i,\text{Re}})^2 + (d\alpha^{i,\text{Im}})^2
\end{equation}
where Re and Im indicates real and imaginary part of $\alpha_k$ respectively. The geodesic is again a straight line in the manifold given by:
\begin{equation}
    \alpha^{i,p}(\tau) = \alpha^{i,p}(\tau =1)+\alpha^{i,p}(\tau =0)
\end{equation}
for each $(i\in k,-k)$ and $(p = \text{Re and Im})$. Given the boundary conditions, we will get,
\begin{align}
\begin{split}
        \alpha^{i,\text{Re}}(\tau = 0) &= \alpha^{i,\text{Im}}(\tau = 0) = 0 \\
        \alpha^{i,\text{Re}}(\tau = 1) &= \frac{1}{2}\text{ln}\left| \frac{\Sigma _{\vec{i}}}{\omega _{\vec{i}}}\right| \\
        \alpha^{i,\text{Im}}(\tau = 1) &= \frac{1}{2}\text{tan}^{-1}\frac{\text{Im}(\Sigma_{\vec{i}})}{\text{Re}(\Sigma _{\vec{i}})}
\end{split}
\end{align}
for each $(i\in k,-k)$.
Now, the circuit complexity for linear $C_1(\Omega_k)$ and quadratic cost $C_2(\Omega_k)$ functions can be derived as follows:

\begin{widetext}
\begin{align}
\nonumber
  C_1(\Omega_k) &= \alpha^{k,\text{Re}}(\tau = 1) + \alpha^{-k,\text{Re}}(\tau = 1)+ \alpha^{k,\text{Im}}(\tau = 1) + \alpha^{-k,\text{Im}}(\tau = 1)\\
  &=  \frac{1}{2} \Bigg( \text{ln}\left| \frac{\Sigma _{\vec{k}}}{\omega _{\vec{k}}}\right| +  \text{ln}\left| \frac{\Sigma _{-\vec{k}}}{\omega _{-\vec{k}}}\right| + \text{tan}^{-1}\frac{\text{Im}(\Sigma _{\vec{k}})}{\text{Re}(\Sigma _{\vec{k}})} + \text{tan}^{-1}\frac{\text{Im}(\Sigma_{-\vec{k}})}{\text{Re}(\Sigma _{-\vec{k}})}\Bigg) 
 \end{align} 
 \end{widetext}
 \begin{widetext}
\begin{align}
\nonumber
      C_2(\Omega_k) &= \sqrt{(\alpha^{k,\text{Re}}(\tau = 1))^2 + (\alpha^{-k,\text{Re}}(\tau = 1))^2+ (\alpha^{k,\text{Im}}(\tau = 1))^2 + (\alpha^{-k,\text{Im}}(\tau = 1)})^2\\
      &=  \frac{1}{2} \sqrt{ \Bigg(  \text{ln}\left| \frac{\Sigma _{\vec{k}}}{\omega _{\vec{k}})}\right|\Bigg)^2  + \Bigg(\text{ln}\left| \frac{\Sigma _{-\vec{k}}}{\omega _{-\vec{k}})}\right|  \Bigg)^2 
   + \Bigg(\text{tan}^{-1}\frac{\text{Im}(\Sigma _{\vec{k}})}{\text{Re}(\Sigma _{\vec{k}})}  \Bigg)^2 + \Bigg(\text{tan}^{-1}\frac{\text{Im}(\Sigma _{-\vec{k}})}{\text{Re}(\Sigma _{-\vec{k}})}  \Bigg)^2}
\end{align}
\end{widetext}
Using explicit values of $\Sigma _{\vec{k}}$, $\Sigma _{-\vec{k}}$, $\omega _{\vec{k}}$ and $\omega _{-\vec{k}}$ from Eq. \eqref{eq:omegas}, we can get general circuit complexity form:
\begin{widetext}
\begin{eqnarray}
 C_1(\Omega_k, \tau) &=&  \left|\text{ln}\left| \frac{1+\text{exp}(-2i\phi_k(\tau))\text{tanh}r_k(\tau)}{1-\text{exp}(-2i\phi_k(\tau))\text{tanh}r_k(\tau)}\right|\right| + \left| \text{tanh}^{-1}(\text{sin}(2\phi_k(\tau))\text{sinh}(2r_k(\tau))) \right| \\
C_2(\Omega_k, \tau) &=& \frac{1}{\sqrt{2}}\sqrt{\left(\text{ln}\left| \frac{1+\text{exp}(-2i\phi_k(\tau))\text{tanh}r_k(\tau)}{1-\text{exp}(-2i\phi_k(\tau))\text{tanh}r_k(\tau)}\right|\right)^2 + \left( \text{tanh}^{-1}(\text{sin}(2\phi_k(\tau))\text{sinh}(2r_k(\tau))) \right)^2}  
\end{eqnarray}
\end{widetext}
We can also obtain approximate expressions for different limiting conditions. These expressions are discussed below:
\begin{itemize}
    \item \textbf{Small $r_k(\tau)$ and Small $\phi_k(\tau)$}
    
    For small $r_k(\tau)$ and $\phi_k(\tau)$, we can use approximating expressions:
    $\text{exp}(-2i\phi_k(\tau)) \approx 1$, $\text{sin}(2\phi_k(\tau)) \approx 2\phi_k(\tau) $, $\text{tanh}r_k(\tau) \approx r_k(\tau)$ and $\text{sinh}(2r_k(\tau)) \approx 2r_k(\tau)$ to get 
    \begin{align}
        C_1(\Omega_k, \tau) &\approx 2|r_k(\tau)|(1+2|\phi_k(\tau)|) \\
        C_2(\Omega_k, \tau) &\approx \sqrt{2}|r_k(\tau)|\sqrt{1+4(\phi_k(\tau))^2}
    \end{align}
    
    \item \textbf{Large $r_k(\tau)$ and Large $\phi_k(\tau)$}
    
    For Large $r_k(\tau)$ and $\phi_k(\tau)$, we can use approximating expressions: $\text{exp}(-2i\phi_k(\tau)) \approx 0$, and obtain the approximated circuit complexity form:
    \begin{align}
        C_1(\Omega_k, \tau) &\approx |\text{tanh}^{-1}(\text{sin}(2\phi_k(\tau))\text{sinh}(2r_k(\tau))) |\\
        C_2(\Omega_k, \tau) &\approx \frac{1}{\sqrt{2}}(\text{sin}(2\phi_k(\tau))\text{sinh}(2r_k(\tau)))
    \end{align}
    These two cost functions are related by: $|C_1(\Omega_k, \tau)|\approx \frac{1}{\sqrt{2}}C_2(\Omega_k, \tau) $
    \item \textbf{Small $r_k(\tau)$ and Large $\phi_k(\tau)$}
    
    For small $r_k(\tau)$ and large$\phi_k(\tau)$, we can use approximating expressions:
    $\text{exp}(-2i\phi_k(\tau)) \approx 0$, $\text{tanh}r_k \approx r_k(\tau)$ and $\text{sinh}(2r_k(\tau)) \approx 2r_k(\tau)$. This leads to approximated circuit complexity of form:
    \begin{align}
        C_1(\Omega_k, \tau) &\approx 2|r_k(\tau)\text{sin}(2\phi_k(\tau))|\\
        C_2(\Omega_k, \tau) &\approx \sqrt{2}r_k(\tau)\text{sin}(2\phi_k(\tau))
    \end{align}
    These two cost functions are then related by:
    $|C_2(\Omega_k, \tau)|\approx \frac{1}{\sqrt{2}}C_1(\Omega_k, \tau) $
    
\end{itemize}

Also for an example, let's see the structure of cost functions for large squeezing parameter $r_k$ and $\phi_k \rightarrow -\frac{\pi}{2}$:
\begin{equation}
    C_1(\Omega_k) \approx \sqrt{2}C_2(\Omega_k) \approx \left| \text{ln} \left( \frac{1-\text{tanh}r_k}{1+\text{tanh}r_k} \right) \right| \approx r_k
\end{equation}

\section*{\textcolor{Sepia}{\textbf{ \Large Entanglement entropy of two mode squeezed states}}}
\label{sec:ee}
In this section, we will compute the entanglement entropy for the two-mode squeezed states and compare it to the Circuit complexity. Not only our states are entangled, there is also a strong correlation between the two modes. $\ket{\psi_{\text{sq}}}_{\vec{k},\vec{-k}}$ is also an eigenstate of the number difference number operator $\hat{n}_k-\hat{n}_{-k}$ with eigenvalue 0, where $\hat{n}_k = \hat{c}_{\vec{k}}^\dagger\hat{c}_{\vec{-k}}$ and $\hat{n}_{-k}= \hat{c}^\dagger_{\vec{-k}}\hat{c}_{\vec{k}}$. Due to this strong correlation and symmetry between two modes, average photon number in each mode is same:
\begin{equation}
    \langle\hat{n_k}\rangle= \langle\hat{n}_{-k}\rangle = \text{sinh}^2r_k
\end{equation}
The reduced density operators for the individual modes are given by:
\begin{eqnarray}
  && \hat{\rho}_k = \sum_{n = 0}^\infty \frac{1}{(\text{cosh }r_k)^2}(\text{tanh }r_k)^{2n}\langle n_k|n_k\rangle,\\
   &&     \hat{\rho}_{-k} = \sum_{n = 0}^\infty \frac{1}{(\text{cosh }r_{-k})^2}(\text{tanh }r_{-k})^{2n}\langle n_{-k}|n_{-k}\rangle.
\end{eqnarray}

The probability of having $n$ photons in a single mode $k$ or $-k$ is:
\begin{equation}
    P_n^{(i)} = \frac{(\text{tanh}r_k)^{2n}}{(\text{cosh}r_k)^2}, \textit{}{ i = k,-k}
\end{equation}
Commonly used entanglement entropies are von-Neumann and Rényi Entanglment entropies. For a density operator $\hat{\rho}$, von-Neumann entropy is given by:
\begin{equation}
    S(\hat{\rho}) = -\text{Tr}[\hat{\rho}\text{ln}\hat{\rho}]
\end{equation}
If the density operator $\hat{\rho}$ is pure, then $S(\hat{\rho}_{\text{pure}})= 0$, while for mixed states $S(\hat{\rho}_{\text{mixed}})> 0$. It is usually not trivial to calculate the entropy. However, for the basis in which density operator is diagonal such as in Schmidt basis, entropy can be calculated simply from the diagonal elements as:
\begin{align}
    S(\hat{\rho}) = -\text{Tr}[\hat{\rho}\text{ln}\hat{\rho}] = -\sum_k \rho_{kk} \text{ln}\rho_{kk}
\end{align}
Since our two mode squeezed state \ref{eq:two-mode state} is already in the form of Schmidt decomposition, and we also have the form of reduced density operators of individual modes $a$ and $b$, we can calculate the Von-Neumann entanglement entropy by realizing that the diagonal elements $\rho_{kk}$ is $P_n^{(i)}$. Then, the Von-Neumann entropy, a measure of degree of entanglement is:
\begin{align}
\begin{split}
\label{eq:entanglementEntropy}
S(\hat{\rho}_k) &= -\text{Tr}[\hat{\rho_k}\text{ln}\hat{\rho_k}]  = S(\hat{\rho}_{-k})\\
    &= - \sum_{n = 0}^\infty P_n \text{ ln}P_n\\
    &= - \sum_{n = 0}^\infty \frac{\text{tanh}^{2n}r_k}{\text{cosh}^2r_k} \text{ln}\frac{\text{tanh}^{2n}r_k}{\text{cosh}^2r_k} \\
    &= -\sum_{n = 0}^\infty \frac{\text{tanh}^{2n}r_k}{\text{cosh}^2r_k}  \big({\text{ln}(\text{tanh}^{2n}r_k) - \text{ln}(\text{cosh}^{2}r_k)  \big) }\\
    &= \text{ln(cosh}^2r_k) \text{cosh}^2r_k - \text{ln(sinh}^2r_k) \text{sinh}^2r_k
\end{split}
\end{align}
\begin{figure}[htb]
	\centering
	\includegraphics[width=9cm,height=8cm]{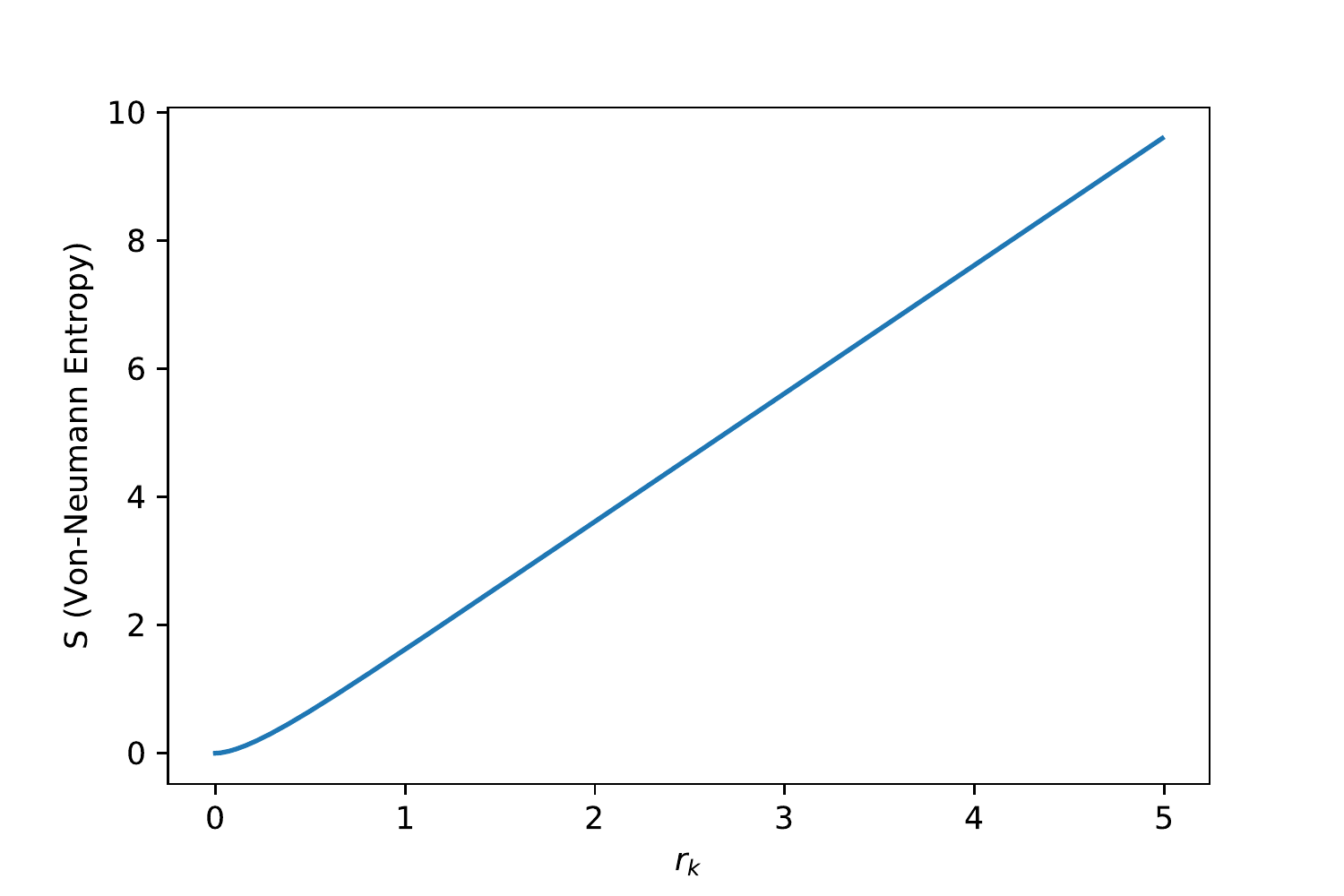}
	\caption{Von-Neumann entanglement entropy $S$ as a function of a squeezing magnitude $r_k$ }
	\label{fig_vonNeumannEntropy}
\end{figure}
We have plotted the Von-Neumann entanglement entropy in fig: \ref{fig_vonNeumannEntropy}. It can be seen that entanglement entropy increases with increasing squeezing parameter $r_k$. Note that we didn't calculate the entropy corresponding to the squeezed state Eq. \ref{eq:two-mode state} because naturally this entropy is going to be zero as it is a pure state. Instead, we have calculated entropy for the reduced density matrix.

\begin{figure}[htb]
	\centering
	\includegraphics[width=9cm,height=8cm]{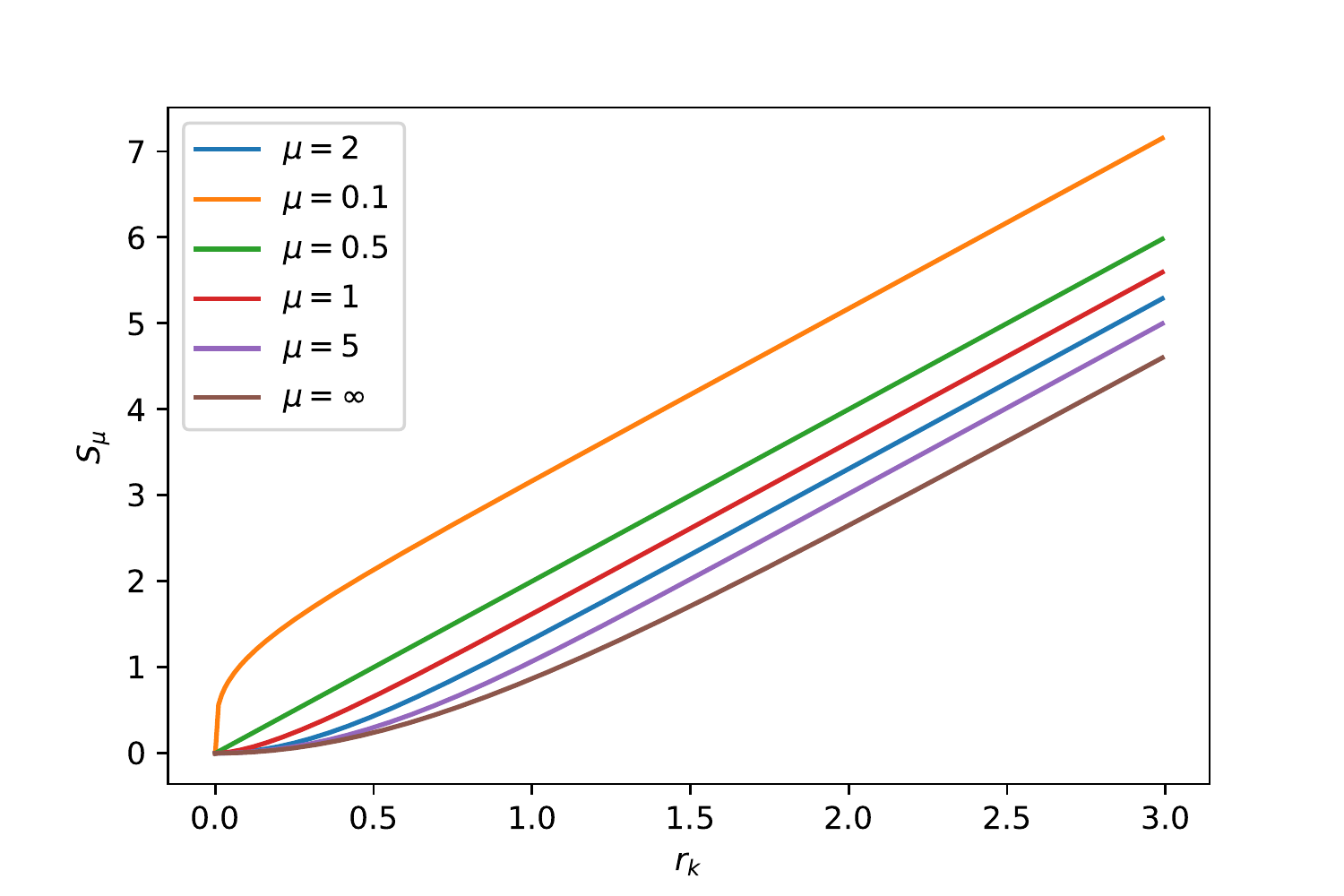}
	\caption{Rényi-Entanglement entropy $S_\mu$ as a function of squeezing magnitude $r$. For large $r$, they grow linearly with increasing $r$.}
	\label{fig_renyiEntropy}
\end{figure}

We can now generalize Von-Neumann entropy to get Rényi entropy for the reduced density operator:
\begin{align}
\begin{split}
   \label{eq:renyi-entropy eqn}
    S_\mu  &= \frac{1}{1-\mu} \text{ln} \sum_{n= 1}^d P_n  \\
    &= \frac{2\mu \text{ ln coshr}_k + \text{ln}(1- \text{tanh}^{2\mu}r_k)}{\mu -1}   
\end{split}
\end{align}
where $\mu \geq 0$ is the Rényi Parameter and $d$ is the Schmidt rank of the squeezed state Eq. \ref{eq:two-mode state} which is infinity. Again, we can see that Rényi entropy increases with increasing squeezing parameter $r_k$. In fig: \ref{fig_renyiEntropy}, we have plotted Rényi-entanglement entropy for various Rényi parameters $\mu$.
\begin{figure}[h!]
	\centering
	\includegraphics[width=9cm,height=8cm]{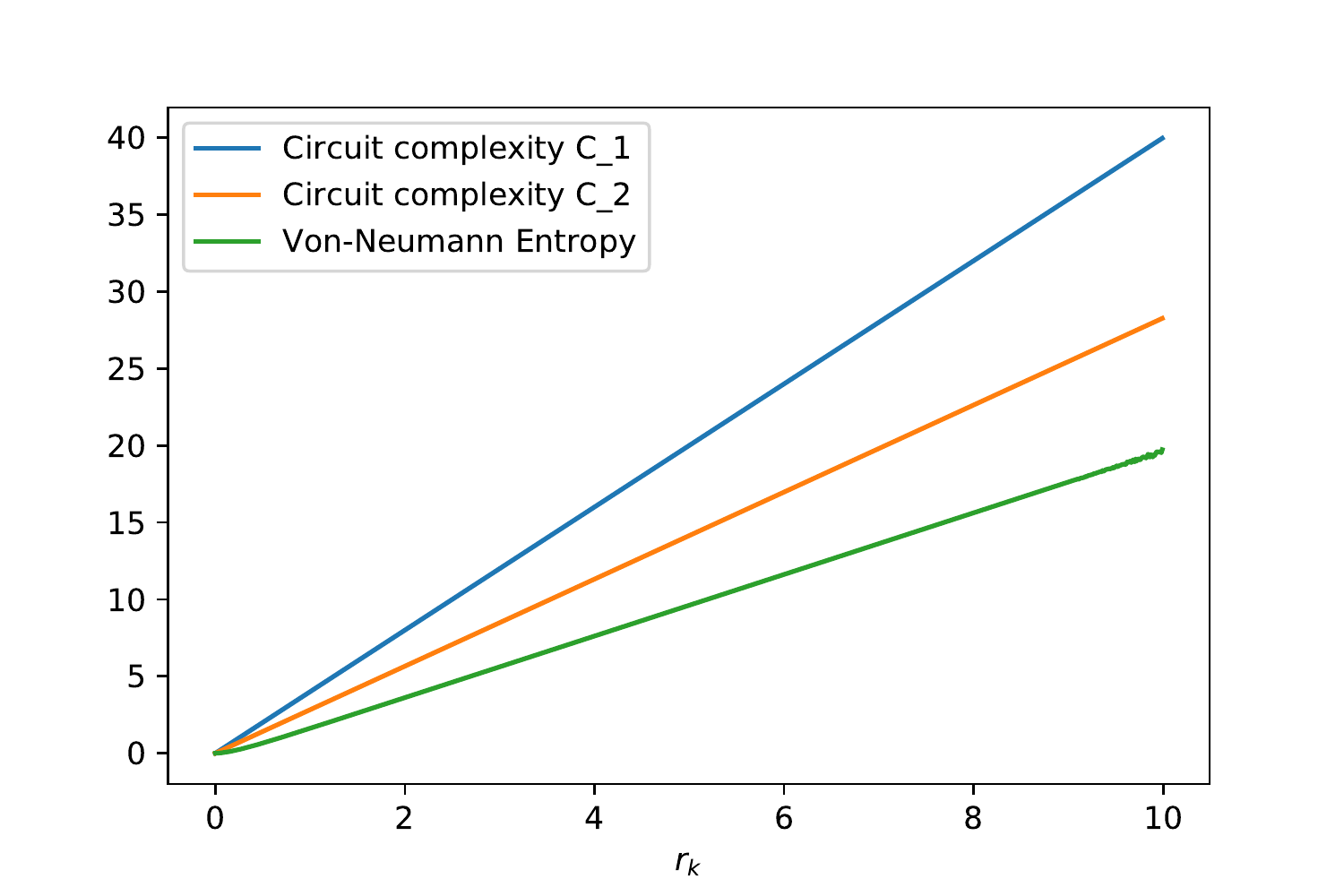}
	\caption{Comparision of the Von-Neumann Entanglement entropy with Circuit complexity }
	\label{complexityVsEntropy}
\end{figure}

\begin{figure}[h!]
	\centering
	\includegraphics[width=9cm,height=8cm]{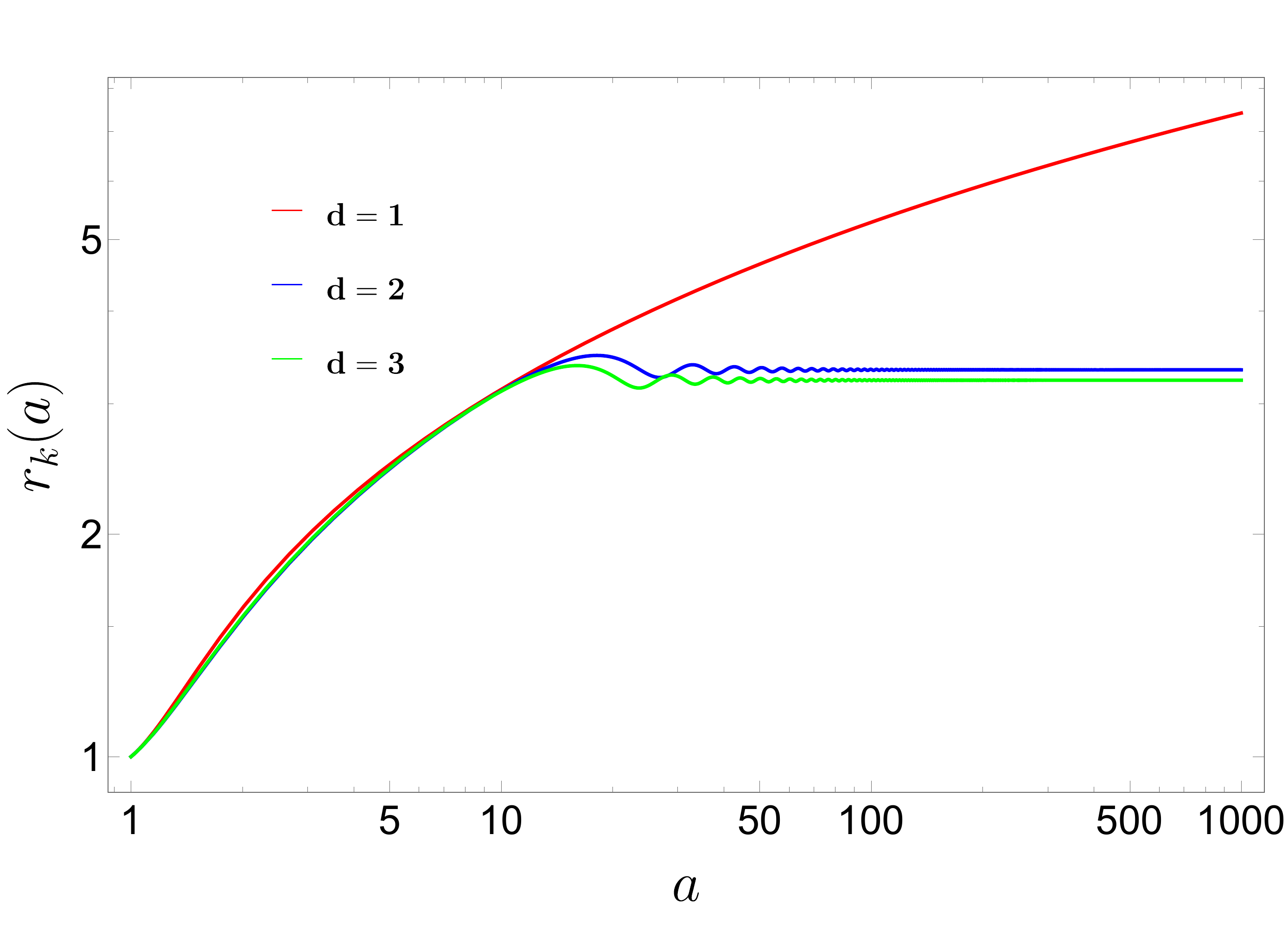}
		\caption{Behaviour of the squeezed state parameter $r_k$ with the scale factor. }
	\label{fig_t_1}
\end{figure}

\begin{figure}[h!]
	\centering
	\includegraphics[width=9cm,height=8cm]{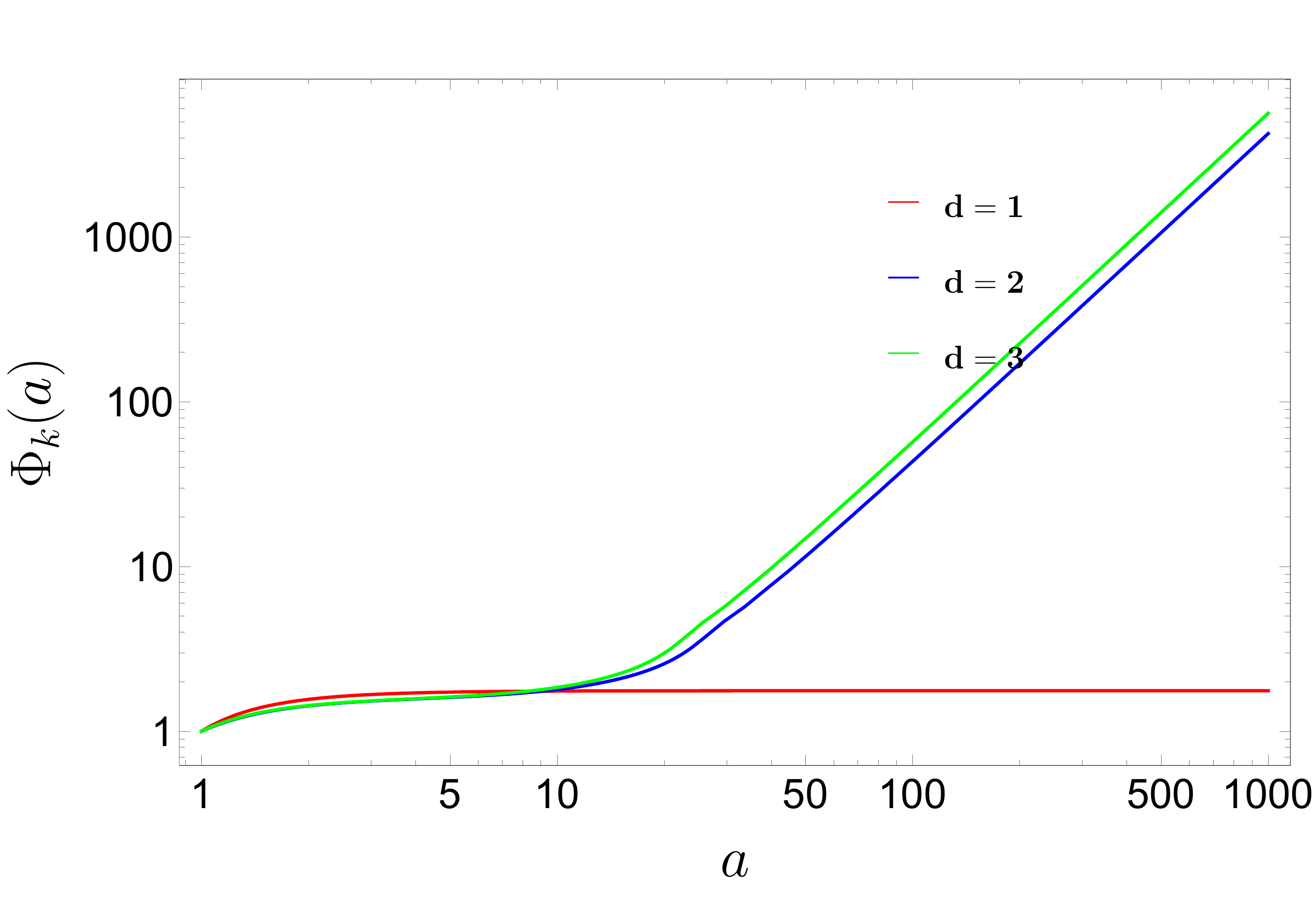}
	\caption{Behaviour of the squeezed state parameter $\phi_k$ with the scale factor. }
	\label{fig_t_2}
\end{figure}
\begin{figure*}[htb]
	\centering
	\subfigure[$\mathcal{C}_1$ from Nielsen's method]{
		\includegraphics[width=8.5cm,height=8.7cm] {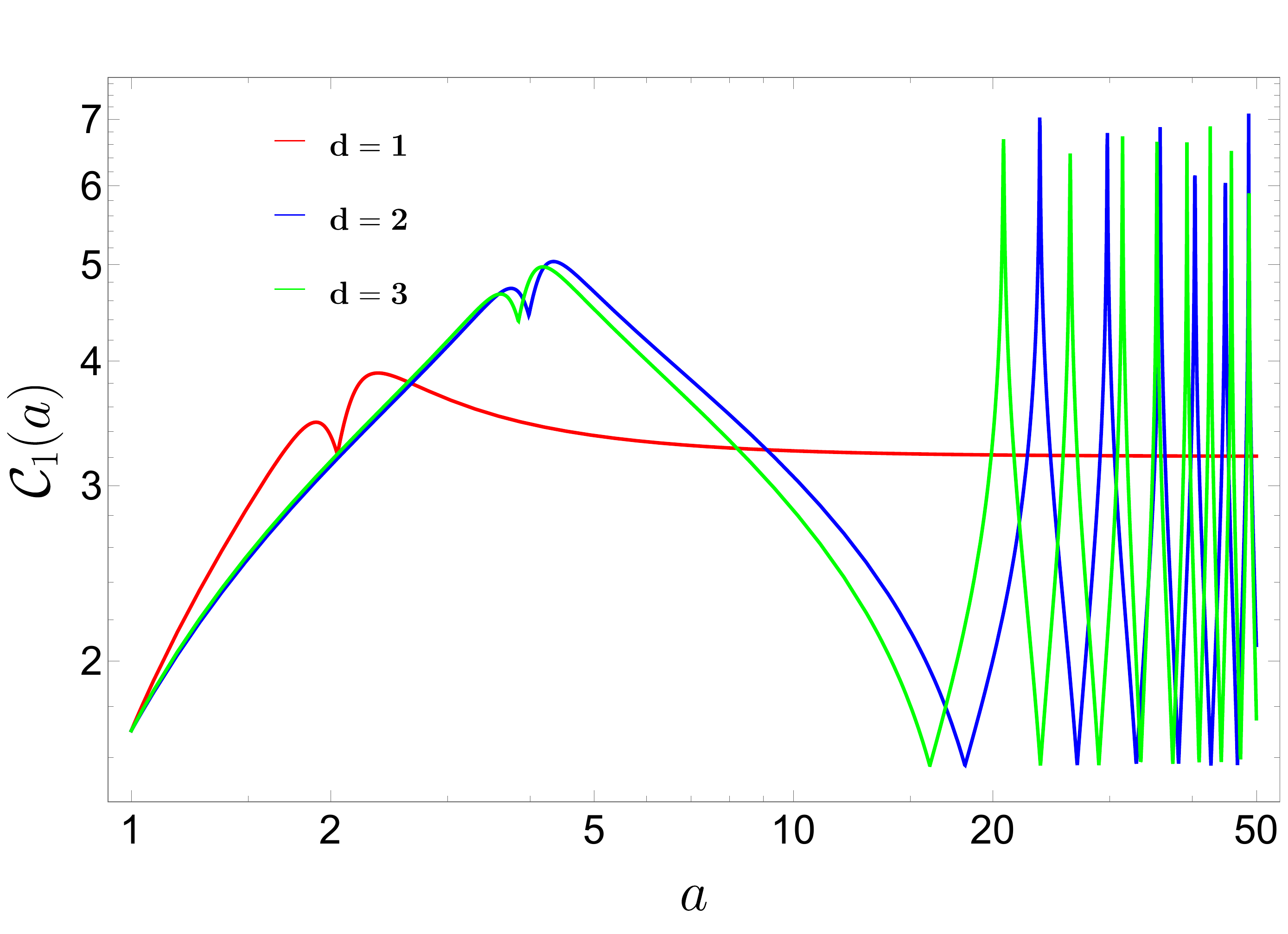}
	}
	\subfigure[$\mathcal{C}_1$ from Covariance matrix method]{
		\includegraphics[width=8.5cm,height=8.7cm] {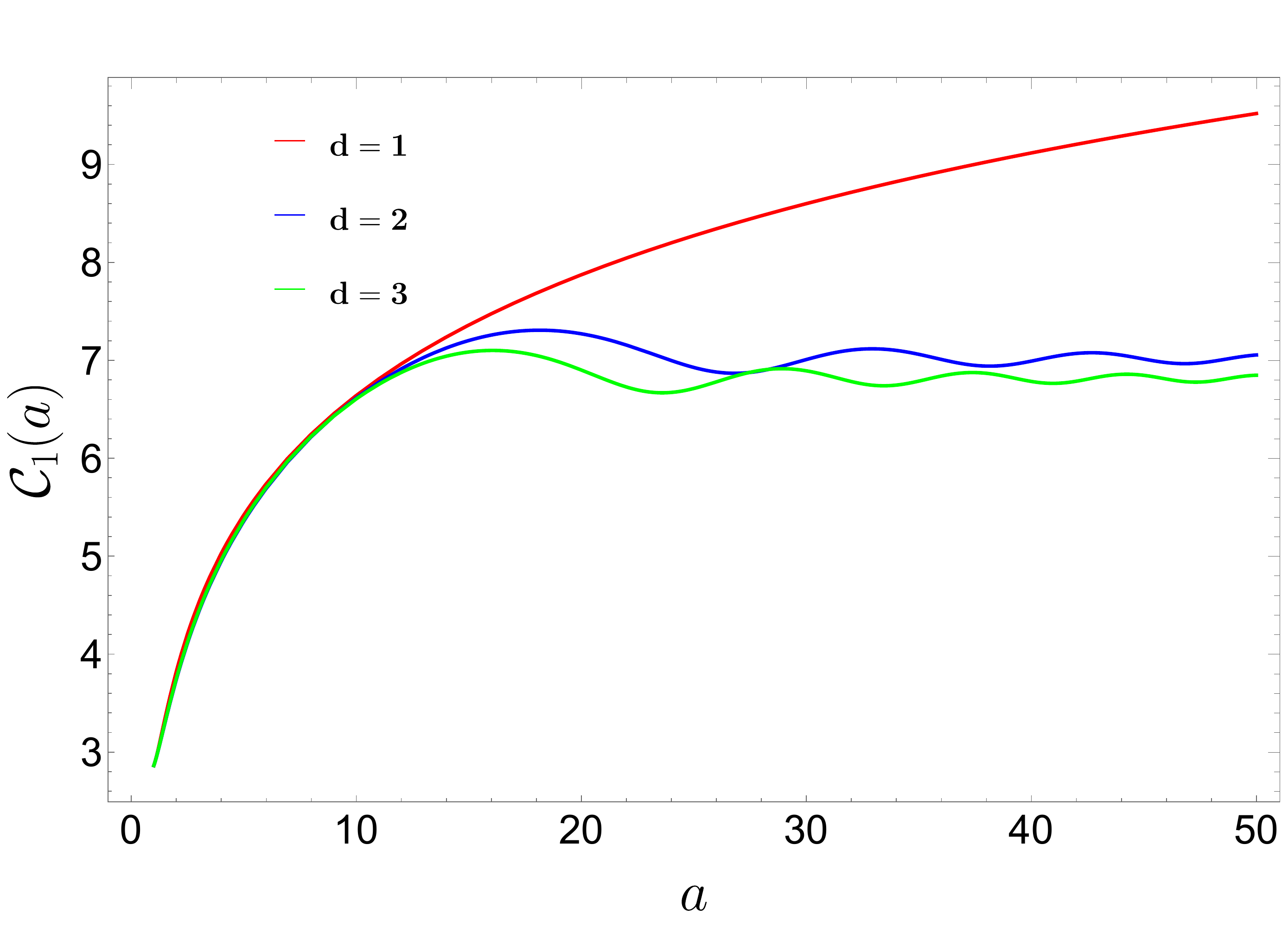}
	}
	\caption{Behavior of the linearly weighted Circuit complexity $\mathcal{C}_1$ with respect to the black hole gas scale factor.}
	\label{fig_t_3}
\end{figure*}

\begin{figure*}[htb]
	\centering
	\subfigure[$\mathcal{C}_2$ from Nielsen's method]{
		\includegraphics[width=8.5cm,height=8.7cm] {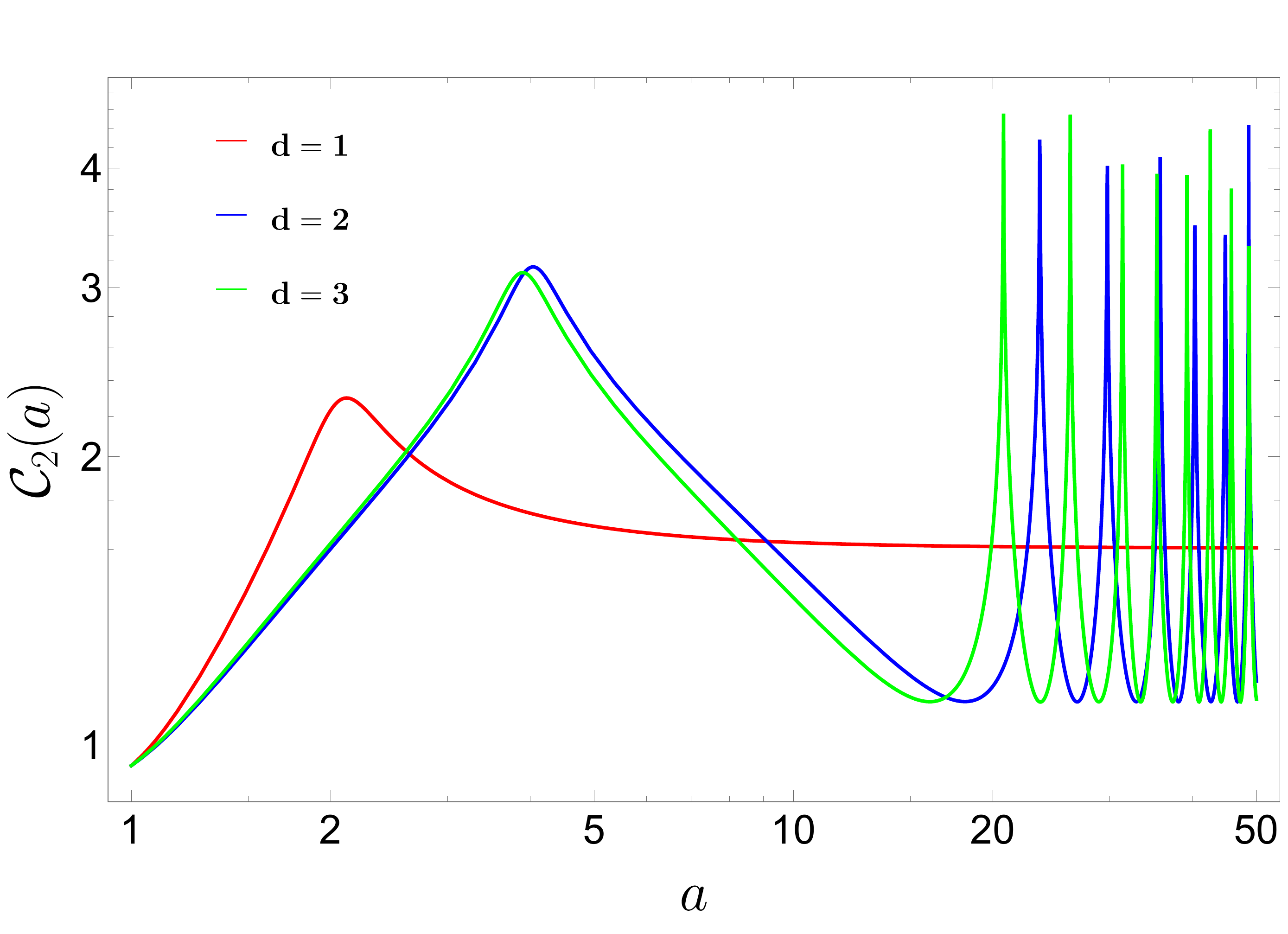}
	}
	\subfigure[$\mathcal{C}_2$ from Covariance matrix method]{
		\includegraphics[width=8.5cm,height=8.7cm] {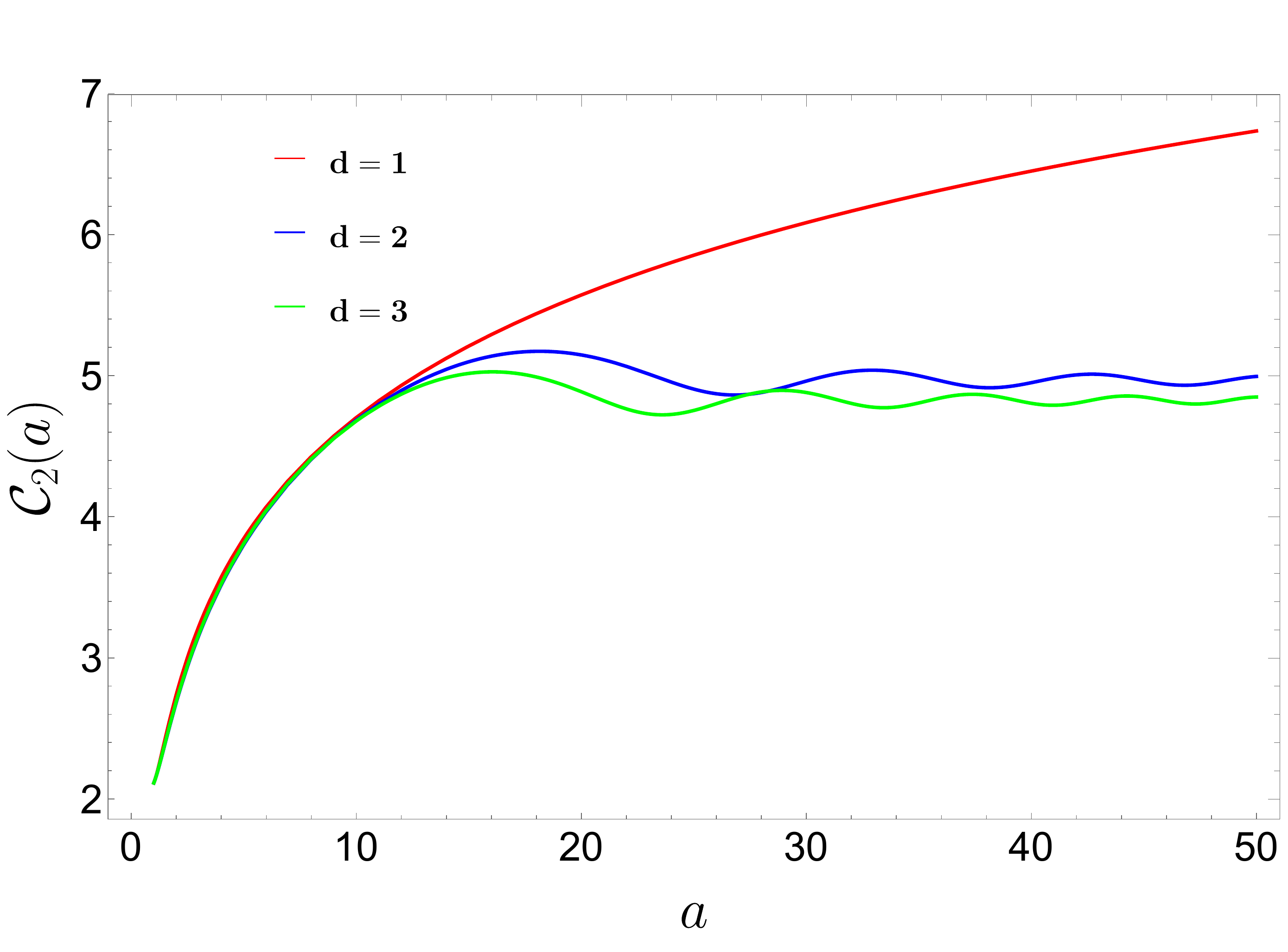}
	}
	\caption{Behavior of the geodesically weighted Circuit complexity $\mathcal{C}_2$ with respect to the black hole gas scale factor.}
	\label{fig_t_4}
\end{figure*}

 \begin{figure}[h!]
 	\centering
 	\includegraphics[width=9cm,height=8.7cm]{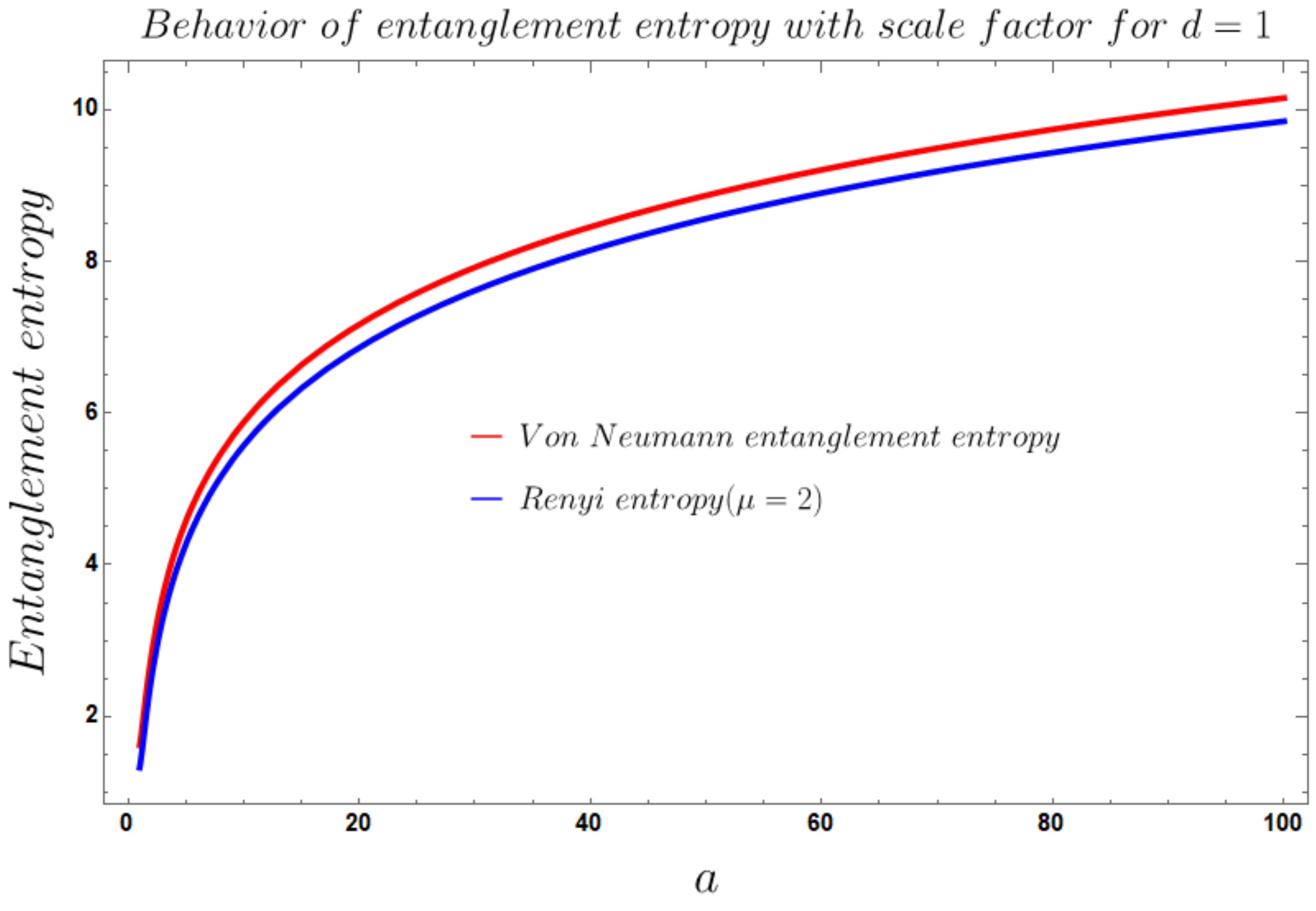}
 	\caption{Behaviour of entanglement entropy vs the scale factor for the black hole gas in d=1 spatial dimension.}
 	\label{fig_Svsa1}
 \end{figure}

\begin{figure}[h!]
	\centering
	\includegraphics[width=9cm,height=8.7cm]{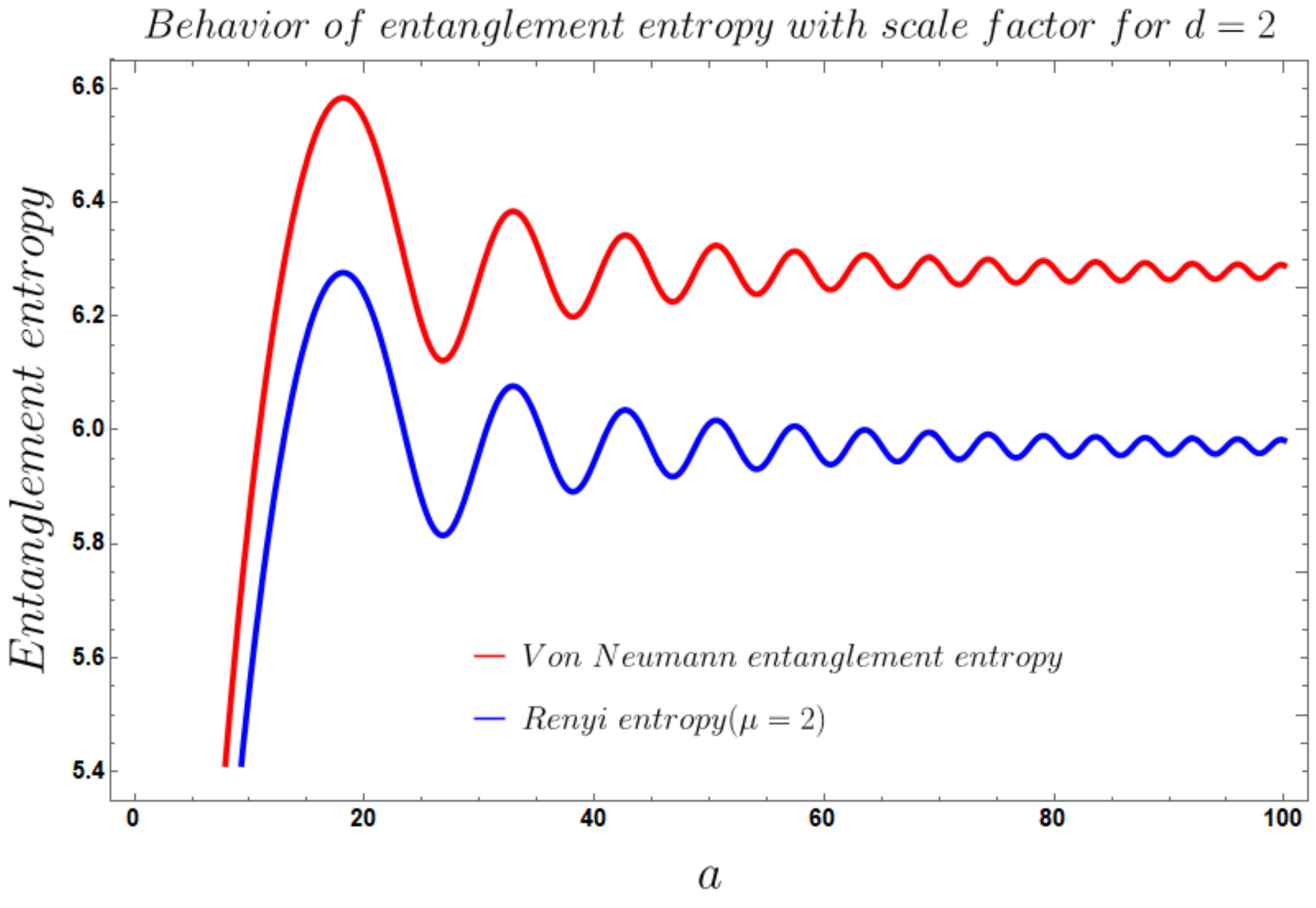}
	\caption{Behaviour of entanglement entropy vs the scale factor for the black hole gas in d=2 spatial dimension.}
	\label{fig_Svsa2}
\end{figure}

\begin{figure}[h!]
	\centering
	\includegraphics[width=9cm,height=8.7cm]{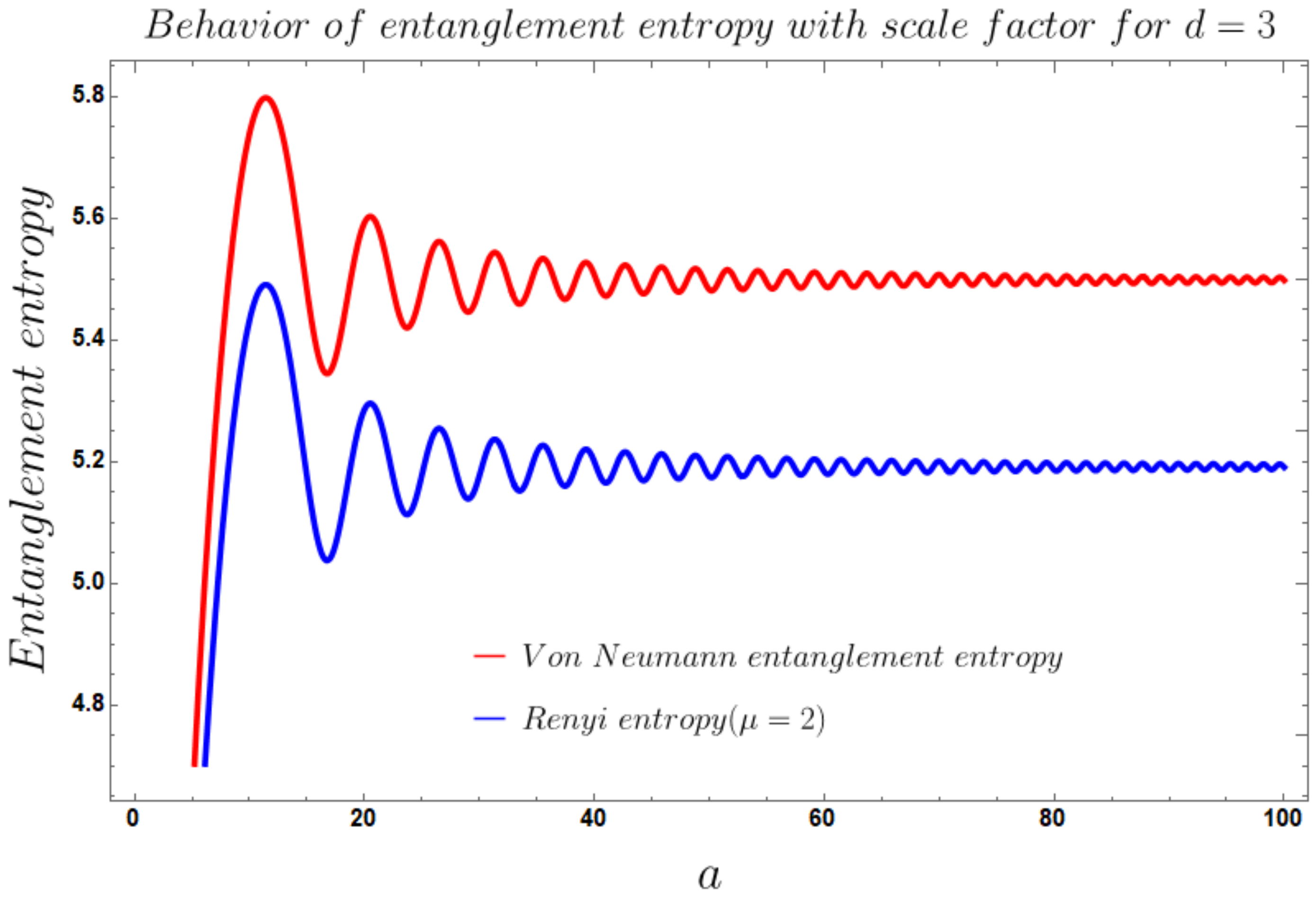}
	\caption{Behaviour of entanglement entropy vs the scale factor for the black hole gas in d=3 spatial dimension.}
	\label{fig_Svsa3}
\end{figure}

\begin{figure*}[htb]
	\centering
	\subfigure[Using Nielsen's method]{
		\includegraphics[width=8.5cm,height=8.7cm] {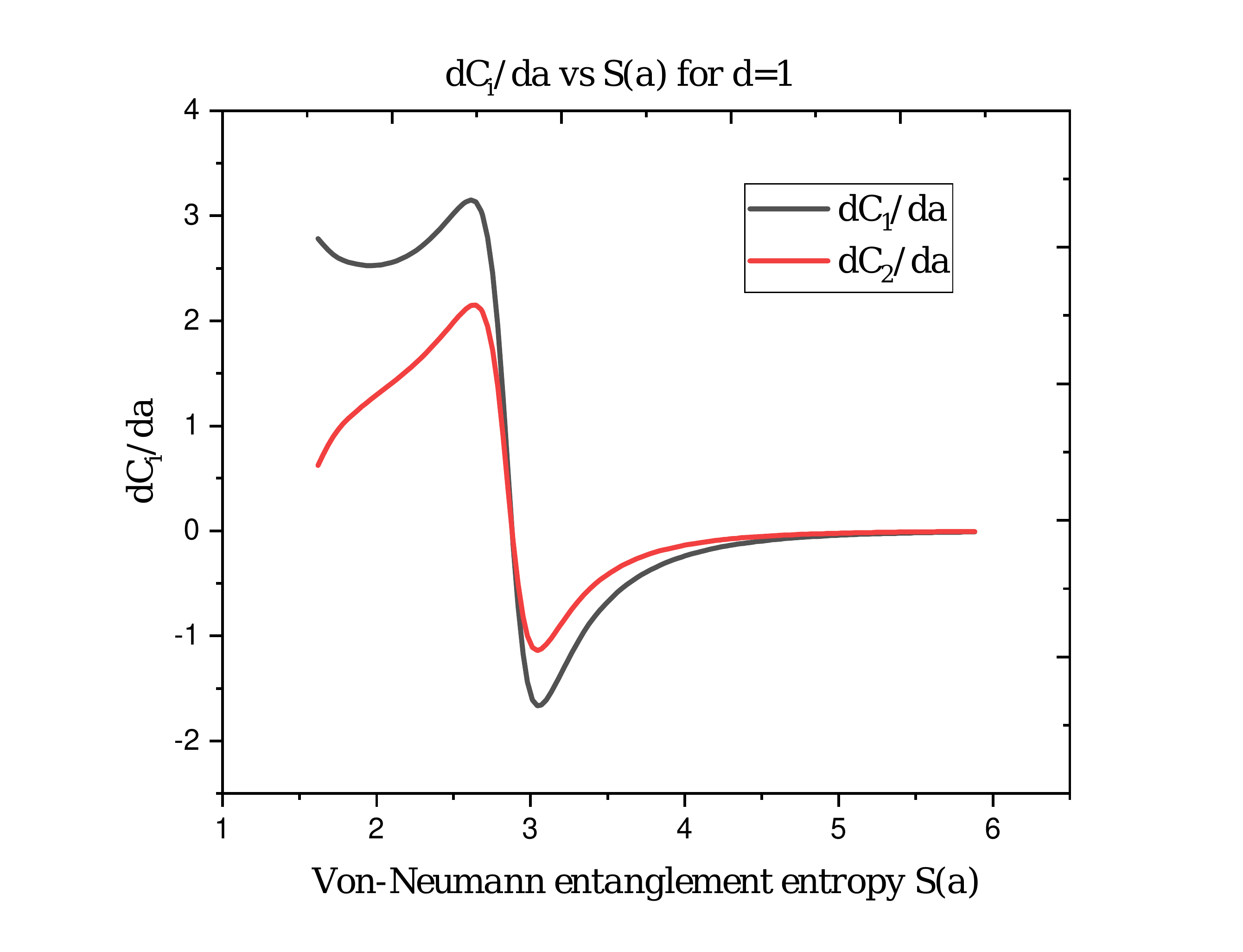}
	}
	\subfigure[Using Covariance matrix method]{
		\includegraphics[width=8.5cm,height=8.7cm] {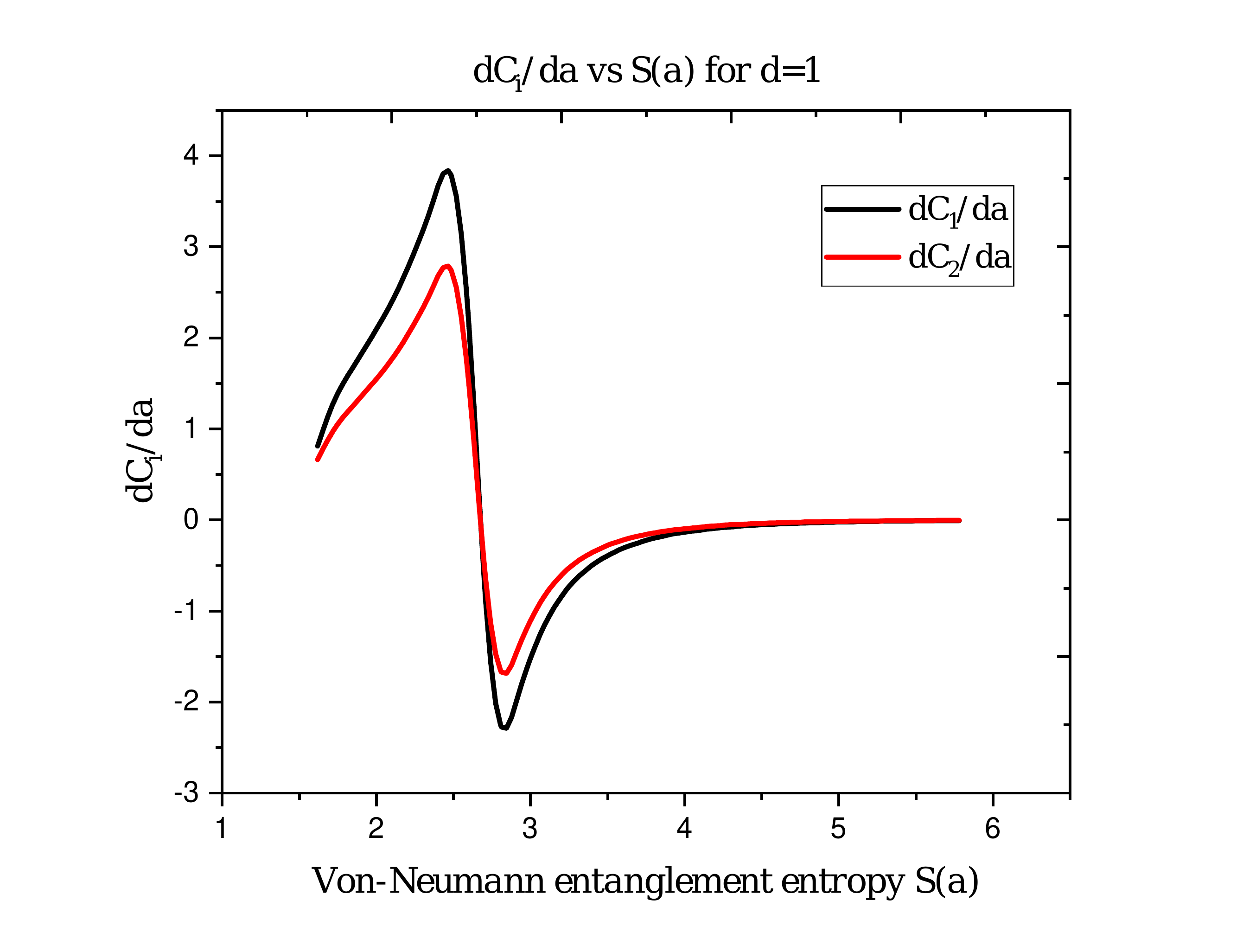}
	}
	\caption{Behavior of $dC_i/da$ vs Von Neumann entanglement entropy for the black hole gas in d=1 spatial dimension. }
	\label{fig_t_5}
\end{figure*}

\begin{figure*}[htb]
	\centering
	\subfigure[Using Nielsen's method.]{
		\includegraphics[width=8.5cm,height=8.7cm] {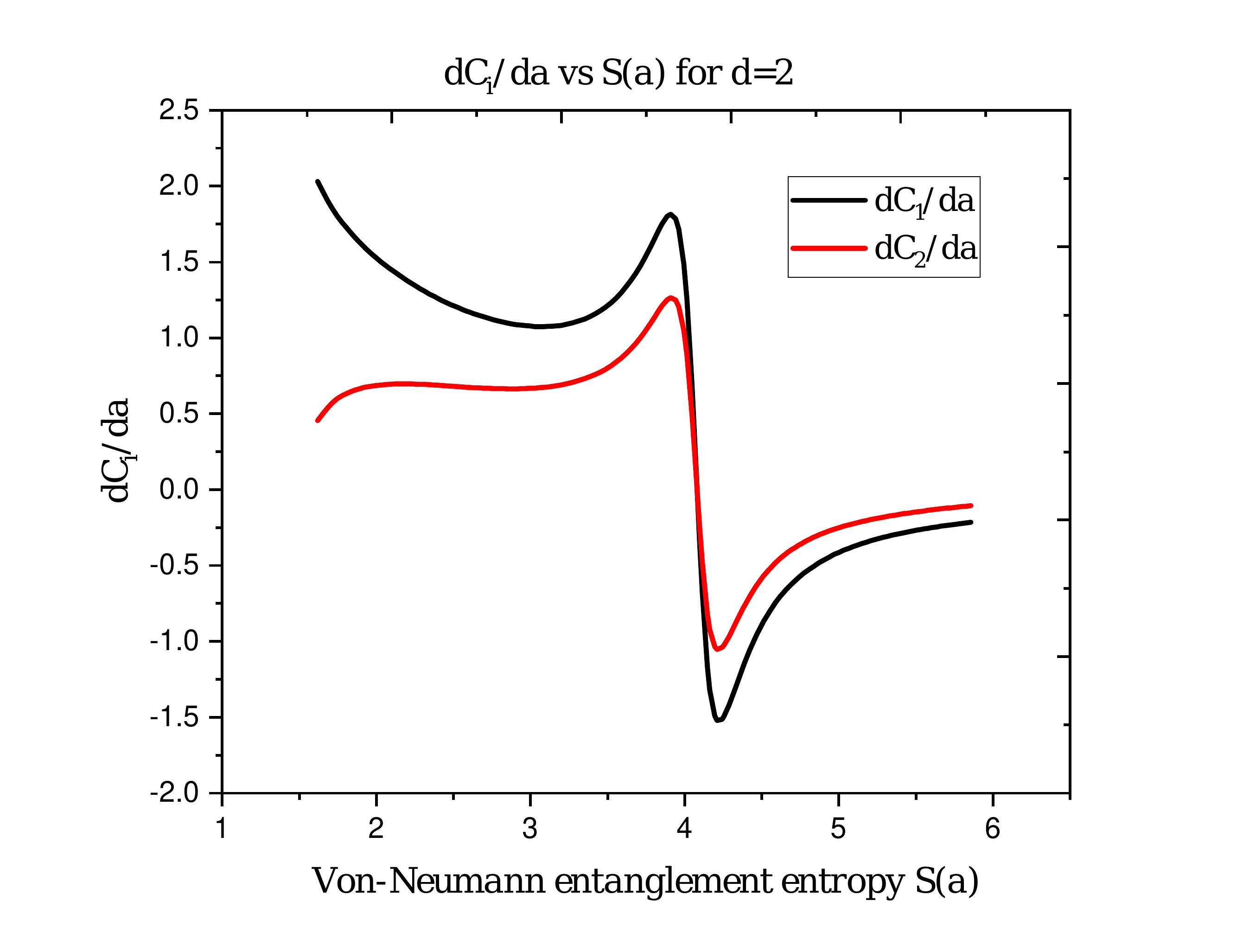}
	}
	\subfigure[Using Covariance matrix method.]{
		\includegraphics[width=8.5cm,height=8.7cm] {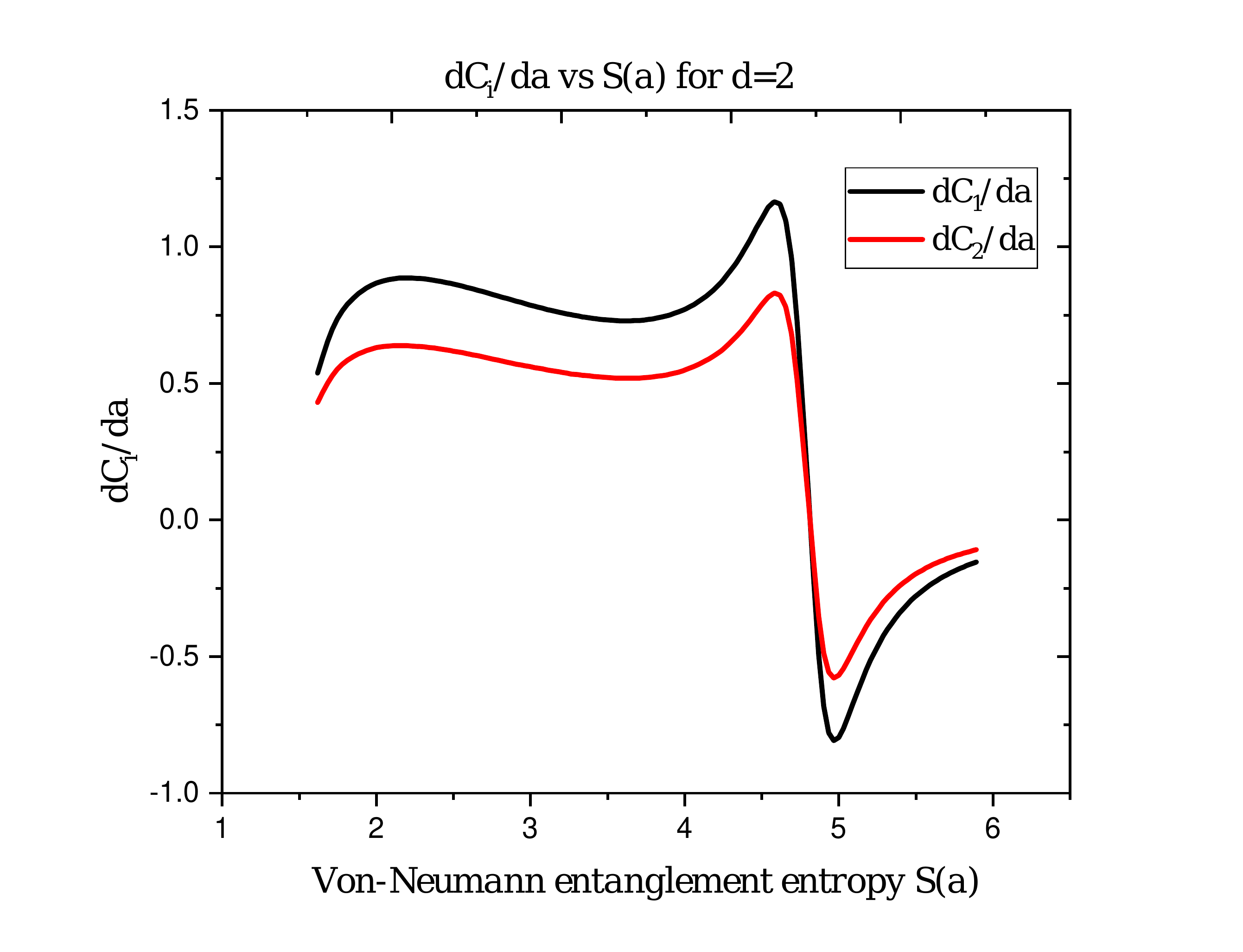}
	}
	\caption{Behavior of $dC_i/da$ vs Von Neumann entanglement entropy for the black hole gas in d=2 spatial dimension }
	\label{fig_t_6}
\end{figure*}

\begin{figure*}[htb]
	\centering
	\subfigure[Using Nielsen's method.]{
		\includegraphics[width=8.5cm,height=8cm] {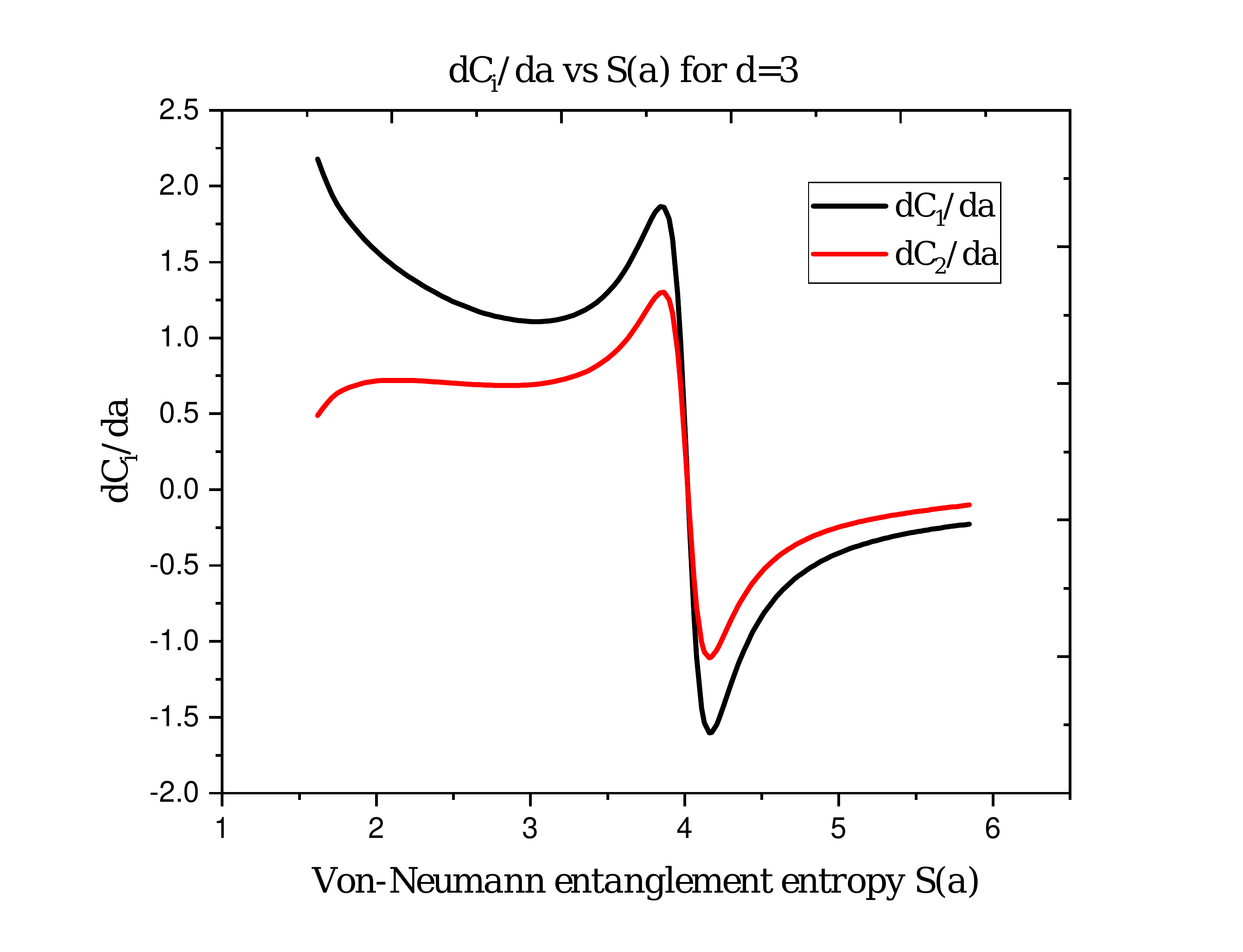}
	}
	\subfigure[Using Covariance matrix method.]{
		\includegraphics[width=8.5cm,height=8cm] {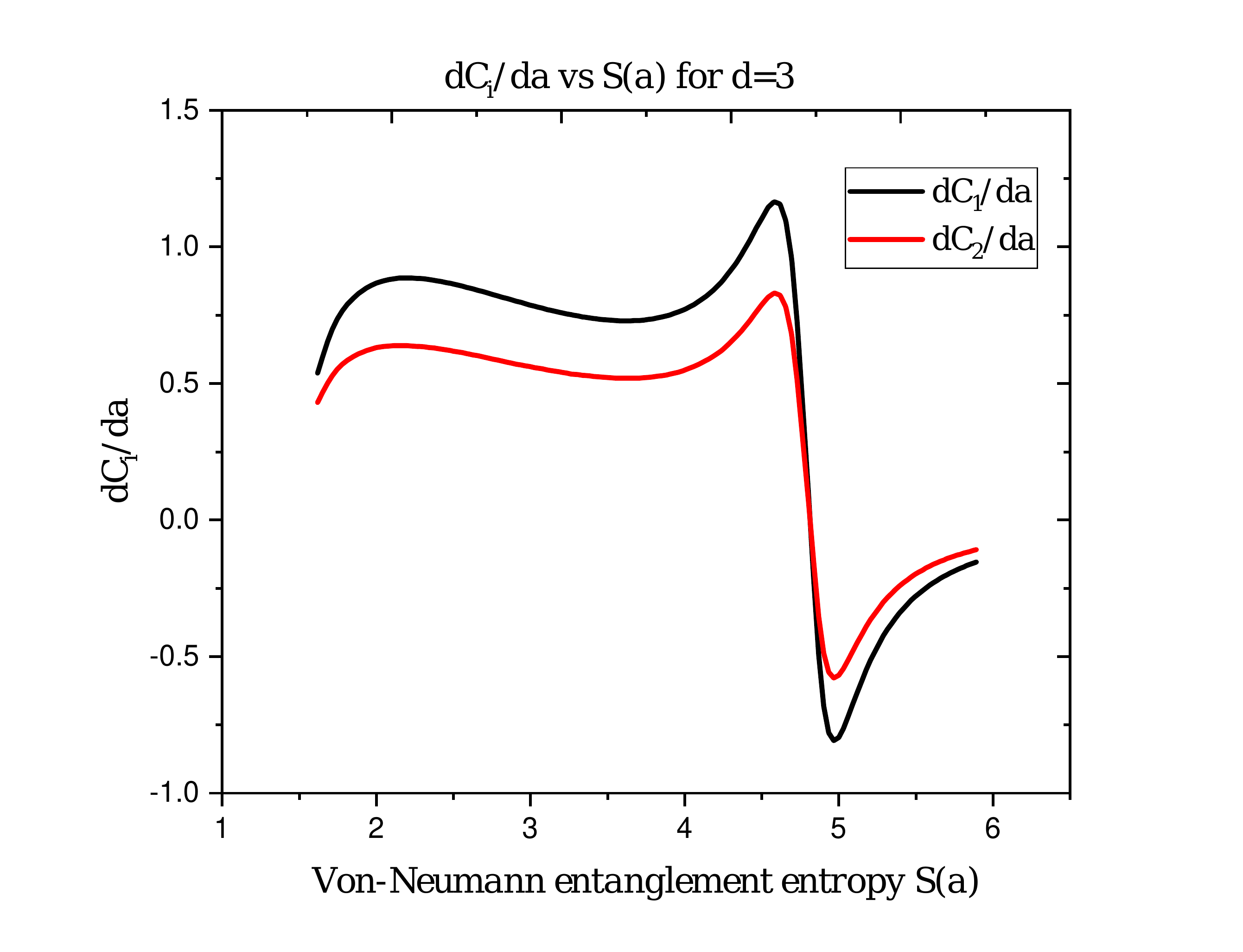}
	}
	\caption{Behavior of $dC_i/da$ vs Von Neumann entanglement entropy for the black hole gas in d=3 spatial dimension }
	\label{fig_t_7}
\end{figure*}

\begin{figure*}[htb]
	\centering
	\subfigure[Using Nielsen's method.]{
		\includegraphics[width=8.5cm,height=8cm] {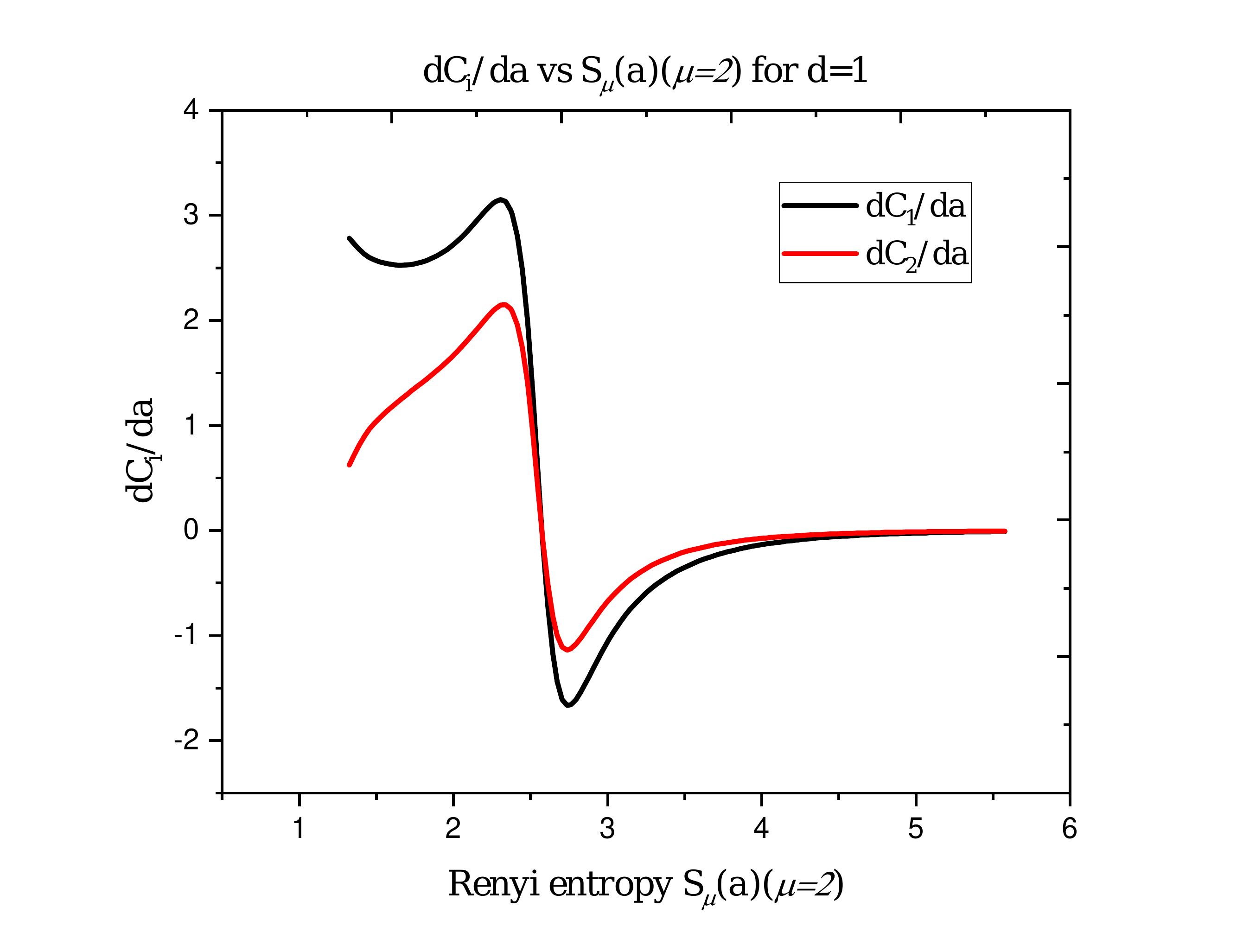}
	}
	\subfigure[Using Covariance matrix method.]{
		\includegraphics[width=8.5cm,height=8cm] {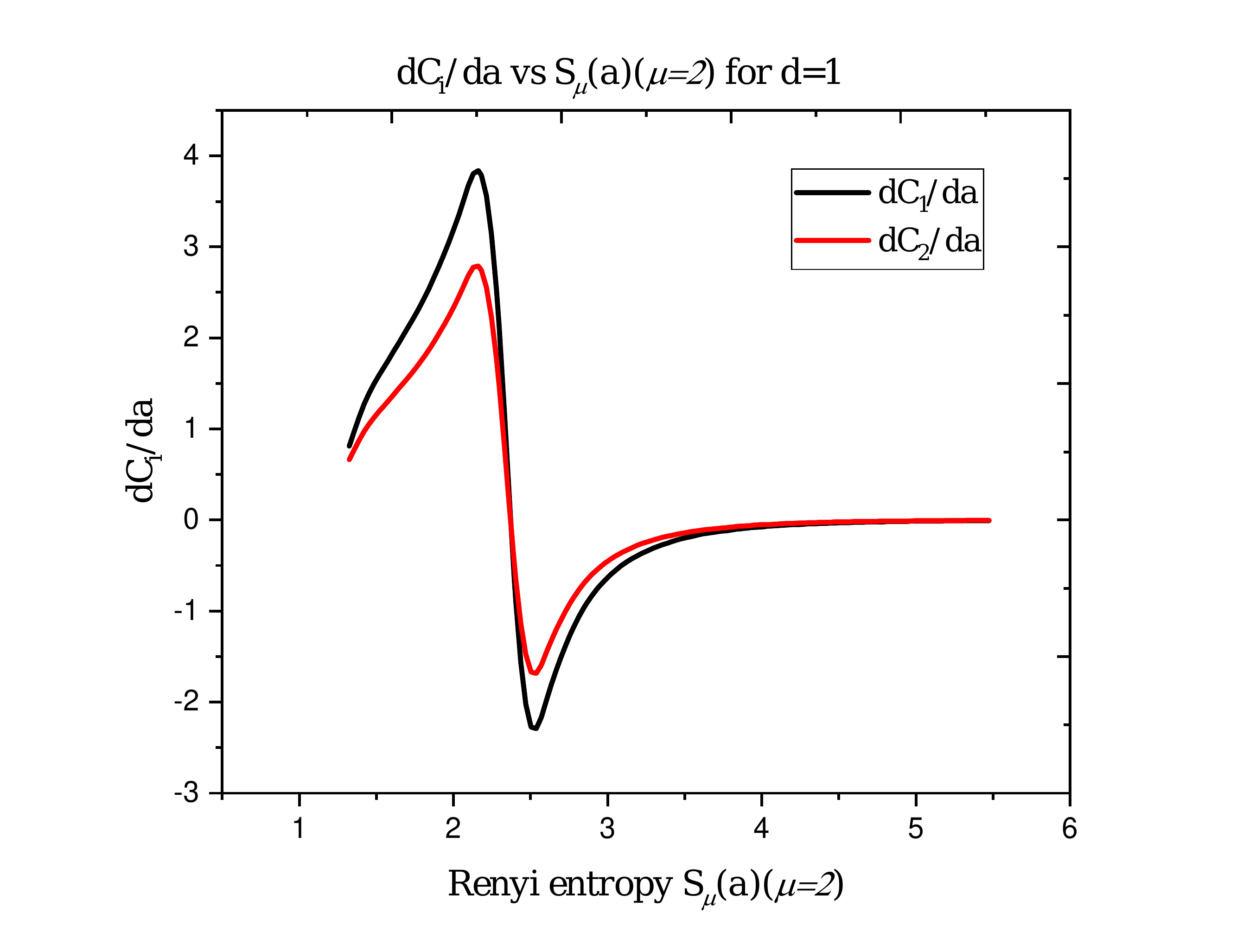}
	}
	\caption{Behavior of $dC_i/da$  vs Rényi entropy for the black hole gas in d=1 spatial dimension. }
	\label{fig_t_8}
\end{figure*}

\begin{figure*}[htb]
	\centering
	\subfigure[Using Nielsen's method.]{
		\includegraphics[width=8.5cm,height=8cm] {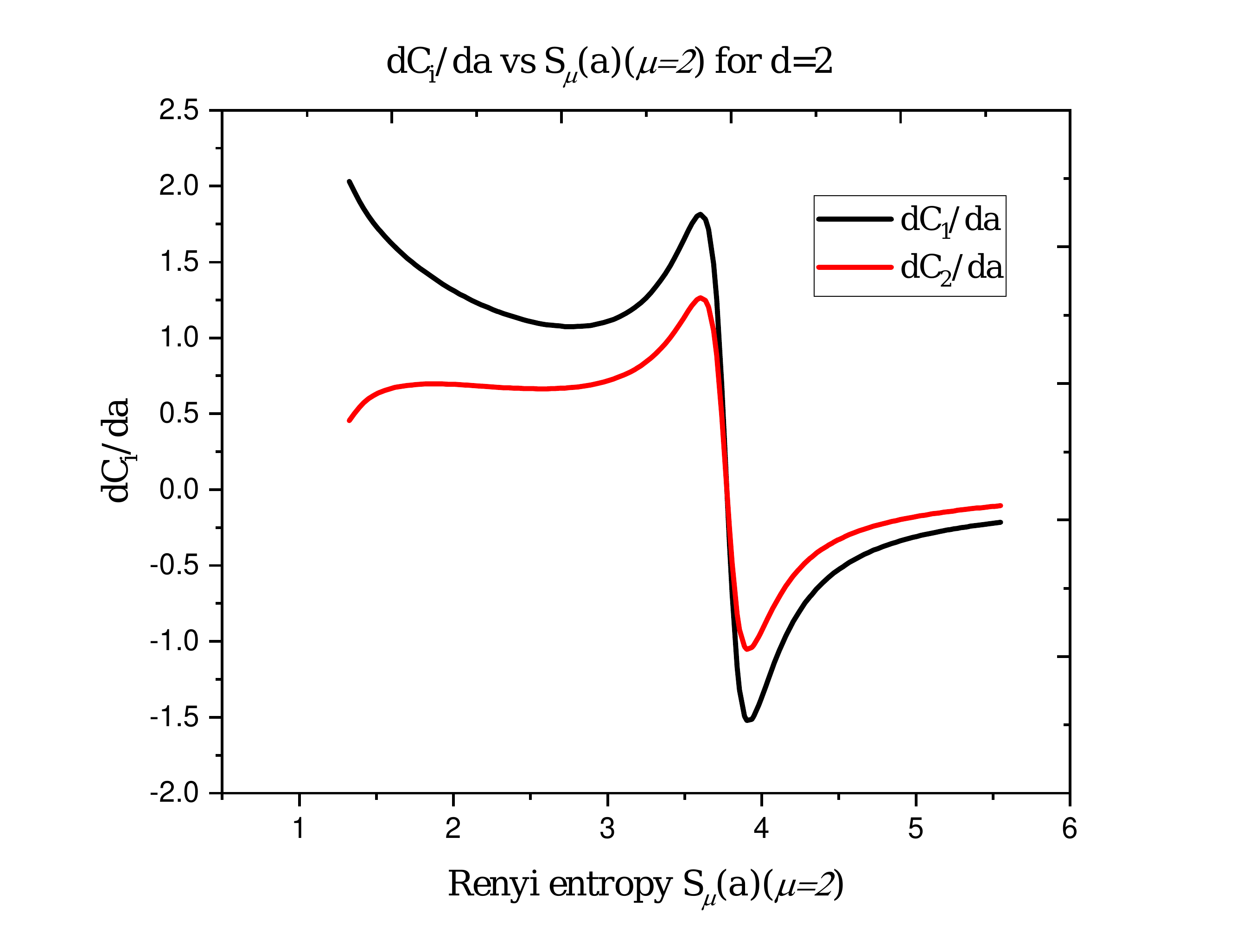}
	}
	\subfigure[Using Covariance matrix method.]{
		\includegraphics[width=8.5cm,height=8cm] {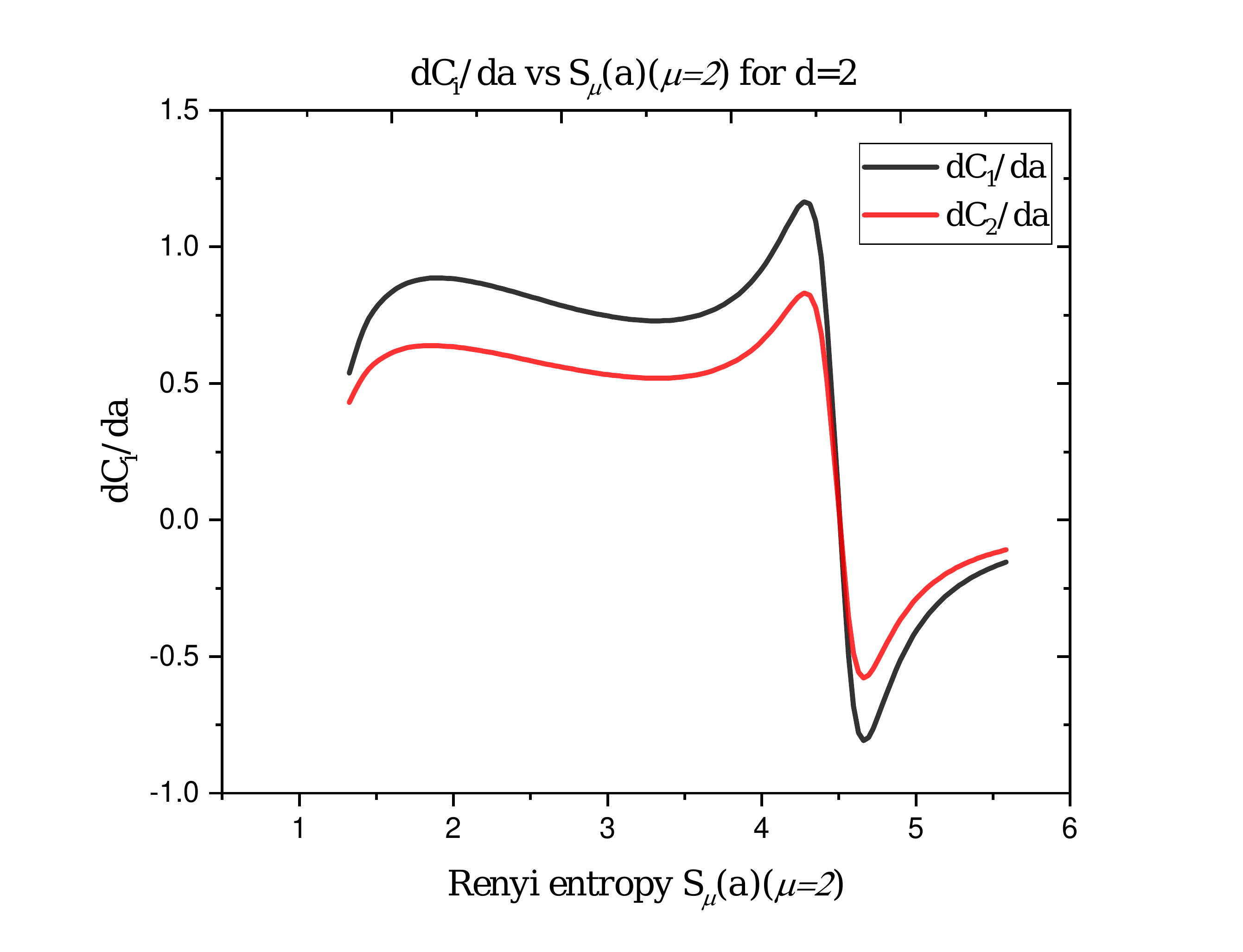}
	}
	\caption{Behavior of $dC_i/da$ vs Rényi entropy for the black hole gas in d=2 spatial dimension. }
	\label{fig_t_9}
\end{figure*}

\begin{figure*}[htb]
	\centering
	\subfigure[Using Nielsen's method.]{
		\includegraphics[width=8.5cm,height=8cm] {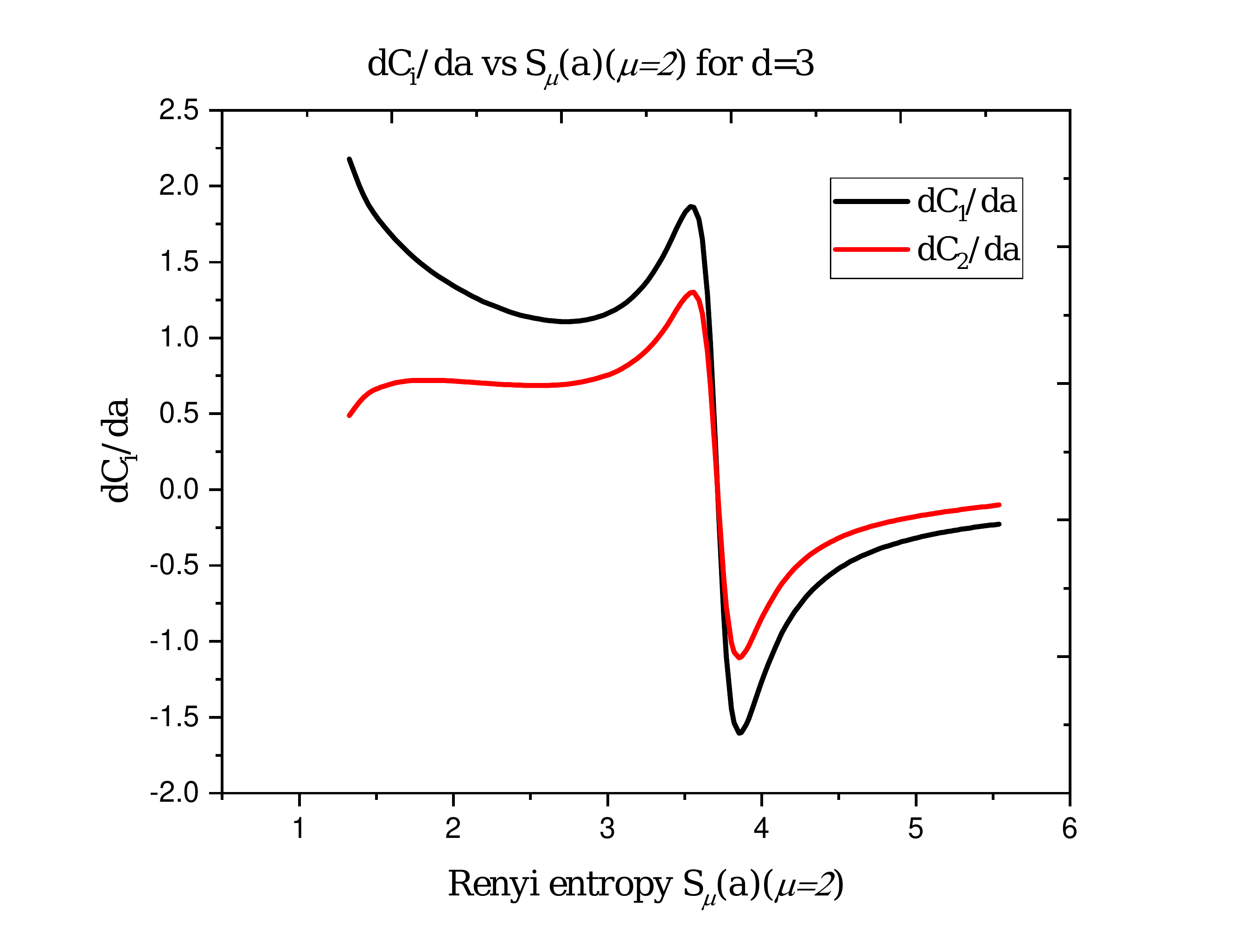}
	}
	\subfigure[Using Covariance matrix method.]{
		\includegraphics[width=8.5cm,height=8cm] {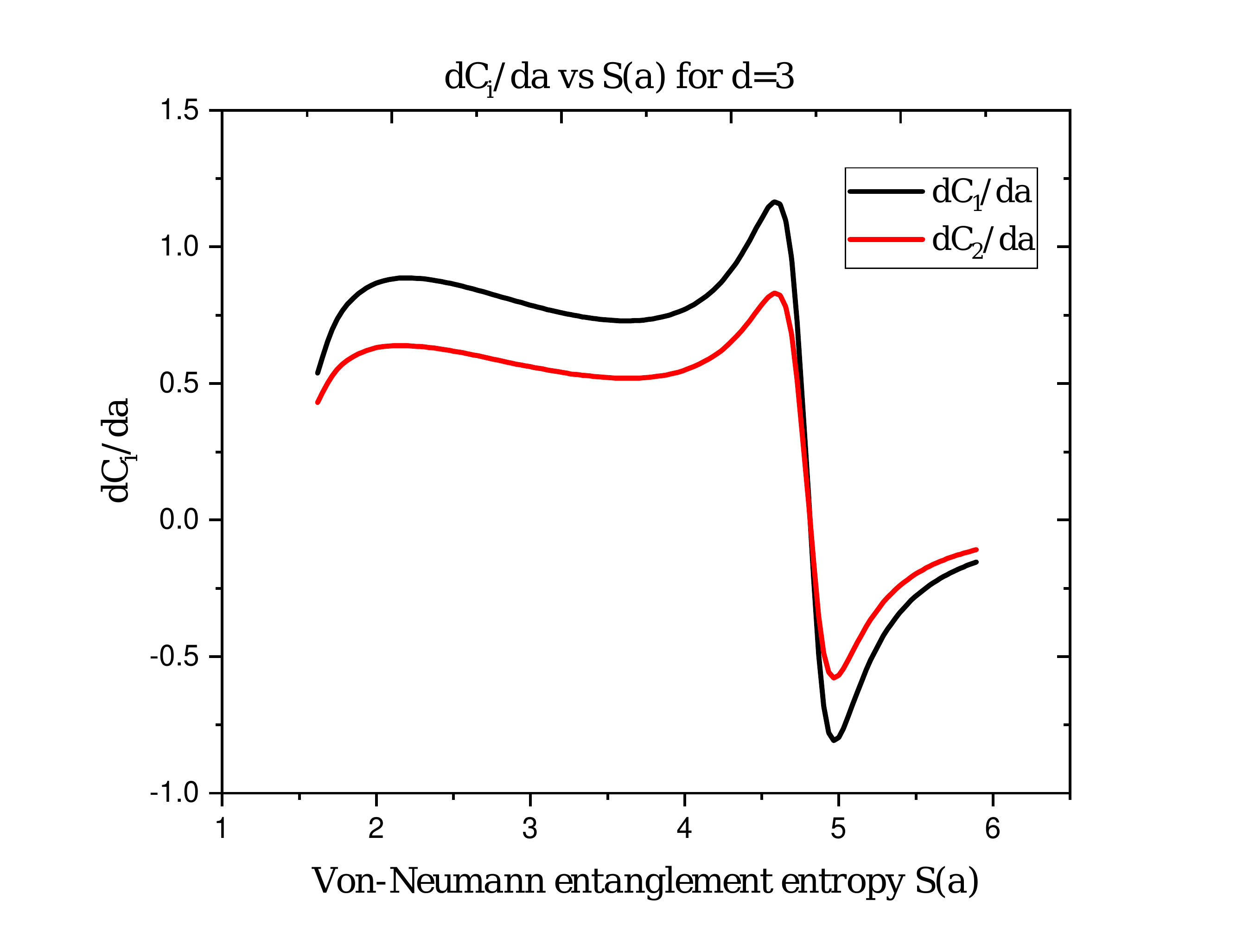}
	}
	\caption{Behavior of $dC_i/da$ vs Rényi entropy for the black hole gas in d=3 spatial dimension. }
	\label{fig_t_10}
\end{figure*}


\begin{figure*}[htb]
	\centering
	\subfigure[Using Nielsen's method.]{
		\includegraphics[width=8.5cm,height=8cm] {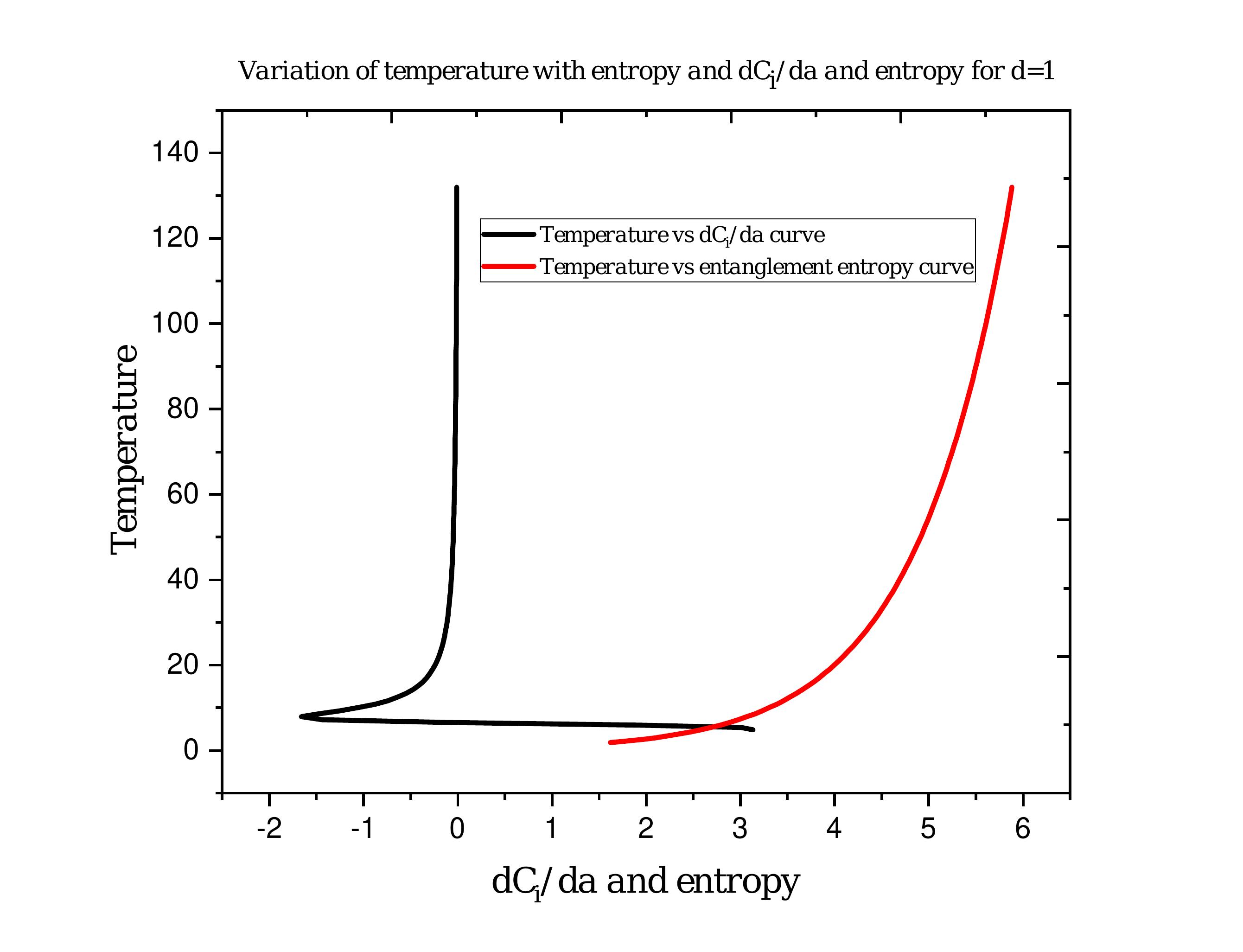}
	}
	\subfigure[Using Covariance matrix method.]{
		\includegraphics[width=8.5cm,height=8cm] {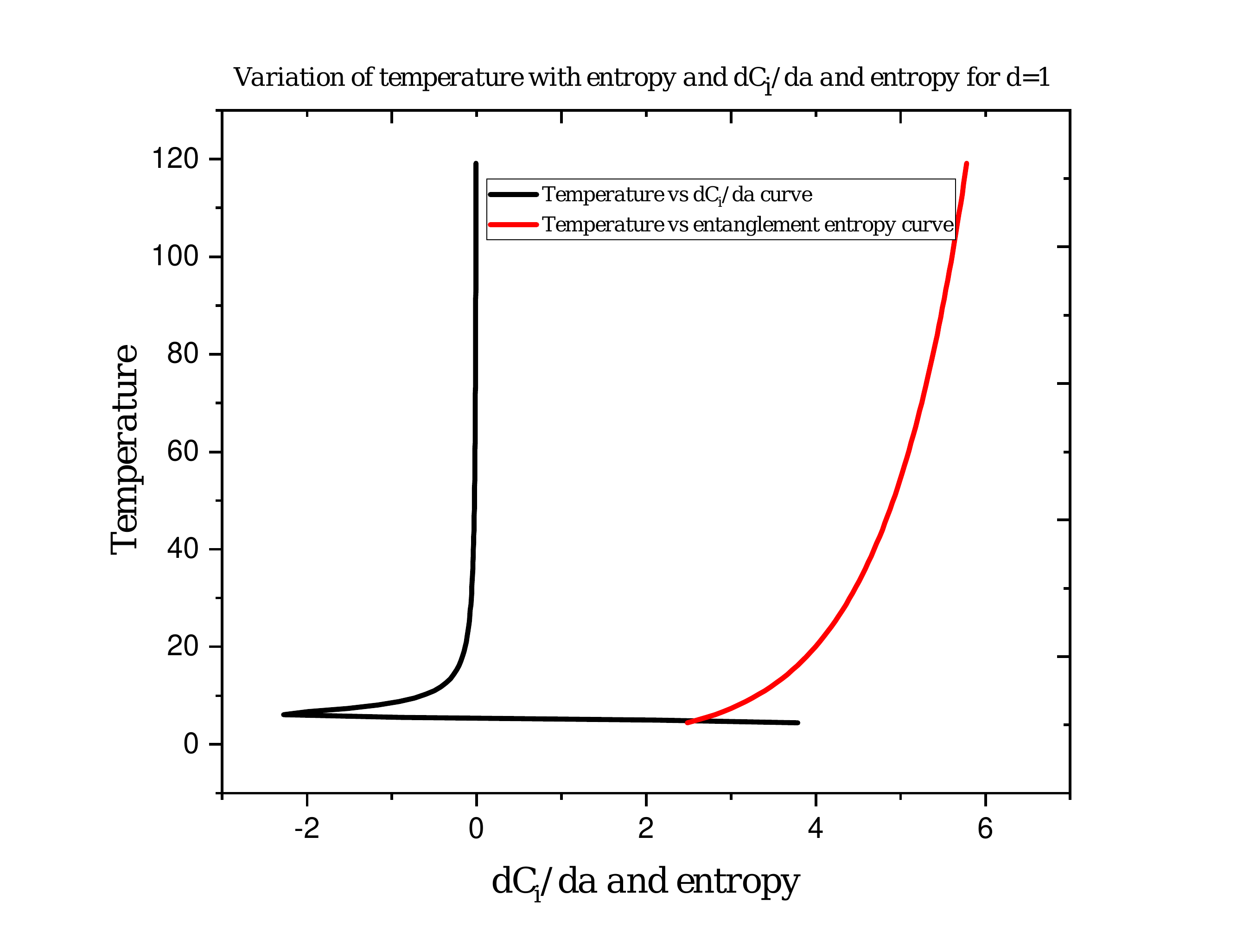}
	}
	\caption{Behaviour of equilibrium temperature of the black hole gas w.r.t.  $dC_i/da$ and entanglement entropy in d=1 spatial dimension. }
	\label{fig_t_11}
\end{figure*}

\begin{figure*}[htb]
	\centering
	\subfigure[Using Nielsen's method.]{
		\includegraphics[width=8.5cm,height=8cm] {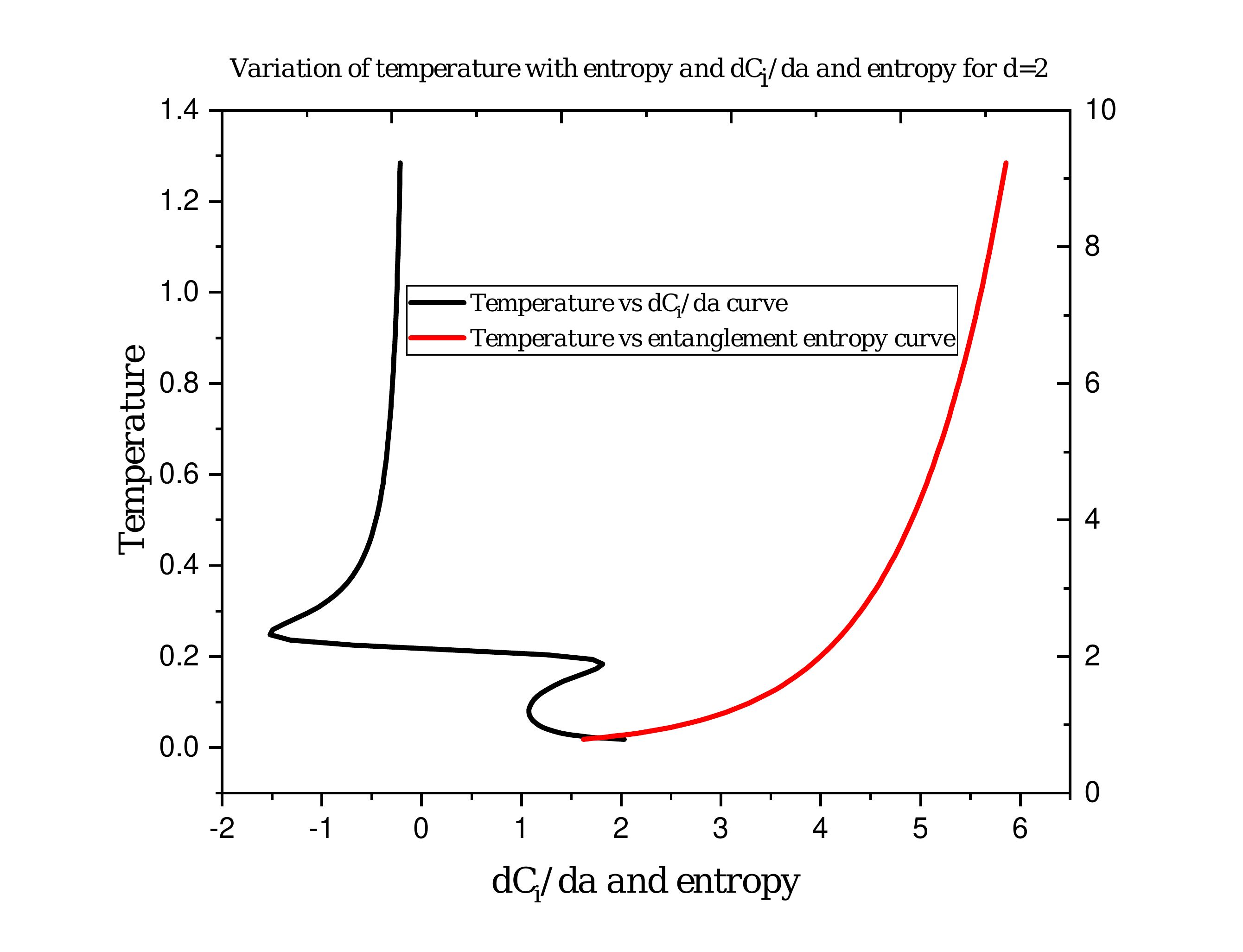}
	}
	\subfigure[Using Covariance matrix method.]{
		\includegraphics[width=8.5cm,height=8cm] {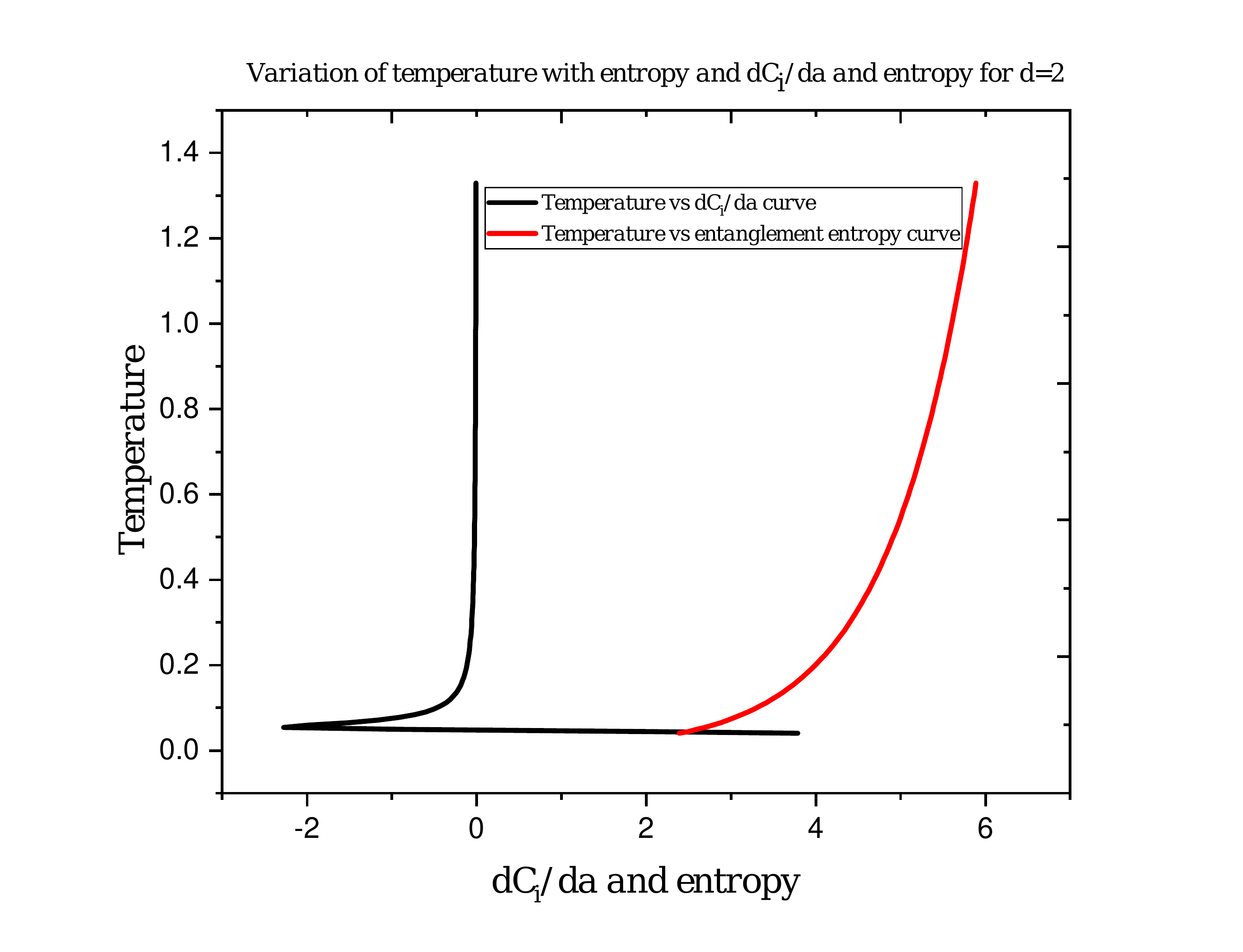}
	}
	\caption{Behaviour of equilibrium temperature of the black hole gas w.r.t.  $dC_i/da$ and entanglement entropy in d=2 spatial dimension. }
	\label{fig_t_12}
\end{figure*}
\begin{figure*}[htb]
	\centering
	\subfigure[Using Nielsen's method.]{
		\includegraphics[width=8.5cm,height=8cm] {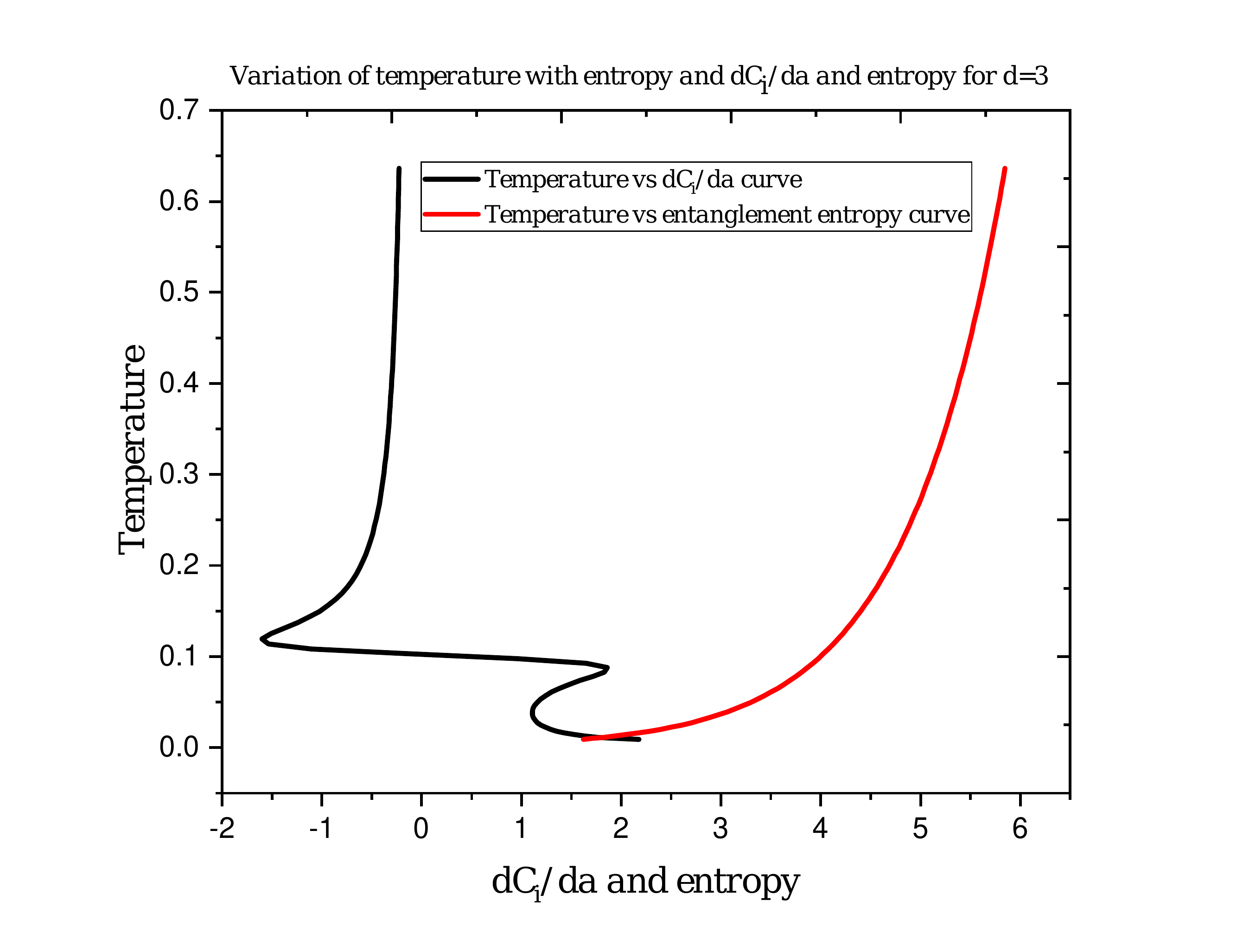}
	}
	\subfigure[Using Covariance matrix method.]{
		\includegraphics[width=8.5cm,height=8cm] {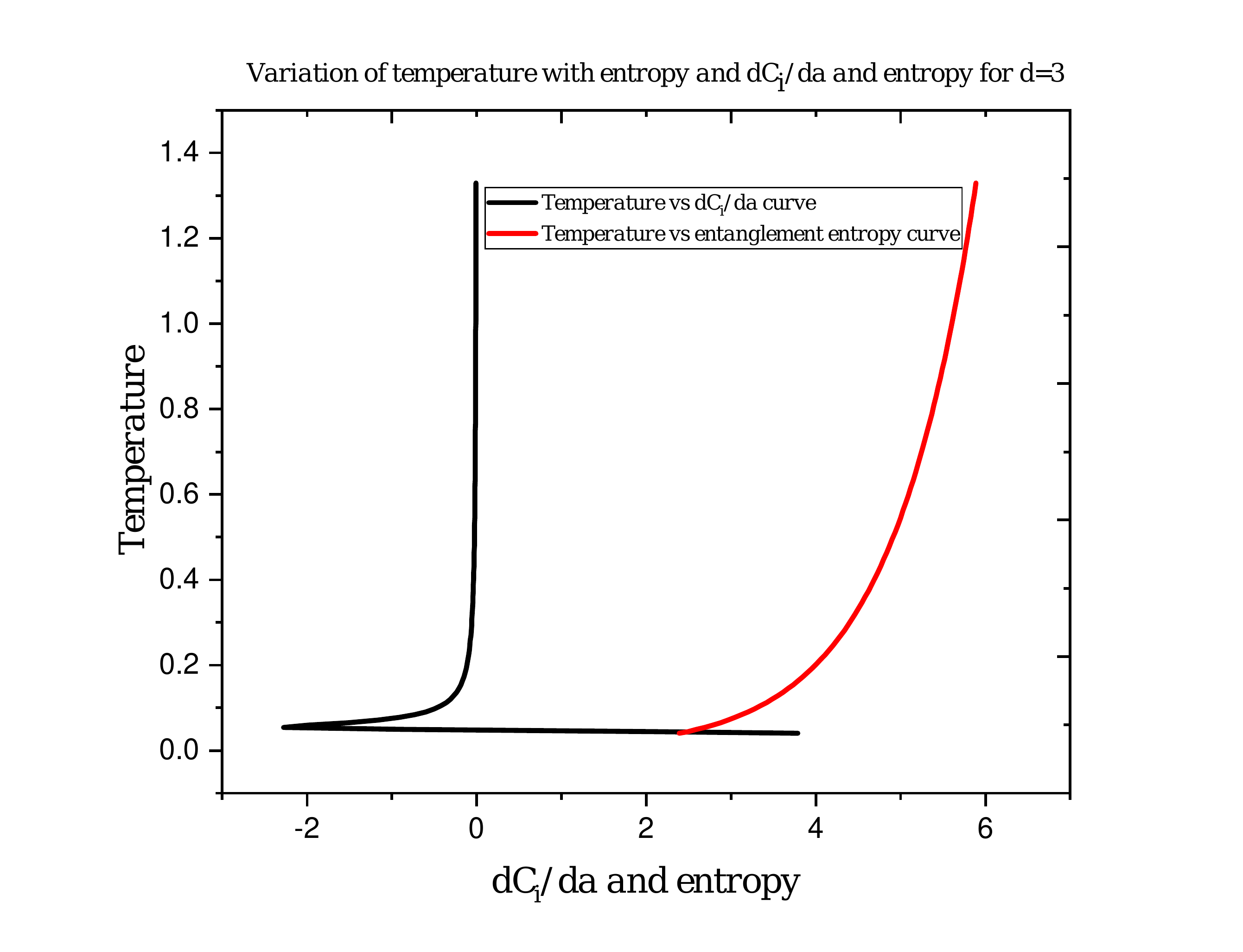}
	}
	\caption{Behaviour of equilibrium temperature of the black hole gas w.r.t.  $dC_i/da$ and entanglement entropy in d=3 spatial dimension. }
	\label{fig_t_13}
\end{figure*}
For very large squeezing parameter, we get:
\begin{equation}
    S_\mu(r_k \rightarrow \infty) \approx \frac{2 \mu r_k}{(\mu - 1)}
\end{equation}

If we take the limit $\mu \rightarrow 1$, we get the Von-Neumann entropy \ref{eq:entanglementEntropy}. Meanwhile, Rényi-2 entropy is given by $S_2(r_k) = \text{ln cosh}2r_k$. 

One can also calculate effective temperature of the source by computing the thermal distribution with an average photon number $\langle\hat{n}_i\rangle = \text{sinh}^2 r_k$. The average photon number of the thermal field is given by: 
\begin{equation}
    \langle\hat{n}_i\rangle = \Bar{n} = \frac{1}{\text{exp}(\hbar\omega/k_BT)-1}
\end{equation}
Then, one can compute the effective temperature as:
\begin{align}
\nonumber
    T &= \frac{\hbar\omega_i}{k_B} \text{ln}\left( \frac{\langle\hat{n}_i\rangle}{ \langle\hat{n}_i\rangle +1 } \right) \\ \nonumber
    &=  \frac{\hbar\omega_i}{k_B} \text{ln}\left( \frac{\text{sinh}^2 r_k }{\text{sinh}^2 r_k +1} \right) \\
    &= \frac{\hbar\omega_i}{2k_B\text{ln(coth}r_k)}
\end{align}
where, $\omega_i = i/c$ is the frequence of the mode and $i \in (k,-k)$.
\section*{\textcolor{Sepia}{\textbf{ \Large Quantum Circuit Complexity Vs Entanglement}}}
Now, the comment on comparison of entanglement entropy with circuit complexity is in order. It was very recently shown in \cite{Eisert:2021mjg} that there exists some relationship between entangling power and circuit complexity.  Most importantly, if the entanglement entropy grows linearly with time, the geometric circuit complexity also grows linearly.

Generally, quantum circuit complexity and entanglement are different quantities. However, for small values of circuit cost and entanglement, one can use the entanglement entropy to bound the circuit complexity. The argument presented in \cite{Eisert:2021mjg} is that quantum gates that are close to the identity performs little entanglement from product or entangled states. One of the interesting corollary presented in \cite{Eisert:2021mjg} is that whenever entanglement entropy grows linearly in time, the circuit complexity also grows linearly. Linear growth of entanglement entropy is a generic feature of several quenched many-body systems.

Our analysis of complexity and entanglement entropy of two-mode squeezed states is in agreement with the result in \cite{Eisert:2021mjg}. Since both entanglement entropy and circuit complexity computed with Covariance matrix method are independent of squeezing angle $\phi_k$, the comparison is clear than with Nielsen's method of wavefunctions. The explicit form of circuit complexity with covariance matrix method is obtained in eq \ref{eq:circuitComplexityCovariance}:
\begin{align}
    C_1(k) &= 4r_k\\
    C_2(k) &= 2\sqrt{2}r_k
\end{align}
 While comparing the form of entanglement entropy  eq. \ref{eq:entanglementEntropy} and \ref{eq:renyi-entropy eqn} with this circuit complexity, we get:
\begin{equation}
   C_1(r_k) =\sqrt{2} C_2(r_k) = 4r_k \geq S(r_k) \approx r_k
\end{equation}
In fig: \ref{complexityVsEntropy}, we have plotted the comparison between Von-Neumann entropy and Circuit complexity (computed using covariance matrix method). The circuit complexity $C_1$ and $C_2$ grows linearly just like entanglement entropy. Up to these distances, circuit complexity is indeed lower bounded by entanglement entropy. This result can have physical interpretation. Since entanglement entropy for the two mode squeezed states increases with increasing $r_k$, entanglement entropy from the vacuum to distant states is large. Therefore, in the context of two-mode squeezed states, with proper circuit complexity cost, entanglement entropy could be used as a measure of complexity.

So far, we have only compared circuit complexity obtained via Covariance approach. A more detailed numerical comparison of circuit complexity via Nielsen's approach with entanglement entropy and temperature will be discussed in the numerical analysis section.

\section*{\textcolor{Sepia}{\textbf{ \Large Numerical Results}}}
\label{sec:numerical}

In this section, we do the numerical analysis of the circuit complexity calculated for the model of "Black Hole gas". To provide a wholesome and physically relevant discussion, we do the analysis in terms of the scale factor. We begin by solving the evolution equations of the squeezed state parameters given in eqn: \ref{eq:evolution}. 

To recast the above differential equations and study the time evolution in terms of scale factor, a simple change of variable is implemented, which transforms the above equation. This change of variable is sometimes called as field redefinition. 

In fig: \ref{fig_t_1}, we have plotted the evolution of the squeezed state parameter with respect to the scale factor. The behaviour of the squeezed state parameter $r_{k}$ is crucial for understanding the behaviour of the circuit complexity and its evolution with the scale factor. From the behaviour of the squeezed state parameters, we see a widely different behaviour of the model in (1+1) dimensions i.e d=1 in the plots. The behaviour for the higher dimensions however looks to be pretty similar. The squeezing is large and growing at early times, however after a certain scale, the squeezing freezes and saturates at a constant value of squeezing. The increase in the squeezing grows up to a very large scale for spatial dimension 1 (d=1 in the plots) and the freezing of the squeezing effect is not observed even for extremely high scales. This makes the spatial dimension 1 markedly different from the higher spatial dimensions where the freezing effect in the squeezing is explicitly observed. 

In fig: \ref{fig_t_2}, we have plotted the squeezing angle $\phi_{k}$ with respect to the scale factor. For the model considered in this paper, we observe that for the spatial dimension 1, the squeezed angle rises for initial scales and is frozen and saturated at intermediate and late scales. However for higher dimension, the squeezed angle increases at the initial scales but  shows a fall after a certain characteristic scale.

In fig: \ref{fig_t_3} and fig: \ref{fig_t_4}, we have plotted the circuit complexity with respect to the scale factor calculated from the two different cost functionals using both Nielsen's and covariance approach. Let us make a comparative analysis of complexity obtained from Nielsen's and covariance approach. The structure of circuit complexity in the covariance approach has the similar pattern as the squeezing parameter $r_k$ in fig: \ref{fig_t_1} and has almost no feature coming out of the squeezing angle $\phi_k$ of fig: \ref{fig_t_2}. This makes sense as the circuit complexity obtained from the covariance approach is independent of the squeezing angle. Irrespective of the spatial dimension, the circuit complexity $C_1$ amd $C_2$ gradually increases and saturates after some values of $a$. 

In contrast to the covariance approach, Nielsen's approach gives a different story of circuit complexity. This is mainly due to the reasoning that the circuit complexity in Nielsen's approach is dependent on both squeezing parameters: $r_k$ and $\phi_k$ from fig: \ref{fig_t_1} and \ref{fig_t_2}. This gives one to look at the detail of evolution of wave-function uniquely. 
As already pointed out in the previous discussion, the speciality of the spatial dimension 1 can be clearly understood from the complexity plots as well. The initial rise in the complexity measures is observed irrespective of the spatial dimension though the scale factor upto which the rise is observed is influenced by the spatial dimension. With the increase in spatial dimension, the rise in the complexities is observed till the lower scale factors. After a critical value of the scale factor the complexity measures show a gradual fall in the values. This rate of fall is found to be extremely less for the spatial dimension 1 where even at large value of the scale factor, only a small fall in the value of the complexity is observed. It can be noted that for higher dimension, complexity measure falls off quickly. The faster a complexity measure falls to a certain minima, it starts to oscillate soon after as seen in the graph. For $d=3$ the oscillation starts early compared to $d=2$. Also, such oscillatory behaviour is saturated at higher value of scale factor. The oscillations in fig \ref{fig_t_3} and \ref{fig_t_3} for higher spatial dimensions could be a hint of the quantum gravity corrections in the very early universe in terms of vaccum fluctuations of "virtual black holes" of radii R. Such fluctuations could effectively resolve the cosmological constant puzzle. The 'vecro component' which describes the part of wavefunctional associated to virtual black hole fluctuations could alter the overall vaccum energy giving us an effective value of cosmological constant $\Lambda = (GR^{2})^{-1}$ and hence resolving the issue. The other way to look at the oscillation of complexity is that at the minimum complexity regions, the distance in initial and evolved states are low as one state can be evolved to next with low number of quantum gates while it is opposite in the maximum complexity regions. So, the structure of the wavefunction in lower complexity regions are more close to the intial states than the one in high complexity region.

In fig: \ref{fig_Svsa1}, fig: \ref{fig_Svsa2} and fig: \ref{fig_Svsa3}, we have plotted the behavior of entanglement entropy with respect to the scale factor. The two curves in the plots correspond to the two types of entanglement entropy we have considered in this paper $viz.$ von-Neumann entanglement entropy and Rényi entropy. Even though the overall behavior of both forms of entanglement entropy are identical, we still a minute difference. It can be seen that the von-Neumann entropy rises faster to a higher value compared to Rényi entropy. This feature is observed for all spatial dimensions. For spatial dimension $d=1$, we observe an increasing behavior of the entropy through the entire range of the scale factor. But for the spatial dimension $d=2,3$, we observe an initial increase in the entropy which then starts to oscillate with its amplitude decaying for higher value of scale factor. It can be noted that with rise in the number of spatial dimension, the rise in entropy decreases and hence saturates to a lower value. We would like to relate the entropy calculated from the squeezed state formalism with the entropy of the black hole gas. One can comment about the entropy of the black hole gas from the entropy calculated using the squeezed state formalism because the information about the black hole gas is itself encoded in the squeezed parameter $r_k$. To be more precise, the evolution equations for the squeezed state parameters written in eqn \eqref{eq:evolution} has been solved using the solution of the scale factor of the black hole gas model as the dynamical variable. Hence, the information about the black hole gas model propagates through the squeezed state parameters to any quantity we calculate. Thus, the entnaglement entropy calculated from the squeezed state parameter is intimately related with the entropy of the black hole gas model. 

In fig: \ref{fig_t_5}, we have plotted $dC_i/da$ computed with both Nielsen's and Covariance approach w.r.t the von Neumann entanglement entropy to inspect the validity of the conjectured relation proposed by Susskind between complexity and entanglement entropy. We observe that for the spatial dimension d=1, in the initial values of entanglement entropy, the behavior $dC_i/da$ shows an increasing behaviour. However, at the intermediate scales, $dC_i/da$ shows a sharp fall followed by a rise and saturation at large values of entanglement entropy. Thus we observe a non-linear relation between $\frac{dC_i}{da}$ and entropy. For low values of entanglement entropy, the difference in amplitude of $dC_1/da$ and $dC_2/da$ is higher in Nielsen's approach than in the covariance-approach. This can be because in Covariance approach $C_1$ and $C_2$ are related by $\mathcal{C}_1$= $\sqrt{2}\mathcal{C}_2= 4\sqrt{2}r_k$ while in Nielsen's approach $C_1$ and $C_2$ has complicated relation. \

In fig: \ref{fig_t_6} and fig: \ref{fig_t_7}, we study the behaviour of $dC_i/da$ with Von Neumann entanglement entropy for the spatial dimension d=2 and d=3.  We observe an almost identical behavior for the higher spatial dimensions with the behavior shown in spatial dimension 1.

In fig: \ref{fig_t_8}, fig: \ref{fig_t_9}, fig: \ref{fig_t_10}, we have plotted $dC_i/da$ vs Rényi entropy ($\mu=2$) for different spatial dimensions. It is observed that the overall behaviour of $dC_i/da$ with respect to the Rényi entropy is identical to what we observe in the Von-Neumann entanglement entropy case. This identical nature in the behaviour of $\frac{dC_i}{da}$ is observed for all spatial dimensions. 

In fig: \ref{fig_t_11}, \ref{fig_t_12}, \ref{fig_t_13} we have plotted the behaviour of the equilibrium temperature of the black hole gas with respect to $dC_i/da$ and entanglement entropy for the spatial dimension d=1,2,3. The reason we have plotted the behaviour of the equilibrium temperature with respect to $dC_i/da$ and the entanglement entropy on the same plot was to get an idea of how it behaves with two most important quantity in our analysis i.e $dC_i/da$ and entanglement entropy. The motivation came from Susskind's conjectured relation where he connected the rate of change of complexity with the entanglement entropy and the equilibrium temperature. However, instead of using $dC_i/dt$, we have used $dC_i/da$ as we have used the scale factor, which is cosmologically much more relevant quantity, as the dynamical variable of our analysis. 
The red curve in the plot shows the behaviour of the equilibrium temperature with respect to the entanglement entropy, whereas the black curve shows the behaviour with respect to $dC_i/da$. It is clearly evident that irrespective of the spatial dimension, the behaviour of the equilibrium temperature shows an increasing behaviour. One can approximate the behaviour as follows:
$$T \propto S^4$$. However, it can be seen that the behaviour of the equilibrium temperature is overall not identical in nature with $dC_i/da$, for different spatial dimension and the measure to compute complexity, though some of the features do match. In Nielsen's approach, it can be observed that for negative values of $dC_i/da$, for two values of $dC_i/da$ the black hole gas model attains same value of the equilibrium temperature. For spatial dimension d=1, in the intermediate and positive values of $dC_i/da$, the equilibrium temperature is almost constant, but for higher spatial dimension the multivalue nature of $dC_i/da$ with the equilibrium temperature returns. However, in the covariance approach in all three spatial dimensions, the behavior is identical.

Thus we see that irrespective of the spatial dimension and the approach of computing complexity in which the black hole model is considered, neither $dC_i/da$ nor entanglement entropy has a linear relationship with the equilibrium temperature.
\subsection*{\textcolor{Sepia}{\textbf{Quantum extremal islands vs Black hole gas}}}
In this portion, we are going to give a comparative analysis of the quantum extremal islands with the black hole gas model from the perspective of circuit complexity.
\begin{itemize}
	\item Circuit complexity calculated from the solution of Cosmological islands resembled the page curve in a specific parameter space \cite{Choudhury:2020hil} but for the black hole gas model we observe different behavior of the circuit complexity for different spatial dimension.
	\item In another parameter space the behaviour of the circuit complexity for the island model showed only a rising behavior which is also different from the one we observe for the black hole gas model. 
	\item The entanglement entropy predicted from the circuit complexity in the cosmological island model again resembled page curve in a particular parameter space and showed a decreasing behavior in another parameter space whereas for the black hole gas model, the entropy showed a increasing behavior for the spatial dimension 1 and 2 and an increasing behavior followed by an oscillation for the spatial dimension 3.
	\item The oscillatory behavior of the circuit complexity at large values of scale factor, which is observed for the higher spatial dimensions for the black hole gas model is absent in the cosmological island model, even when probed to very high scales.
\end{itemize}
\subsection*{\textcolor{Sepia}{\textbf{Comparative analysis of Circuit Complexity from Nielsen's method and Covariance matrix method}}}

\begin{center}
	\scalebox{0.8}{
		\begin{tabular}{|c | c | c |}
			\hline
			\thead{Parameters}&\thead{Covariance \\ approach} & \thead{Nielsen's \\ approach} \\
			\hline
			\makecell{Dependence on \\squeezing angle $\phi_k$}&\makecell{Does not depend \\ on $\phi_k$}  & \makecell{Depends on $\phi_k$} \\
			\hline
			\makecell{Dependence on \\squeezing \\parameter $r_k$}&\makecell{Always linearly\\dependent on\\squeezing \\parameter $r_k$}  & \makecell{May have\\ non-linear\\ dependence on\\ squeezing\\ parameter $r_k$} \\
			\hline
			\makecell{Sensitivity to\\ details of wavefunction}&\makecell{Since it is\\ independent of $\phi_k$, \\ it is not so\\ sensitive to the\\ details of the \\wavefunction }  & \makecell{Since it depends\\ on both $r_k$ and $\phi_k$\\ it is sensitive\\ to the \\details of the\\ wavefunction.} \\
			\hline
			\makecell{Limiting conditions}&\makecell{Only one condition\\ exists $\mathcal{C}_1$= $\sqrt{2}\mathcal{C}_2= 4\sqrt{2}r_k$}  & \makecell{$\mathcal{C}_1$ and $\mathcal{C}_2$\\ are vastly\\ different} \\
			\hline
			\makecell{Structure of \\Circuit complexity \\ in Black Hole\\ Gas model}&\makecell{For all spatial dimensions \\ it grows until certain $a$ \\ then, it saturates}  & \makecell{Depending on the\\ spatial dimension,\\ it can oscillate} \\
			\hline 
			\makecell{Comparison with\\ genral form of \\entanglement \\ entropy}&\makecell{Easier to\\ compare with \\entanglement entropy\\as both are\\independent of $\phi_k$}  & \makecell{Due to dependence\\ on $\phi_k$, it is\\difficult to \\compare with\\ entanglement entropy.} \\
			\hline
			\makecell{Entanglement entropy\\ and \\ Circuit Complexity \\in Black hole Gas \\ Model}&\makecell{Complexity has \\ the same growth\\ pattern as entanglement\\ entropy in all three\\spatial dimensions\\ which is expected.  } & \makecell{It is not trivial\\ to compare complexity\\ with entanglement\\ entropy.So, one\\ has to do \\ case by case analysis.} \\
			\hline
			\makecell{$dC_i/da$ and\\ Entanglement entropy($S$) \\in Black hole Gas \\ Model}&\makecell{The difference in \\ amplitude of\\ $dC_1/da$ and $dC_2/da$ \\is lower.  } & \makecell{The difference in \\ amplitude of\\ $dC_1/da$ and $dC_2/da$ \\is higher.} \\
			\hline
			\makecell{$dC_i/da$, Temperature\\ and entropy($S$) \\in Black hole Gas \\ Model}&\makecell{The behavior of \\Temperature with $dC_i/da$ \\and entropy is different\\ in three different \\spatial dimensions. } & \makecell{The behavior of \\temperature with\\ $dC_i/da$ and\\ entropy is same\\ in three different \\spatial dimensions.} \\
			\hline
	\end{tabular}}
\end{center}

\section*{\textcolor{Sepia}{\textbf{ \Large Conclusions}}}
\label{sec:Conclusions}

Through analysis of the black hole gas model from the perspective of circuit complexity and entanglement entropy using the language of squeezed state formalism we arrive at the following conclusions:
\begin{itemize}
	\item The circuit complexity computed from Nielsen's wave function approach provides a much better understanding than that computed from the covariance matrix method as it depends on both the squeezing angle and the squeezing parameter and hence can be related to the entanglement entropy.
	\item The behaviour of the circuit complexity for the spatial dimension d=1 is significantly different from higher spatial dimensions. Whereas complexity saturates or changes significantly slowly at large scale factors for d=1, it falls off rapidly and has an oscillatory behaviour for higher spatial dimensions. 
	\item The behaviour of the entanglement entropy w.r.t the scale factor for different spatial dimension shows different features. For d=1, it is just an increasing function whereas for d=2 and 3, we observe an oscillatory behavior with the frequency of oscillation increasing with the increase in spatial dimension.
	\item We observe that for no spatial dimensions the quantity $dC_i/da$ varies linearly with the Von-Neumann entanglement entropy or Rényi entropy. 
	\item For different spatial dimensions, the behaviour of the equilibrium temperature with $dC_i/da$ is peculiar and it is not possible to predict an approximate relation and one has to study different ranges of $dC_i/da$ separately to understand the behavior of the equilibrium temperature.
	\item For different spatial dimensions, from the behavior of equilibrium temperature with entanglement entropy, it can be understood that the relation between entanglement entropy and equilibrium temperature is not linear but goes as: $$ T \propto S^4$$.
	\item From the comparative analysis of the black hole gas model with that of the cosmological islands from the perspective of circuit complexity, we can conclude that Circuit complexity can be used as a useful tool to discover the underlying features of a model which are otherwise difficult to analyse.
	 
\end{itemize}

The future prospects of the work can be written:
\begin{itemize}
	\item Circuit complexity has been studied for thermofield double states \cite{Chapman:2018hou}. The process of thermalization can also be realized by a process known as quantum quench, where the states are expressed as the generalised Calabrese Cardy form. Hence one can explore the thermalization phenomenon using circuit complexity.
	\item People have studied circuit complexity as a deformation in the euclidean path integral for CFT's. This is mainly known as path integral optimization \cite{Camargo:2019isp}. However, these deformations appear in the context of cosmological perturbation theory as well and one can try to extend this circuit complexity using path integral optimization in de Sitter space. 
\end{itemize}

\textbf{Acknowledgement:}
~~~The research fellowship of SC is supported by the J.  C.  Bose National Fellowship of Sudhakar Panda.  SC also would line to thank School of Physical Sciences, National Institute for Science Education and Research (NISER),  Bhubaneswar for providing the work friendly environment.  SC would like to sincerely thank Professor Samir D.  Mathur from Ohio State University,  USA for various helpful discussions and suggesting the ref.~\cite{Mathur:2020ivc} regarding the Black Hole Gas.  SC also thank all the members of our newly formed virtual international non-profit consortium Quantum Structures of the Space-Time \& Matter (QASTM) for elaborative discussions. Kiran Adhikari would like to thank TTK, RWTH and JARA, Institute of Quantum Information for fellowships. Satyaki Chowdhury and K. Shirish would like to thank NISER Bhubaneswar and VNIT Nagpur respectively, for providing fellowships.  
Last but not least,  we would like to acknowledge our debt to the people belonging to the various part of the world for their generous and steady support for research in natural sciences.

\bibliography{referencesnew}
\bibliographystyle{utphys}

\end{document}